\documentclass[twocolumn]{aastex631}

\newcommand{\discminer}{\textsc{discminer}}
\newcommand{\disksurf}{\texttt{disksurf}}

\newcommand{\gofish}{\texttt{GoFish}}
\newcommand{\zr}{$\langle z / r \rangle$} 
\newcommand{\linmix}{\texttt{linmix}}
\newcommand\dvphi[1]{$\delta\upsilon_{\phi}$}

\newcommand{\maria}[1]{#1}

\newcommand{\mariatwo}[1]{#1}

\usepackage{multirow}
\usepackage{float}
\usepackage{makecell}
\usepackage{graphicx}
\usepackage{amsmath}
\usepackage{verbatim}

\begin{document}

\title{exoALMA V: Gaseous Emission Surfaces and Temperature Structures}

\author[0000-0002-5503-5476]{Maria Galloway-Sprietsma}
\affiliation{Department of Astronomy, University of Florida, Gainesville, FL 32611, USA}

\author[0000-0001-7258-770X]{Jaehan Bae}
\affiliation{Department of Astronomy, University of Florida, Gainesville, FL 32611, USA}

\author[0000-0001-8446-3026]{Andrés F. Izquierdo}
\altaffiliation{NASA Hubble Fellowship Program Sagan Fellow}
\affiliation{Department of Astronomy, University of Florida, Gainesville, FL 32611, USA}
\affiliation{Leiden Observatory, Leiden University, P.O. Box 9513, NL-2300 RA Leiden, The Netherlands}
\affiliation{European Southern Observatory, Karl-Schwarzschild-Str. 2, D-85748 Garching bei München, Germany }

\author[0000-0002-0491-143X]{Jochen Stadler}
\affiliation{Universit\'{e} C\^{o}te d'Azur, Observatoire de la C\^{o}te d'Azur, CNRS, Laboratoire Lagrange, France}


\author[0000-0003-4663-0318]{Cristiano Longarini}
\affiliation{Institute of Astronomy, University of Cambridge, Madingley Rd, CB30HA, Cambridge, UK }
\affiliation{Dipartimento di Fisica, Universit\`a degli Studi di Milano, Via Celoria 16, 20133 Milano, Italy}

\author[0000-0003-1534-5186]{Richard Teague}
\affiliation{Department of Earth, Atmospheric, and Planetary Sciences, Massachusetts Institute of Technology, Cambridge, MA 02139, USA}

\author[0000-0003-2253-2270]{Sean M. Andrews}
\affiliation{Center for Astrophysics | Harvard \& Smithsonian, Cambridge, MA 02138, USA}

\author[0000-0002-7501-9801]{Andrew J. Winter}
\affiliation{Universit\'{e} C\^{o}te d'Azur, Observatoire de la C\^{o}te d'Azur, CNRS, Laboratoire Lagrange, France}

\affiliation{Max-Planck Institute for Astronomy (MPIA), Königstuhl 17, 69117 Heidelberg, Germany}

\author[0000-0002-7695-7605]{Myriam Benisty}
\affiliation{Universit\'{e} C\^{o}te d'Azur, Observatoire de la C\^{o}te d'Azur, CNRS, Laboratoire Lagrange, France}
\affiliation{Max-Planck Institute for Astronomy (MPIA), Königstuhl 17, 69117 Heidelberg, Germany}

\author[0000-0003-4689-2684]{Stefano Facchini}
\affiliation{Dipartimento di Fisica, Universit\`a degli Studi di Milano, Via Celoria 16, 20133 Milano, Italy}

\author[0000-0003-4853-5736]{Giovanni Rosotti}
\affiliation{Dipartimento di Fisica, Universit\`a degli Studi di Milano, Via Celoria 16, 20133 Milano, Italy}

\author[0000-0001-9319-1296]{Brianna Zawadzki}
\affiliation{Department of Astronomy, Van Vleck Observatory, Wesleyan University, 96 Foss Hill Drive, Middletown, CT 06459, USA}
\affiliation{Department of Astronomy \& Astrophysics, 525 Davey Laboratory, The Pennsylvania State University, University Park, PA 16802, USA}

\author[0000-0001-5907-5179]{Christophe Pinte}
\affiliation{Univ. Grenoble Alpes, CNRS, IPAG, 38000 Grenoble, France}
\affiliation{School of Physics and Astronomy, Monash University, VIC 3800, Australia}

\author[0000-0003-4679-4072]{Daniele Fasano}
\affiliation{Universit\'{e} C\^{o}te d'Azur, Observatoire de la C\^{o}te d'Azur, CNRS, Laboratoire Lagrange, France}




\author[0000-0001-6378-7873]{Marcelo Barraza-Alfaro}
\affiliation{Department of Earth, Atmospheric, and Planetary Sciences, Massachusetts Institute of Technology, Cambridge, MA 02139, USA}


\author[0000-0002-2700-9676]{Gianni Cataldi} 
\affiliation{National Astronomical Observatory of Japan, Osawa 2-21-1, Mitaka, Tokyo 181-8588, Japan}

\author[0000-0003-3713-8073]{Nicolás Cuello} 
\affiliation{Univ. Grenoble Alpes, CNRS, IPAG, 38000 Grenoble, France}

\author[0000-0003-2045-2154]{Pietro Curone} 
\affiliation{Dipartimento di Fisica, Università degli Studi di Milano, Via Celoria 16, 20133 Milano, Italy}
\affiliation{Departamento de Astronom\'ia, Universidad de Chile, Camino El Observatorio 1515, Las Condes, Santiago, Chile}



\author[0000-0002-1483-8811]{Ian Czekala}
\affiliation{School of Physics \& Astronomy, University of St. Andrews, North Haugh, St. Andrews KY16 9SS, UK}

\author[0000-0002-9298-3029]{Mario Flock} 
\affiliation{Max-Planck Institute for Astronomy (MPIA), Königstuhl 17, 69117 Heidelberg, Germany}

\author[0000-0003-1117-9213]{Misato Fukagawa} 
\affiliation{National Astronomical Observatory of Japan, 2-21-1 Osawa, Mitaka, Tokyo 181-8588, Japan}

\author[0009-0003-8984-2094]{Charles H. Gardner} 
\affiliation{Department of Physics and Astronomy, Rice University, 6100 Main St, Houston, TX 77005, USA}
\affiliation{Los Alamos National Laboratory, Los Alamos, NM 87545, USA}

\author[0000-0002-5910-4598]{Himanshi Garg}
\affiliation{School of Physics and Astronomy, Monash University, Clayton VIC 3800, Australia}

\author[0000-0002-8138-0425]{Cassandra Hall} 
\affiliation{Department of Physics and Astronomy, The University of Georgia, Athens, GA 30602, USA}
\affiliation{Center for Simulational Physics, The University of Georgia, Athens, GA 30602, USA}
\affiliation{Institute for Artificial Intelligence, The University of Georgia, Athens, GA, 30602, USA}

\author[0000-0001-6947-6072]{Jane Huang} 
\affiliation{Department of Astronomy, Columbia University, 538 W. 120th Street, Pupin Hall, New York, NY, USA}

\author[0000-0003-1008-1142]{John~D.~Ilee} 
\affiliation{School of Physics and Astronomy, University of Leeds, Leeds, UK, LS2 9JT}


\author[0000-0001-7235-2417]{Kazuhiro Kanagawa} 
\affiliation{College of Science, Ibaraki University, 2-1-1 Bunkyo, Mito, Ibaraki 310-8512, Japan}

\author[0000-0002-8896-9435]{Geoffroy Lesur} 
\affiliation{Univ. Grenoble Alpes, CNRS, IPAG, 38000 Grenoble, France}


\author[0000-0002-2357-7692]{Giuseppe Lodato} 
\affiliation{Dipartimento di Fisica, Università degli Studi di Milano, Via Celoria 16, 20133 Milano, Italy}

\author[0000-0002-8932-1219]{Ryan A. Loomis}
\affiliation{National Radio Astronomy Observatory, 520 Edgemont Rd., Charlottesville, VA 22903, USA}

\author[0000-0002-1637-7393]{Francois Menard} 
\affiliation{Univ. Grenoble Alpes, CNRS, IPAG, 38000 Grenoble, France}


\author[0000-0003-4039-8933]{Ryuta Orihara} 
\affiliation{College of Science, Ibaraki University, 2-1-1 Bunkyo, Mito, Ibaraki 310-8512, Japan}


\author[0000-0002-4716-4235]{Daniel J. Price} 
\affiliation{School of Physics and Astronomy, Monash University, Clayton VIC 3800, Australia}




\author[0000-0002-3468-9577]{Gaylor Wafflard-Fernandez} 
\affiliation{Univ. Grenoble Alpes, CNRS, IPAG, 38000 Grenoble, France}


\author[0000-0003-1526-7587]{David J. Wilner} 
\affiliation{Center for Astrophysics | Harvard \& Smithsonian, Cambridge, MA 02138, USA}

\author[0000-0002-7212-2416]{Lisa W\"olfer} 
\affiliation{Department of Earth, Atmospheric, and Planetary Sciences, Massachusetts Institute of Technology, Cambridge, MA 02139, USA}

\author[0000-0003-1412-893X]{Hsi-Wei Yen} 
\affiliation{Academia Sinica Institute of Astronomy \& Astrophysics, 11F of Astronomy-Mathematics Building, AS/NTU, No.1, Sec. 4, Roosevelt Rd, Taipei 10617, Taiwan}

\author[0000-0001-8002-8473	]{Tomohiro C. Yoshida} 
\affiliation{National Astronomical Observatory of Japan, 2-21-1 Osawa, Mitaka, Tokyo 181-8588, Japan}
\affiliation{Department of Astronomical Science, The Graduate University for Advanced Studies, SOKENDAI, 2-21-1 Osawa, Mitaka, Tokyo 181-8588, Japan}


\begin{abstract}

Analysis of the gaseous component in protoplanetary disks can inform us about their thermal and physical structure, chemical composition, and kinematic properties, all of which are crucial for understanding various processes within the disks.
By exploiting the asymmetry of the line emission, or via line profile analysis, we can locate the emitting surfaces. 
Here, we present the emission surfaces of the exoALMA sources in $^{12}$CO $J=3-2$, $^{13}$CO $J=3-2$, and CS $J=7-6$. We find that $^{12}$CO traces the upper disk atmosphere, with mean \zr{} values of $\approx$ 0.28, while $^{13}$CO and CS trace lower regions of the disk with mean \zr{} values of $\approx$ 0.16 and $\approx$ 0.18, respectively. 
We find that $^{12}$CO \zr{} and the disk mass are positively correlated with each other; this relationship offers a straightforward way to infer the disk mass.
We derive 2-D $r-z$ temperature distributions of the disks. Additionally, we search for substructure in the surfaces and radial intensity profiles; we find evidence of localized substructure in the emission surfaces and peak intensity profiles of nearly every disk, with this substructure often being co-incident between molecular tracers, intensity profiles, and kinematic perturbations. Four disks display evidence of potential photo-desorption, implying that this effect may be common even in low FUV star-forming regions. For most disks, we find that the physical and thermal structure is more complex than analytical models can account for, highlighting a need for more theoretical work and a better understanding of the role of projection effects on our observations.

\end{abstract}

\keywords{Protoplanetary disks (1300) — Planet formation (1241) — High angular resolution (2167) - CO Line emission (262)}

\section{Introduction}\label{sec:intro}

Planets are born within the gaseous and dusty environments of circumstellar disks. Despite the many observations of fully formed planets, our knowledge of the preceding planet formation process is still largely incomplete. In order to best understand this process, we must first characterize the physical and chemical properties of protoplanetary disks. In recent years, the high angular and spectral resolution offered by \maria{the} Atacama Large Millimeter/submillimeter Array (ALMA) has allowed us to map planetary nurseries in deeper detail than ever before. In particular, molecular line observations of protoplanetary disks can inform us about the gas structure, which holds a wealth of information about the disk and potential planets forming within \citep{miotello_2023ASPC..534..501M}.

Protoplanetary disks are composed of two main components: dust and gas. The \maria{millimeter and} larger dust grains generally settle towards the midplane due to gravity exerted by the central star, aerodynamical gas drag, and a vertical pressure gradient \citep{2005A&A...443..185B}. 
In contrast, the bulk of the gaseous disk component extends beyond the dust both radially and vertically \mariatwo{\citep[e.g.,][]{ansdell_2018ApJ...859...21A, MAPS_emission_surfaces_2021ApJS..257....4L, long_2022ApJ...931....6L}}.
Although the gas is primarily composed of radio-quiet molecular hydrogen (H$_2$), there are trace amounts of other molecules that we can readily observe with ALMA.
Because these molecules have differing opacities, observations targeting various molecular tracers allow us to probe the 2-D structure of the disk, from the upper atmospheres down to the midplane.
The vertical and radial distribution of these molecules is dependent on the physical conditions found within the disk, particularly the temperature, density, and radiation \citep{2001A&A...377..566V, walsh_2015A&A...582A..88W, oberg_2023ARA&A..61..287O}.
We can directly measure the vertical emission surfaces for the front and back sides of the disk via several methods. \maria{The vertical extent of the surfaces can be directly probed in edge-on disks, as in \citealt{dutrey_2017A&A...607A.130D}, \citealt{villenave_2020A&A...642A.164V}, \citealt{duchene_2024AJ....167...77D}, and \citealt{villenave_2024ApJ...961...95V}}.
For mid-inclination disks (generally 30$^{\circ}$ - 60$^{\circ}$), the geometry of the disk can be exploited to determine isovelocity contours in each channel. The front- and back-side emission will be well-separated for disks of these intermediate inclinations, \maria{thus allowing us access to the full 3-D structure covering the entire azimuth;}
see \cite{dartois_2003A&A...399..773D, pinte_2018A&A...609A..47P, disksurf_2021JOSS....6.3827T, paneque_2023A&A...669A.126P}.
Line-profile analysis also allows for a measurement of the emission surfaces by identifying the multiple components of the line from the front and back sides of the disk \citep{discminer1_2021A&A...650A.179I, discminer2_2023A&A...674A.113I}. These techniques can identify CO snow lines \citep{pinte_2018A&A...609A..47P}, map the 2-D structure of disks \citep{MAPS_emission_surfaces_2021ApJS..257....4L, rich_2021ApJ...913..138R, law_2022ApJ...932..114L, paneque_2023A&A...669A.126P, stapper_2023A&A...669A.158S, hernandez_2024ApJ...967...68H},  identify substructures in emission surfaces that relate to disk physical processes \citep{teague_2021ApJS..257...18T, paneque_2022A&A...666A.168P, galloway_2023ApJ...950..147G, discminer2_2023A&A...674A.113I, urbina_2024A&A...686A.120U, paneque_2025arXiv250108294P}, and infer turbulence levels at multiple vertical locations \citep{paneque_carreno_2024A&A...684A.174P}.
Importantly, knowing the radial and vertical location of molecular line emission is critical to interpreting the kinematic signatures in the gas (see review by \citealt{pinte_review_2023ASPC..534..645P}). 

The ALMA Large Program exoALMA aims to detect plants forming in-situ at sub-millimeter wavelengths, utilizing kinematic planet-detection techniques. To achieve those goals, fifteen protoplanetary disks were observed at high angular and spectral resolution to target the molecular line emission of $^{12}$CO $J=3-2$, $^{13}$CO $J=3-2$, and CS $J=7-6$. In this paper, we present the emission surfaces, intensity radial profiles, and two-dimensional temperature profiles of the fifteen sources. In Section \ref{sec:obs}, we describe the calibration and imaging of the data utilized in this paper. In Section \ref{sec:surfaces}, we outline the methods used to derive the emission surfaces and the results. In Section \ref{sec:temps}, we detail the methods used to obtain radial temperature profiles and the subsequent 2-D temperature profile calculation.
In Section \ref{sec:substucture} we identify substructures in the surfaces and radial profiles, and compare these to continuum structures and kinematic perturbations.
We discuss the results and implications of this study in Section \ref{sec:discussion}, and summarize our work in Section \ref{sec:conclusions}.

\section{Data} \label{sec:obs}
The data used for this analysis were obtained for exoALMA\footnote{ \url{https://www.exoalma.com}} , an ALMA Large Program (2021.1.01123.L; \citealt{exoALMA_main}).
This program observed 15 protoplanetary disks, most with previously observed continuum substructures indicative of ongoing planet formation processes. Observations of $^{12}$CO $J=3-2$, 
$^{13}$CO $J=3-2$, and CS $J=7-6$, at 0$\farcs$1 angular resolution and a 35.5 kHz (26 m~s$^{-1}$) spectral resolution with a sensitivity of 3 K over 150 m s$^{-1}$ were requested to ensure a detection of line emission in the outer disk where we expect the gas temperature, T$_B$, to be $\approx$ 20 K for all molecules. 
A complete description of observations, source selection, and imaging techniques used are described in \cite{exoALMA_main} and \cite{exoALMA_imaging}. 
Figure \ref{fig:fnu} presents the exoALMA disk sample in $^{12}$CO $J=3-2$ moment eight maps, depicting the peak intensity. These moment maps were made using \verb|bettermoments| \citep{bettermoments_2018zndo...1419754T}, \maria{with the quadratic method and assuming the full Planck function.} The Fiducial Images are shown in in blue contour ($0\farcs15$, 100 m s$^{-1}$ resolution) and
the High Surface Brightness Sensitivity Images in the background in gray ($0\farcs30$, 100 m s$^{-1}$ resolution). A 5$\sigma$ signal-to-noise ratio (SNR) clip has been applied.

\section{Emission Surfaces} \label{sec:surfaces}
In this section, we discuss the method used to constrain the emission surfaces of the fifteen target disks and then present the results. 
\maria{The non-parametric surfaces discussed represent the radial and vertical points, and the parametric surfaces represent the tapered power law fits.}
We report surfaces for the three target molecules: $^{12}$CO $J=3-2$, $^{13}$CO $J=3-2$, and CS $J=7-6$. 

\subsection{Non-Parametric Surfaces} \label{subsec:non_para_surfaces}

To derive the non-parametric surfaces, we employ \disksurf{}\footnote{\url{https://github.com/richteague/disksurf}}  \citep{disksurf_2021JOSS....6.3827T}, which utilizes the method outlined in \cite{pinte_2018A&A...609A..47P}. This method relies on the asymmetry of the line emission above the disk midplane to  infer an emission height, which allows us to locate emission arising from specific locations in the disk. Using this method, we obtain the deprojected radius $r$ and emission height $z$, along with the brightness temperature $T_B$ and the SNR for each point which will be used to derive two-dimensional temperature structure later in Section \ref{sec:temps}. 

Unless otherwise specified, all of the non-parametric surfaces \maria{were derived using} the non-continuum subtracted cubes with 100 m/s resolution and 0\farcs15 beamsize for $^{12}$CO and $^{13}$CO, and the continuum-subtracted cubes with 200 m~s$^{-1}$ resolution and 0\farcs15 beamsize for CS. One exception to this is DM Tau, where we use the  0\farcs30 beamsize cubes for both $^{12}$CO and $^{13}$CO surface extraction to \mariatwo{retrieve the emission surface out to a larger radial extent due to the low SNR in the 0\farcs15 cubes (see DM Tau in Figure \ref{fig:fnu}).}

We opt for the non-continuum subtracted cubes for $^{12}$CO and $^{13}$CO to ensure we retrieve accurate temperature measurements (discussed more in Section \ref{sec:temps}). We found no significant differences between the surface structure retrieved from continuum subtracted versus non-continuum subtracted cubes for $^{12}$CO and $^{13}$CO. However, the CS emission tends to have comparable intensities to the continuum, affecting the surface retrievals.
In some cases, the continuum emission was confused for the gaseous emission; therefore, we opt to use the continuum subtracted cubes for the CS surfaces. We use geometric parameters (M$_{*}$, $i$, ${\rm PA}$, $v_{\rm LSR}$, x$_0$, y$_0$) derived from \discminer{} (see \citealt{exoALMA_main, exoALMA_line_profiles}). 

We employ several methods to improve emission surfaces obtained with \disksurf{} by using its updated functionalities. First, we use the iterative surface-finding approach, 
which relies on a previously derived surface and rotation curve, $z(r)$ and $v_{\phi}(r)$.
This surface and rotation curve is then used to draw a mask for the subsequent iteration, \maria{using a specified beamsize and signal-to-noise (SNR). The SNR for each point is measured as the intensity divided by the cube RMS.}
We repeat the iteration five times for each surface, allowing us to \maria{decrease the maximum beamsize from 2 beams for the first iteration to 1 beam on the subsequent iterations}, and increase minimum signal-to-noise (SNR) linearly from 1.0 to 5.0 for $^{12}$CO. For $^{13}$CO we increased the SNR from 1.0 to 3.5, and for CS from 1.0 to 3.0. \maria{Therefore, the final SNR clips for the surfaces are 5.0 for $^{12}$CO, 3.5 for $^{13}$CO, and 3.0 for CS.} This iterative approach does not allow for surface points to fall below z=0; therefore, we do not use it for the back surfaces, and all front surfaces are masked below this point. Despite this, when comparing the surfaces retrieved from the non-iterative versus iterative, there is not significant differences in the height of the surfaces.
We apply additional masks to cut out visually identified noise, as well as points that belong to more diffuse CO emission in the outer disk which does not clearly belong to either the front or back surfaces (see Section \ref{sec:diffuse_surfaces}). These steps were done on a per-disk basis, and were visually justified by looking at the channel maps with the detected points overlaid. The individual $r-z$ points derived via this process comprise the non-parametric surfaces. These surfaces are presented in Figure \ref{fig:binned_surfaces} as both raw and binned datapoints; \maria{all points are for the front surfaces only}. \maria{The points are binned by a quarter of a beamsize.} For individual molecular surfaces, see Figures \ref{fig:surfaces_12co}, \ref{fig:surfaces_13co}, and \ref{fig:surfaces_cs} in Appendix \ref{sec:emission_surf_extra}. 
We have excluded J1604 and HD 135344B from this analysis, as they are too face-on to reliably separate the front and back surfaces using the \disksurf{} method.

\subsection{Parametric Fitting} \label{subsec:parametric_surfaces}
We find a parametric surface using both \disksurf{} and \discminer{}\footnote{\url{https://github.com/andizq/discminer}} (see \citealt{discminer1_2021A&A...650A.179I, discminer2_2023A&A...674A.113I}). \discminer{} fits a Keplerian model to each channel, and has been used to derive the geometric parameters for each exoALMA source \citep{exoALMA_line_profiles}. We assume a tapered power law analytic prescription as the surface model, given by

\begin{equation}\label{eqn:emission_surface}
    z(r) = z_0 \, \left( \frac{r}{100~ \mathrm{ au}} \right)^{\psi}  \exp \left( -\left[ \frac{r}{r_{\rm taper}} \right]^{q_{\rm taper}} \right),
\end{equation}
where \maria{$z_0$ is the normalization factor of the surface at 100 au}, $\psi$ is the power-law exponent, \mariatwo{$r_{taper}$} is the characteristic radius for the exponential taper, and \mariatwo{$q_{taper}$} is the exponent of the taper term. 

This tapered power law model is fit to the non-parametric surfaces using the Markov Chain Monte Carlo (MCMC) method from \disksurf{}, which uses \texttt{emcee} \citep{emcee_2013PASP..125..306F}, a Python implementation of an Affine Invariant Ensemble sampler. We employ 256 walkers, 100 burn-in steps, and 1000 steps to obtain a fit to the non-parametric surface data points. \maria{The best-fit surfaces represent the 50th percentile parameters from the posterior distributions, and the errors are obtained from the 16th and 84th percentiles}.

\begin{figure*}[t]
    \centering
    \includegraphics[width=1.0\textwidth]{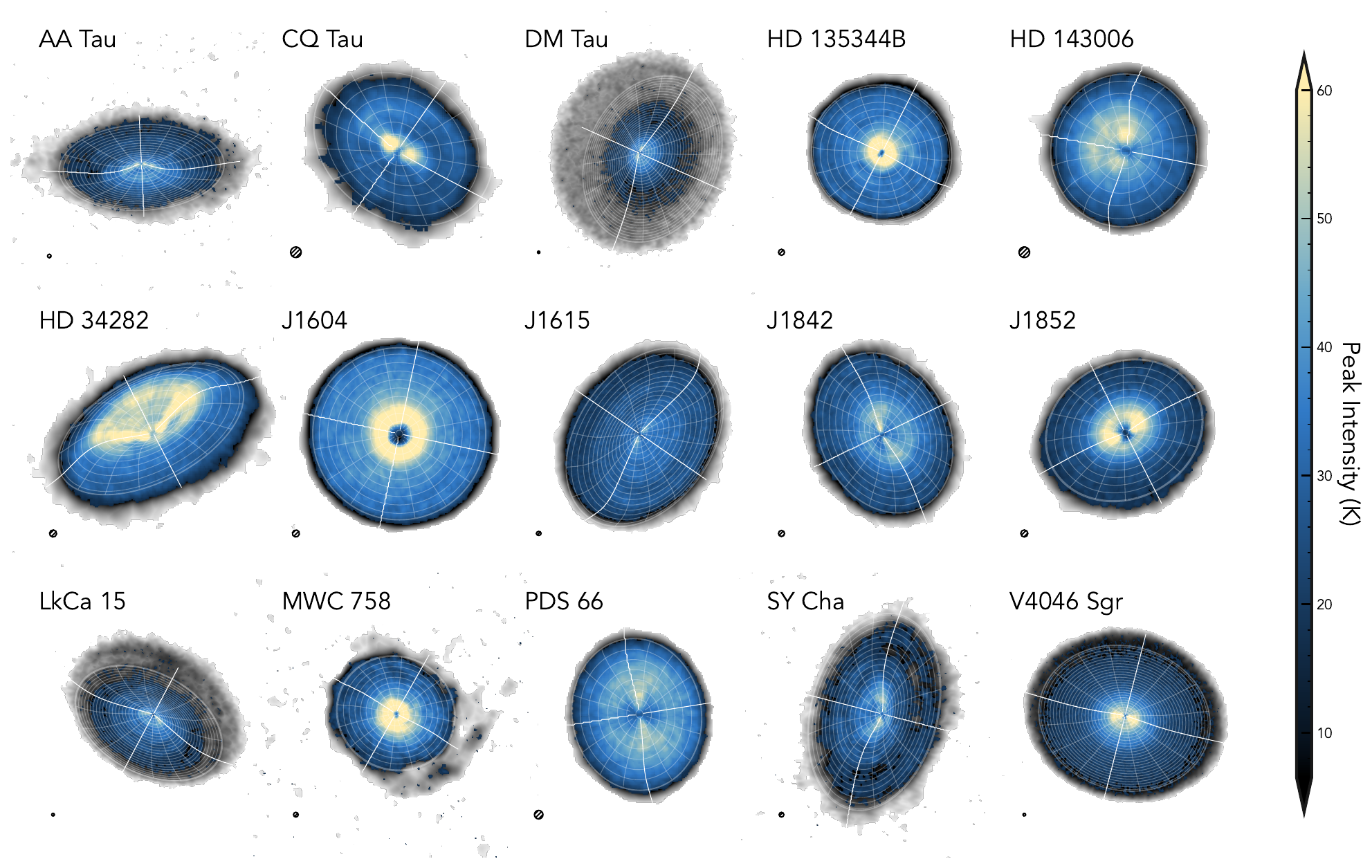}
    \caption{Peak intensity maps of $^{12}$CO $J=3-2$ for all exoALMA sources made with the Fiducial Images. Parametric tapered power-law surfaces from \discminer{} are overlaid in white, with each contour being separated by the beamsize (0$\farcs$15); the bold lines represent the major and minor axes. The synthesized beam is shown in the lower left corner of each plot. The 0$\farcs$30 beam cubes are plotted in the background in gray.}
    \label{fig:fnu}
\end{figure*}

\begin{figure*}[t]
    \centering
    \includegraphics[width=1.0
\textwidth]{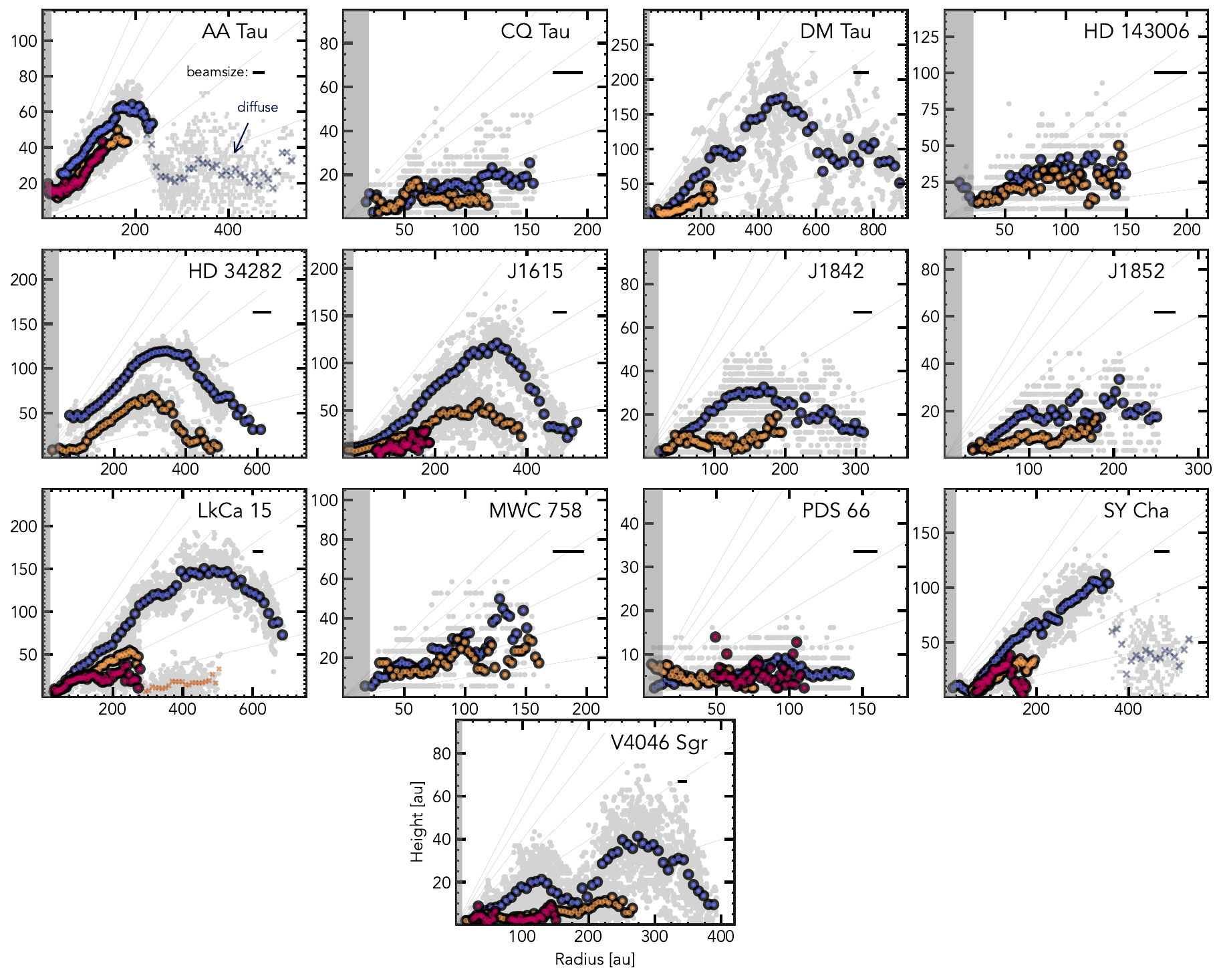}
    \caption{Raw and binned front surfaces for disks with viable emission surfaces in $^{12}$CO J=3-2 (blue), $^{13}$CO J=3-2 (orange), and CS $J=7-6$ (purple). The binned diffuse backside points of AA Tau, SY Cha, and LkCa 15 are shown as `x' points. Lines of constant \zr{} from 0.1 to 0.6 are shown as grey background lines. The beamsize is shown in the upper right corner of each plot; the gray shaded region corresponds to everything interior to one beamsize. 
    }
    \label{fig:binned_surfaces}
\end{figure*}

We present $z_0$, $r_{taper}$, $\psi$, and $q_{taper}$, derived from \disksurf{} and \discminer{}, in Table \ref{tab:parametric_surfaces}. 
Figure \ref{fig:fnu} presents the \discminer{} tapered power law fits overlaid on peak intensity maps of each disk to demonstrate the 3-D geometry of the surfaces. Figures \ref{fig:surfaces_12co}, \ref{fig:surfaces_13co}, and \ref{fig:surfaces_cs} in Appendix \ref{sec:emission_surf_extra} compare the parametric surfaces from \disksurf{} and \discminer{}, overlaid on the non-parametric surfaces.
The differences between the parametric surfaces are due to the underlying methodological differences between \disksurf{} and \discminer{}; the \disksurf{} parametric surfaces are fit to the data points extracted, which are shown in the plot, while \discminer{} fits a model to the line profiles. In some cases, the difference may be due to backside contamination; \discminer{} can readily distinguish between emission stemming from the front versus back surface via line profile modeling, but \disksurf{} relies on finding peaks along isovelocity contours. Additionally, if there is a large deviation from Keplerian rotation, \discminer{} takes this into account during modeling, but \disksurf{} assumes purely circular motion. However, the definite underlying cause of the differences is still unknown; a comparison of the two codes is discussed in Appendix \ref{sec:app_disksurfvsdiscminer}.
Despite the slight parametric differences, both surfaces are broadly consistent.

\subsection{Emission Surface Results} \label{subsec:results_surfaces}

We obtain non-parametric emission surfaces in $^{12}$CO and $^{13}$CO $J=3-2$ for thirteen out of the fifteen exoALMA sources, and six non-parametric emission surfaces in CS $J=7-6$. 
Although emission was detected for each molecule from each disk, we are only able to characterize a surface for those with high enough SNR.
The two disk surfaces we do not report belong to HD 135344B and J1604, both of which are too face-on to retrieve accurate emission heights using the \disksurf{} and \discminer{} methods.
The raw and binned surfaces for each molecule are presented in Figure \ref{fig:binned_surfaces}. 
The surfaces plotted individually by line are shown in Figures \ref{fig:surfaces_12co}, \ref{fig:surfaces_13co}, and \ref{fig:surfaces_cs}. Figure \ref{fig:rolling_surfaces} plots all rolling surfaces together for comparison, where the rolling surface represents a moving average of the binned datapoints. We find that in most disks, the $^{12}$CO surface lies above that of the $^{13}$CO surface, and the $^{13}$CO surface generally lies above that of CS surface, which is expected as reaching an optical depth of $\sim$1 will happen at deeper layers for less abundant molecules, and different molecules will be distributed differently throughout the disk. However, this is not true of all disks; in particular, the $^{13}$CO and CS molecular surfaces of AA Tau, J1615, LkCa 15, SY Cha, and V4046 Sgr overlap and appear to be stemming from similar heights, at least in the inner disk regions (e.g., $\lessapprox$~150 au).
In nearly all disks, the $^{12}$CO surfaces extend radially further out than the $^{13}$CO and CS surfaces, which is due to the higher SNR achieved with $^{12}$CO in those regions, as well as the intrinsic distribution of molecules due to the temperature and density. DM Tau hosts the largest $^{12}$CO surface in both height and radius, extending out to $\sim$850 au and reaching a height of $\sim$250 au, although we note most of the emission in the outer disk is low-SNR and scattered. PDS 66 has the lowest surface among the molecules detected; it is consistent with a mostly flat surface (e.g., consistent with an origin in the disk midplane) in all molecules and extends out to \maria{a radius of} $\sim$150 au. \maria{While nearly all surfaces follow the expected power-law or tapered power-law shape, V4046 Sgr is a notable exception with a double-peaked emission surface in $^{12}$CO. This will be discussed later in Section \ref{sec:discussion}.}

We calculate a representative \zr{} for each surface by taking an average of the binned surface points. 
We cut points past the peak of the emission surface, thus excluding those that belong to the tapered portion of the surfaces. This differs slightly from the method employed in previous works such as \cite{MAPS_emission_surfaces_2021ApJS..257....4L} or \cite{paneque_2023A&A...669A.126P} who use 0.8 $\times$ r$_{taper}$ to make the radial cut. We opt to use the radial location of the peak of the surface due to a few surface fits having r$_{taper}$ values interior to most of the rising emission; these two methods do not give discrepant results and the method employed here still captures only the rising portion of the emission surfaces, which we verified by eye. \maria{We additionally exclude any data interior to one beamsize in the \zr{} calculation.}
We have verified that the \zr{}  value is consistent when measured at the same radial location in each disk (125 au), and is not inherently influenced by the radial extent of the disk.

We find \zr{} values for $^{12}$CO emission ranging from $\sim$0.13 to $\sim$0.4, with an overall average \zr{} of 0.28. AA Tau hosts the most elevated $^{12}$CO surface with a \zr{} of 0.41, whereas PDS 66 hosts the flattest one, with a \zr{} of 0.12. The $^{13}$CO surfaces range in \zr{} from 0.05 (excluding PDS 66 which exhibits a completely flat surface in $^{13}$CO) to 0.32, with an average of 0.16. The $^{13}$CO surfaces generally do not exhibit a taper in the outer disk, likely because the SNR is too low in this portion of the disks. We retrieve CS emission surfaces for six disks. The remaining disks had too low SNR to retrieve a viable surface. 
Of these six, two are consistent with a flat surface close to \textit{z} = 0 (PDS 66 and V4046 Sgr). The remaining CS surfaces are elevated, and that of SY Cha even exhibits the characteristic tapered emission surface. The average \zr{} for CS is 0.18, with AA Tau having the largest at $\sim$0.32. 
In general, the CS surfaces that we can detect are elevated, but span smaller radii due to low abundance and/or low SNR in the outer disk. We note that this result is dictated by the SNR we are able to achieve. 

The error on each individual \textit{r, z} point is dependent on several factors, mostly stemming from the \maria{SNR} and the geometric parameters M$_{*}$, $i$, ${\rm PA}$, $v_{\rm LSR}$, x$_0$, and y$_0$ \citep[see][]{andrews_2024ApJ...970..153A}, derived in \cite{exoALMA_line_profiles}. \maria{Of course, a lower SNR will lead to higher errors, due to the increased difficulty of correctly identifying isovelocity contours. The effect of low SNR will lead to scatter in the surface, or non-detection entirely.}
\maria{Of the geometric parameters,} inclination is one of the largest contributors to the error of the emission surfaces.
By nature, disks that are more face-on will have a larger scatter in the z direction, due to the fact that the front and back surfaces become less separated from our viewing angle. The molecular layers of low-inclination sources will therefore also be less distinguishable, causing layers to appear to be overlapping.
Disks with large-scale non-Keplerian motion also introduce error, as this emission does not follow the traditional `butterfly' pattern seen in channel maps, and can lead to emission being incorrectly vertically classified. While we do not report explicit errors on each \textit{r, z} point, the amount of scatter in the surface reflects the confidence of the emission surfaces. This scatter can be seen as the gray background points in Figures \ref{fig:binned_surfaces}, \ref{fig:surfaces_12co}, \ref{fig:surfaces_13co}, and \ref{fig:surfaces_cs}. This is also reflected in the upper and lower shaded regions of the rolling surfaces, which represent the standard deviation within each bin, shown in Figure \ref{fig:rolling_surfaces}. In particular, the surfaces of HD 143006, J1852, MWC 758, and PDS 66 exhibit large vertical scatter due to either a face-on orientation or large-scale non-Keplerian motion, or a combination of the both.

\section{Temperature Radial Profiles \& 2-D Structure} \label{sec:temps}

\maria{In Section \ref{subsec:disk_size} we describe how the gas disk size was determined.}
In Section \ref{sec:gas_temps} we outline the methods used to obtain the azimuthally averaged temperature profiles and the gas temperatures for each emission surface point. In Section \ref{subsec:met_temp_structure} we discuss how we fit an analytical model to the temperatures to obtain a 2-D temperature distribution for each disk. 

\subsection{Gas Disk Size}\label{subsec:disk_size}
\begin{figure}[!ht]
    \includegraphics[width=0.44\textwidth]{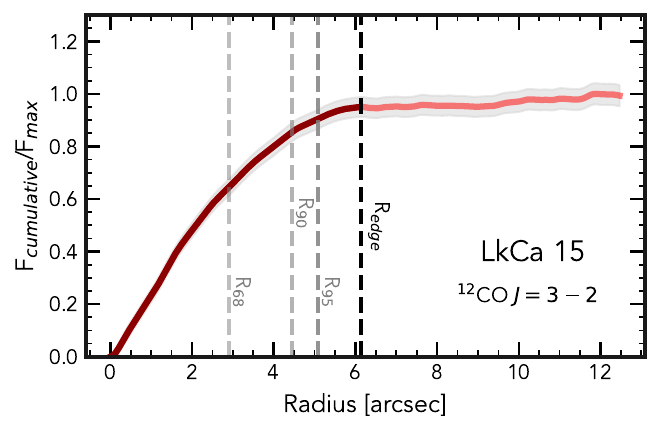}
    \caption{Normalized integrated intensity for LkCa 15. The dashed vertical lines show R$_{\rm edge}$, R$_{\rm 95}$, R$_{\rm 90}$, and R$_{\rm 68}$. Only intensity inward of R$_{\rm edge}$ is used in the calculation of the total flux and respective 95th, 90th, and 68th percentile radii.}
    \label{fig:r_edge}
\end{figure}

We compute the size of the gaseous disk utilizing the integrated intensity radial profiles of $^{12}$CO via a procedure that sums the emission so that the total flux can be determined at various radii, broadly consistent with literature methods \citep{ansdell_2018ApJ...859...21A, trapman_2023ApJ...954...41T}.
When first calculating the radial profiles we do not define an outer cutoff radius and instead rely on a hand-selected field-of-view that encompasses all the emission, plus an extent of a few arcseconds with just noise. We then take a cumulative sum of the intensity, F$_{\rm cuml}$, which is calculated as a function of radius (see Equation A.1 in \citealt{sanchis_2020A&A...633A.114S}). \maria{This cumulative sum is then normalized by its maximum flux, F$_{\rm max}$}. We define an outer edge, R$_{\rm edge}$, as the first peak of F$_{\rm cuml}$/F$_{\rm max}$, found using \verb|scipy| with the \verb|peak_finder| function \citep{2020SciPy-NMeth}. We make a radial cut of all intensity past R$_{\rm edge}$; we use the remaining intensity to calculate the total flux, F$_{\rm tot}$. We additionally calculate R$_{\rm 95}$, R$_{\rm 90}$, and R$_{\rm 68}$ by taking the 95th, 90th, and 68th percentiles of the total flux. We report R$_{\rm edge}$ and R$_{\rm 90}$ in Table \ref{tab:gas_disk_radii}. The flux here differs slightly from that reported in \cite{exoALMA_main} due to differences in methods, but our results are broadly consistent. 
Figure \ref{fig:r_edge} demonstrates this procedure for LkCa 15. 

\subsection{Brightness Temperatures} \label{sec:gas_temps}

We use the \verb|radial_profile| function from \gofish{} \citep{GoFish} to obtain the integrated and peak intensity radial profiles from the 3-D data cubes in Jy beam$^{-1}$ km s$^{-1}$ and K \maria{respectively}, assuming the full Planck function. This method shifts and stacks the spectra, utilizing the known rotation of the disk, to better resolve the line profile (see also \citealt{yen_2016ApJ...832..204Y, teague_2016A&A...592A..49T}). We assume the geometric parameters and parametric surfaces as determined by \discminer{} for each fit; the differences between using the  \discminer{} surface versus the \disksurf{} surface are negligible. For the integrated intensity radial profiles, we use the continuum subtracted cubes for all three molecules. For the peak intensity radial profiles, we use the non-continuum subtracted fiducial image cubes for $^{12}$CO and $^{13}$CO. \cite{weaver_2018ApJ...853..113W} finds that continuum subtraction leads to an underestimation of the gas temperature when the line is optically thick and absorbs the underlying continuum. Therefore, we opt to use the non-continuum subtracted cubes as to not underestimate the temperatures. 
The radial profiles from \gofish{} are determined by azimuthally averaging radial bins that are binned by one-quarter of the FWHM of the synthesized beam. We report uncertainties of the radial profiles which account for the noise in each channel as described in \cite{yen_2016ApJ...832..204Y}. 

We also obtain peak intensity radial profiles from the emission surface data. Each \textit{r, z} point found with \disksurf{} has an associated emission intensity, \textit{I}, in Jy beam$^{-1}$. We obtain the intensity for each of these $r-z$ points, and convert that to a brightness temperature using the full Planck function:   
\begin{equation}\label{eqn:planck}
    T_B = \frac{h \nu}{k_B} \left[\mathrm{ln} \left(\frac{2h \nu^3}{c^2 I_\nu} +1\right)  \right]^{-1} . 
\end{equation}

In Equation \ref{eqn:planck}, \textit{h} is the Planck constant, $\nu$ is the frequency, $k_B$ is the Stefan-Boltzmann constant, \textit{c} is the speed of light, and $I_{\nu}$ is the emission intensity. Again, we opt to use the non-continuum subtracted cubes for surface extraction to properly estimate the temperatures. We have repeated the emission surface analysis with the continuum subtracted cubes and do not find statistically significant differences between them. For a more in-depth discussion on the imaging procedure of continuum subtraction on the exoALMA sources, see \cite{exoALMA_main} and \cite{exoALMA_imaging}.

\begin{figure*}
    \centering
    \includegraphics[width=1.0\textwidth]{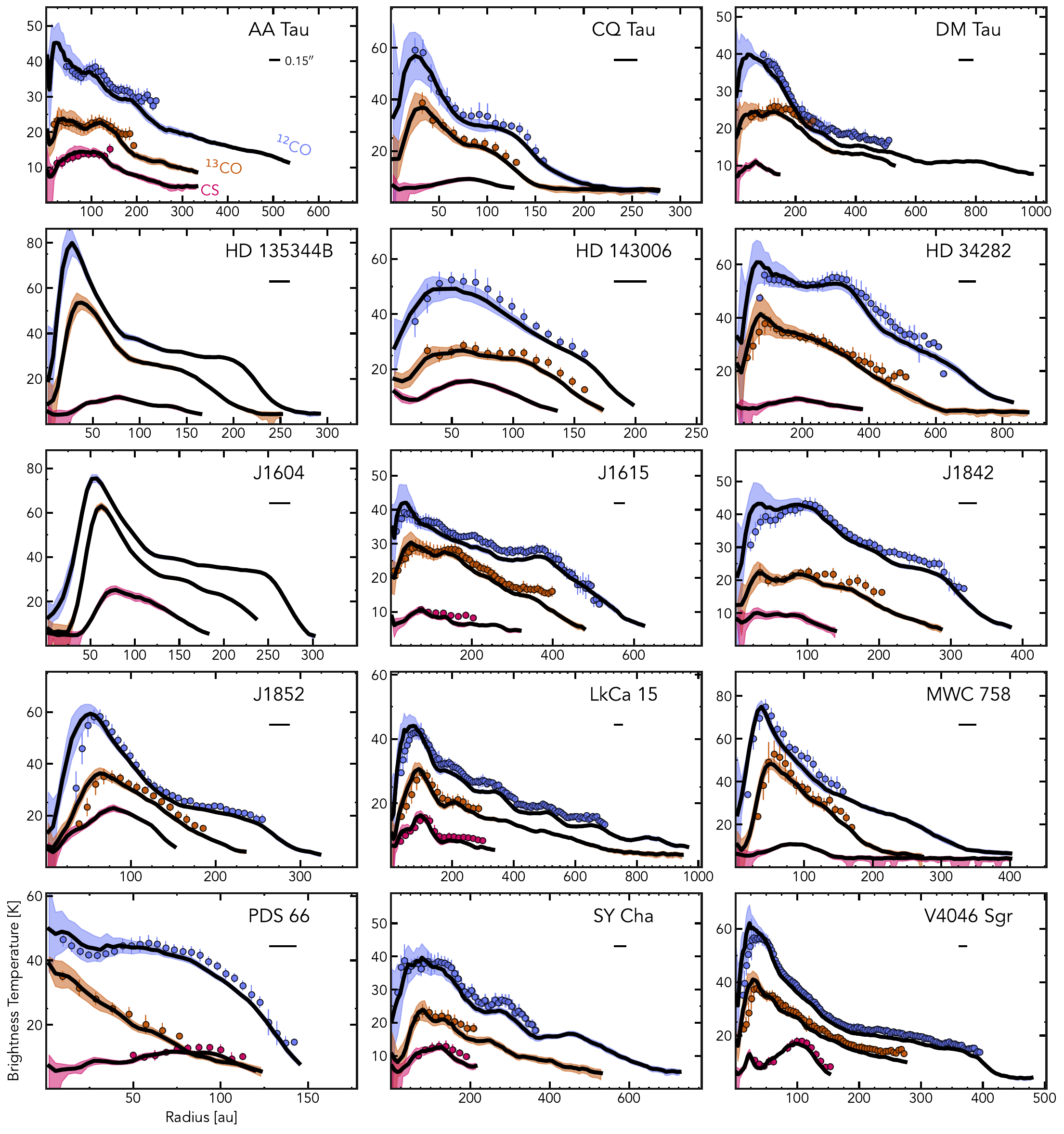}
    \caption{Peak intensities of $^{12}$CO $J=3-2$ (blue), $^{13}$CO $J=3-2$ (orange), and CS$J=7-6$ (pink) in K for all exoALMA sources. The solid black lines show the peak intensities found with \gofish{}, and the points are the temperature points found using \disksurf{}, binned by a quarter of the beamsize. The errorbars on the points represent the standard deviation if the temperature within each bin. The radial profiles are reported out to the molecular emission $R_{\rm edge}$ radius.}
    \label{fig:radial_temp_profile}
\end{figure*}

The resulting peak intensity radial profiles for each molecule are shown as solid lines (for \gofish{} values) and scatter points (for \disksurf{} values) in Figure \ref{fig:radial_temp_profile}. The error bars on these points represent the standard deviation of the temperature within each bin. 
The integrated intensity profiles are shown in Appendix \ref{sec:app_radprofiles} in Figures \ref{fig:integrated_intensity_12co}, \ref{fig:integrated_intensity_13co}, and \ref{fig:integrated_intensity_cs}. We also compare radial profiles derived directly from \discminer{} and find they are broadly consistent with those derived with \verb|go_fish| and \disksurf{}; Figure \ref{fig:peak_intensity_compare} in Appendix \ref{sec:app_radprofiles} shows the comparison. 

The brightness temperatures obtained via \gofish{} and \disksurf{} are in good agreement (Figure \ref{fig:radial_temp_profile}). Nearly all of the binned datapoints from \disksurf{} are within the error bounds of the peak intensities from \gofish{}. 
Some disks have slight discrepancies between \gofish{} and \disksurf{} peak intensities in the outer regions of the disks (e.g., AA Tau, DM Tau, HD 34282, and J1842.) This is likely due to \disksurf{} tracing \maria{discrete vertical and radial locations within} the disk, whereas the shifted and staked radial profiles could face contamination from the backside more easily in the outer disk. Even so, these differences are slight and generally under 5 K.

Figure \ref{fig:temp_profile} shows the emission surfaces with each respective point colored by its brightness temperature. The background contours of this plot will be discussed in Section \ref{subsec:met_temp_structure}. The smaller background points are the raw $r-z$ surfaces. All three molecules have been binned separately to distinguish between the layers, and the color of the binned points represents the temperature of that specific radial bin. Points below 20 K are denoted by a triangle. From Figure \ref{fig:temp_profile} we can see how the temperature decreases as we probe deeper vertical layers and further radial extents. In each disk, we see that the surface layer probed by $^{12}$CO is the hottest, whereas the $^{13}$CO surfaces are generally below 30 K and the CS surface points are almost all below 20 K. For DM Tau the the 0\farcs30 beamsize cubes were used for the emission surface extraction, \maria{causing the measured temperatures to be colder due to beam dilution.}

The mean $^{12}$CO brightness temperature at 100 au across all disks is $\sim$40 K $\pm$ 7 K. For $^{13}$CO, the mean brightness temperature at 100 au is $\sim$26 K $\pm$ 8 K, and for CS, this value is $\sim$12 K $\pm$ 5 K.
The brightness temperatures of $^{13}$CO are generally colder than that of $^{12}$CO, and the same is true for $^{13}$CO and CS.
The average $^{13}$CO temperature is $\sim$22 K, with much of the emission falling below 20 K. If we assume a CO freezeout temperature of 20 K, this implies that at least some of the $^{13}$CO emission is optically thin. Nearly all of the observed CS $J=7-6$ emission is  colder than that of $^{13}$CO, falling below 20 K. The one exception is the outer CS disk of PDS 66, which becomes slightly hotter than $^{13}$CO at $\sim$75 au.
The freeze-out temperature of CS in disks is $\approx$50-60 K \citep{garrod_2006A&A...457..927G, vanderPlas_2014ApJ...792L..25V}, which implies that nearly all of the CS is optically thin.

\subsection{2-D Temperature Structure} \label{subsec:met_temp_structure}
With multiple molecular tracers probing different heights across the disk, we can construct the 2-D temperature structure of the disk in an $r-z$ plane (see \citealt{dartois_2003A&A...399..773D, rosenfeld_2013ApJ...774...16R, dullemond_2020A&A...633A.137D,
MAPS_emission_surfaces_2021ApJS..257....4L, leemker_2022A&A...663A..23L,
law_2023ApJ...948...60L, law_2024ApJ...964..190L}). 
If we assume that the CO lines are optically thick and in local thermodynamic equilibrium (LTE), we can assume that the measured brightness temperature $T_B$ is equal to the gas temperature.
In this analysis, we adopt the two-layer model connecting the atmosphere temperature to the midplane temperature with a sinusoidal equation originally used in 
\cite{dartois_2003A&A...399..773D}. We tested other functional forms, including the hyperbolic tangent function used in \cite{dullemond_2020A&A...633A.137D} and \cite{MAPS_emission_surfaces_2021ApJS..257....4L}. We find that the sinusoidal function described below best reproduces the observations and gives more reasonable estimates of the midplane temperatures with fewer discrepancies between the observed temperatures and the 2-D fit (see Appendix \ref{sec:app_temp} and Figure \ref{fig:temp_profile_errors}).
With this functional form, the midplane temperature ($T_{\rm mid}$) and atmospheric temperature ($T_{\rm atm}$) are described by power-laws:

\begin{equation}
\label{eqn:tmid}
T_{\rm{mid}} (r) = T_{\rm{mid}, 0} \left( r / r_0 \right)^{q_{\rm{mid}}}
\end{equation}

\begin{equation}
\label{eqn:tatm}
T_{\rm{atm}} (r) = T_{\rm{atm}, 0} \left( r / r_0\right)^{q_{\rm{atm}}}
\end{equation}
where  T$_{\rm mid, 0}$ and  T$_{\rm atm, 0}$ represent the atmospheric and midplane temperature at $r_0=100$ au, and $q_{\rm mid}$ and $q_{\rm atm}$ are the slopes to the power laws, respectively. At each radius, the temperature is smoothly connected between the midplane and atmosphere using a sinusoidal function:


\begin{equation}\label{eqn:temperature}
    T(R,z) = \left\{ \begin{array}{ll}
         T_\text{atm} + \left(T_\text{mid} - T_\text{atm} \right) \cos^2 \left(\frac{\pi z}{2Z_q}  \right), & z<Z_q \\
         T_\text{atm}, & z>Z_q
    \end{array}
    \right. 
\end{equation}

Here, $z_q$ is given by a power law of the form $z_q(r) = z_0(r/r_0)^\beta$ \maria{where z$_0$ is the height in the disk at r = r$_0$ and $\beta$ is the power law exponent}.
To determine the 2-D temperature structure, we fit these equations to the non-parametric surface data for six parameters: $T_{\rm atm, 0}$, $T_{\rm mid, 0}$, $q_{\rm atm}$, $q_{\rm mid}$, $z_0$, and $\beta$.
In this fit, we discard the data points with brightness temperatures below 20 K because those likely probe optically thin emission and thus do not reflect the true gas temperature. All CS temperatures used in the temperature fitting fall below 20 K, and are therefore not used.

We have constrained the parameter space with priors as follows: $T_{\rm mid, 0}$ \textless~ T$_{\rm atm, 0}$ \textless~ 150 K, $-2$ \textless~ $q_{\rm atm}$ \textless~ 0, $-2$ \textless~ $q_{\rm mid}$ \textless~ 0, 0 \textless~ $z_0$ \textless~ $r_{\rm max}$ where $r_{\rm max}$ is the largest radial point, and 0 \textless~ $\beta$ \textless~ 2. 
These priors were informed by previous observations of similar disks and modeling efforts; see \citealt{dartois_2003A&A...399..773D, zhang_2021ApJS..257....5Z, calahan_2021ApJS..257...17C, MAPS_emission_surfaces_2021ApJS..257....4L}.
We attempted to keep the priors listed above as broad as possible as to not impart inherent assumptions on the temperature profiles, but the convergence of the fits was contingent on these priors being in place.
The lower bounds of $T_{\rm mid}$ are set by the continuum temperatures $T_{\rm cont}$ derived in \cite{exoALMA_cont}. At radial locations where there is no continuum information, we set the lower bound for $T_{\rm mid}$ to the CMB temperature $T_{\rm CMB} = 2.73$~K. We additionally place an upper bound on $T_{\rm mid}$ using the expected irradiation temperature given by

\begin{equation}
\label{eqn:t_irr}
T_{\rm{irr}} (r) = \left(\frac{\phi L_*}{8 \pi r^2 \sigma_{SB}} \right)^{1/4}
\end{equation}
where $L_*$ is the stellar luminosity, $r$ is the distance from the central star,  $\sigma_{SB}$ is the Stefan-Boltzmann constant, and $\phi$ is an irradiation factor that is dependent on the angle which the stellar radiation impinges onto the disk \citep{2001ApJ...553..321D}. \maria{The midplane temperature bounds are then max($T_{\rm cont}$, $T_{\rm CMB}$) \textless~ $T_{\rm mid}$ \textless~ $T_{\rm irr}$.}
We opt for $\phi = 1$ as a conservative upper bound for stellar irradiation-dominated disk; in practice, we find that the posterior midplane temperature is usually much lower than $T_{\rm irr}$.

To fit Equations \ref{eqn:tatm}, \ref{eqn:tmid}, and \ref{eqn:temperature}, we use \verb|emcee| \citep{emcee_2013PASP..125..306F},
adopting 256 walkers, 1000 steps, and 100 burn-in steps. We start with a randomized initial guess within the bounds of the prior conditions to obtain the first fit. We then use the results of this first fit to inform the initial guess of a second round with the same number of walkers, steps, and burn-in steps. We found that this procedure yielded more converged results due to walkers getting stuck in the first round of fitting, but we note that differences between the first and second round of fitting are not generally large.
The results of the 2-D temperature structure are reported in Table \ref{tab:temperature_profiles} where we report the \maria{50th percentiles as the best-fit values} and the 16th and 84th percentiles as upper and lower errors. These represent only the statistical errors and do not include systematic errors which can be larger than the statistical errors. \maria{Because of this, some of the errors reported in Table \ref{tab:temperature_profiles} are zero, particularly for q$_{\rm mid}$ and $\beta$.} For a discussion of the systematic errors, see Appendix \ref{sec:app_temp}.

\begin{table*}[t]
    \centering
    \caption{2-D temperature structure fits.}
    \label{tab:temperature_profiles}

    \begin{tabular}{ccccccc}
    \hline \hline
        Source  & $T_{\rm atm, 0}$ (K) & $T_{\rm mid, 0}$ (K) & $q_{\rm atm}$ & $q_{\rm mid}$ & $z_0$ (au) & $\beta$ \\ \hline

    DM Tau & 36$^{+0.22}_{-3}$ & 23$^{+0.15}_{-4.9}$ & -0.13$^{+0.015}_{-0.1}$ & -0.5$^{+0.015}_{-0.1}$ & 25$^{+0.79}_{-19}$ & 2$^{+0.0057}_{-0.93}$ \\ \hline
    
    AA Tau & 41$^{+0.43}_{-0.39}$ & 13$^{+0.31}_{-0.33}$ & -0.51$^{+0.017}_{-0.018}$ & -0.21$^{+0.017}_{-0.018}$ & 60$^{+0.87}_{-0.8}$ & 0.075$^{+0.015}_{-0.015}$ \\ \hline
    
    LkCa 15 & 48$^{+0.43}_{-0.4}$ & 20$^{+0.15}_{-0.15}$ & -0.55$^{+0.0075}_{-0.0077}$ & -0.23$^{+0.0075}_{-0.0077}$ & 55$^{+0.69}_{-0.67}$ & 0.59$^{+0.0073}_{-0.0074}$ \\ \hline
    
    HD 34282 & 67$^{+1.3}_{-0.78}$ & 32$^{+0.14}_{-0.14}$ & -0.00093$^{+0.00071}_{-0.0025}$ & -0.25$^{+0.00071}_{-0.0025}$ & 87$^{+2.5}_{-1.6}$ & 0.69$^{+0.0059}_{-0.009}$ \\ \hline
    
    SY Cha & 45$^{+0.37}_{-0.35}$ & 24$^{+0.13}_{-0.15}$ & -0.58$^{+0.0096}_{-0.01}$ & -0.3$^{+0.0096}_{-0.01}$ & 56$^{+0.83}_{-0.82}$ & 0.0069$^{+0.0096}_{-0.005}$ \\ \hline
    
    RXJ1615.3-3255 & 34$^{+0.051}_{-0.044}$ & 24$^{+0.12}_{-0.15}$ & -0.098$^{+0.0019}_{-0.0032}$ & -0.25$^{+0.0019}_{-0.0032}$ & 32$^{+0.22}_{-0.23}$ & 1.1$^{+0.011}_{-0.018}$ \\ \hline
    
    V4046 Sgr & 37$^{+0.036}_{-0.039}$ & 28$^{+0.067}_{-0.067}$ & -0.62$^{+0.0022}_{-0.0021}$ & -0.35$^{+0.0022}_{-0.0021}$ & 10$^{+0.06}_{-0.062}$ & 0.00025$^{+0.00051}_{-0.00019}$ \\ \hline
    
    RXJ1842.9-3532 & 43$^{+0.14}_{-0.13}$ & 25$^{+0.08}_{-0.094}$ & -0.45$^{+0.0062}_{-0.0062}$ & -0.23$^{+0.0062}_{-0.0062}$ & 27$^{+0.25}_{-0.25}$ & 0.00077$^{+0.0015}_{-0.00058}$ \\ \hline
    
    RXJ1852.3-3700 & 40$^{+0.092}_{-0.088}$ & 30$^{+0.15}_{-0.16}$ & -0.87$^{+0.0072}_{-0.0071}$ & -0.36$^{+0.0072}_{-0.0071}$ & 16$^{+0.2}_{-0.2}$ & 0.00081$^{+0.0019}_{-0.00061}$ \\ \hline

    \end{tabular}
\end{table*}

\begin{figure*}
    \centering
    \includegraphics[width=1.0\textwidth]{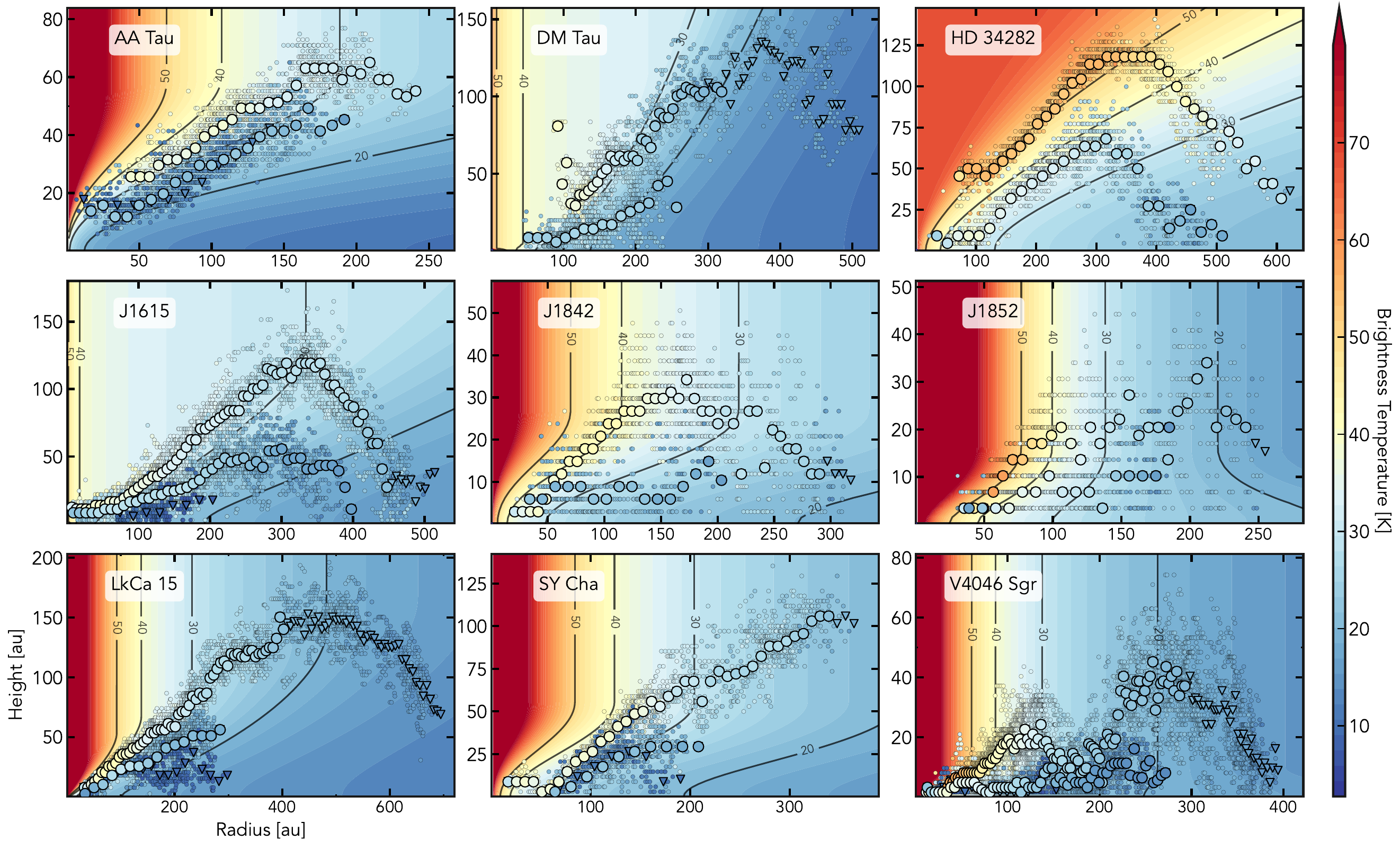}
    \caption{Emission surface points with associated brightness temperatures; \maria{raw surface data points are shown as smaller background points}. The raw data is binned by a quarter of a beamsize for molecules above 20 K (larger circles) and below 20 K (triangles). The contour background shows the 2D temperature profiles calculated with Equation \ref{eqn:temperature}. Lines of constant temperatures at \maria{50 K}, 40 K, 30 K, and 20 K are shown in black.}
    \label{fig:temp_profile}
\end{figure*}

The derived 2-D temperature profiles are shown in Figure \ref{fig:temp_profile} as filled contours. In the figure, the larger circles show the data points binned by 1/4 of the beamsize, and the triangles show the binned data that was not utilized in the fitting procedure because it was below the CO freezeout temperature (\textless~ 20 K). Lines of constant temperature are drawn at \maria{50 K}, 40 K, 30 K and 20 K. 
From the fitting procedure, we obtain an estimate of atmospheric temperature as a function of radius, $T_{\rm atm}(r)$, and the midplane temperature, $T_{\rm mid}(r)$. 
We find atmospheric temperatures (at 100~au) range from $\sim$67 K (HD 34282) to $\sim$34 K (J1615), with a mean atmospheric temperature (T$_{\rm atm}$) at 100 au across all disks of 43.4 K. 
We find midplane temperatures ranging from 13 K (AA Tau) up to 32 K (HD 34282). The average midplane temperature at 100 au across the disk sample is 24 K.
Because there is no optically thick data at the midplane, this value is an extrapolation informed by the priors. 
Due to this, we stipulate that the reported T$_{\rm mid}$ is closer to an upper bound to the true midplane temperature. \maria{In J1852 and in V4046 \mariatwo{Sgr}, the midplane temperature inverts, as shown by the contour at 20 K in Figure \ref{fig:temp_profile}. Both disks have low inclinations around 35$^{\circ}$, which causes highly scattered data and difficulty in separating the  molecular layers, particularly in the outer disk. This leads to larger uncertainties in the temperature profiles.}

\section{Substructure}\label{sec:substucture}

Vertical substructures have been observed in many emission surfaces (see \citealt{MAPS_emission_surfaces_2021ApJS..257....4L}, \citealt{law_2023ApJ...948...60L}, \citealt{paneque_2023A&A...669A.126P}). These localized structures have been found to be co-located with substructures identified in continuum \citep{MAPS_emission_surfaces_2021ApJS..257....4L, paneque_2023A&A...669A.126P}, intensity radial profiles \citep{MAPS_radial_profiles_2021ApJS..257....3L}, or velocity perturbations \citep{teague_2019Natur.574..378T, teague_2021ApJS..257...18T, galloway_2023ApJ...950..147G, paneque_2023A&A...669A.126P}. 
Their identification can provide insight into finer disk structures;
surface modulations are expected to arise during planet formation, which is expected to open gaseous gaps \citep{dong_2015ApJ...809...93D}, or produce changes in molecular abundance and chemical stratification \citep{aikawa_1999A&A...351..233A}.
In the following section, we detail the process of substructure identification for both the emission surfaces and radial intensity profiles, and compare these with those seen in the continuum and the velocity profiles. 

\subsection{Emission Surface Substructures}\label{subsec:results_substructures}

Some of the most prominent emission surface features easily identified by eye are \maria{the sharp cutoff from rising emission} seen in AA Tau, SY Cha, and LkCa 15. In $^{12}$CO, both surfaces of AA Tau and SY Cha sharply drop at 237 au and 359 au, respectively. In $^{13}$CO, LkCa 15 also exhibits a sharp drop in the emission surface at 274 au. \maria{These emission surface morphologies differ from the traditional tapered power law in that the falling portion of the emission is distinctly separated from the rising portion of the surface, and that this outer emission is scattered with no well-defined shape.}
While \disksurf{} has identified these regions as part of the upper emission surface, visual inspection and line profile analysis has shown that they are likely a combination of emission from the upper and lower surfaces, as well as the midplane. For this reason, we exclude these points when fitting the surface with \disksurf{}. The physical reasons for this ambiguous `diffuse' emission is unclear. It may be a projection effect, given that all three disks are highly inclined at around 50$^{\circ}$. It may imply that the disks have different sizes between the upper and lower surfaces, or that the outer region of the upper surface becomes colder than that of the backside. It could also be due to photodesorption in the outer regions of the disks. Sections \ref{sec:diffuse_surfaces} and \ref{sec:projection_effects} discuss these possibilities in more detail. 

For smaller-scale substructure identification, we rely on the rolling surfaces derived using \disksurf{} shown in Figure \ref{fig:rolling_surfaces}. The rolling surfaces shown in this figure have been smoothed with a Savitzky-Golay filter for visual clarity, \maria{with a window length of one arcsecond and a polynomial order of 2}; surface perturbations smaller than the beamsize are smoothed out. To identify surface structure, we utilize the \verb|find_peaks| function from \verb|scipy| \citep{2020SciPy-NMeth}. This function finds maxima or minima in 1-D data sets by comparison of neighboring values. To ensure the substructures are physical, we set the minimum width of each substructure and the minimum distance between substructures to be at least the beamsize (0\farcs15), up to two times the beamsize (0\farcs3). We scale the maximum height, or prominence, of the peak to the beamsize or larger. After running \verb|find_peaks|, we visually confirm by eye the validity of the substructures, as this method can be sensitive to small perturbations, even with the imposed limits. We report each vertical substructure dip in Table \ref{tab:substructures} and in Figure \ref{fig:rolling_surfaces} with dashed lines. Each substructure is labeled with `Z', in the nomenclature of \cite{MAPS_emission_surfaces_2021ApJS..257....4L}, followed by the radial location. For visual clarity, we show only the structures identified as dips, and not the peaks. 

We find evidence of localized substructures in 76\% of disks with viable emission surfaces. Of these, seven of the substructures are in $^{12}$CO emission. Nine of the disks also have vertical substructure in $^{13}$CO emission, which is generally co-located with that seen in $^{12}$CO. Only two definite substructures were identified in CS, although several slope changes in the surfaces are observed. The emission surfaces extracted using the iterative methodology described in Section \ref{sec:surfaces} are generally smoother than the initial surfaces found with just one iteration; we have verified that all surface substructures presented in Figure \ref{fig:rolling_surfaces} are consistent between the iterative and non-iterative surfaces. 
We do not find localized surface perturbations large enough to warrant classification in HD 34282 or PDS 66. 

\subsection{Radial Peak Intensity Substructures}\label{subsec:results_substructures_intensities}

We repeat the structure identification process outlined in the previous section, but for the radial peak intensity profiles, smoothed using a  Savitzky-Golay filter \maria{with a window length of five and a polynomial order of 3}. To identify structure, we again utilize the peak-finding function from \verb|scipy| \citep{2020SciPy-NMeth}. We again set the minimum radial width of each substructure and the minimum distance between substructures to be at least the beamsize. We scale the maximum height, or prominence, of the peak to the temperature, ranging from 7 to 15 K for some disks. We find peak intensity structure in the disks of AA Tau, CQ Tau, HD 34282, J1615, J1842, LkCa 15, and SY Cha, visualized in Figure \ref{fig:peak_intensity_structure}.
The profiles of HD 135344B, HD 143006, J1604, J1852, MWC 758, and PDS 66 are generally smooth, and we do not characterize any substructure for these disks. We do not characterize the first peak in the intensity profiles, as this is an expected feature present in all disks \maria{due to beam dilution in the inner radial regions causing a drop in intensity}. We also do not characterize any structure interior to this first peak.

AA Tau, LkCa 15, and SY Cha disks exhibit a wave-like pattern in their $^{12}$CO intensity profiles. In AA Tau, there are two slight bright bumps at $\sim$120 au and $\sim$195 au. The wave pattern in LkCa 15 is distinct; there is an initial bright spot at 174 au, also seen in $^{13}$CO, with subsequent dips and bright spots at nearly evenly spaced intervals, ending at a dip identified at 781 au. The wave patterns are spaced roughly 175 au apart. SY Cha exhibits two bright peaks in a distinct pattern as well, at 276 au and 447 au, which are observed to a lesser extent in the $^{13}$CO peak intensities. The potential causes of this structure is discussed in Section \ref{sec:discussion_substructure}.

CQ Tau, HD 34282, J1615, J1842, and V4046 Sgr exhibit singular bright spots in multiple molecules. In CQ Tau, a bright `shoulder' is seen at 125 au; the shape of the $^{13}$CO profile is similar, but no definite peak was identified via our process. HD 34282 displays a prominent bright shoulder in $^{12}$CO as well, at 295 au, but the remaining molecules do not show any structure. J1615 exhibits a bright shoulder in $^{12}$CO at 366 au, and there is a co-located bright spot radially inward at 132 au for $^{13}$CO and CS. In J1842, we identified a bright shoulder at 88 au that is co-located between all three molecules. Finally, V4046 Sgr exhibits a bright shoulder only in CS emission, at 98 au.

We do not find a significant correlation between substructures identified in the peak intensity and those identified in the emission surfaces. One exception is LkCa 15, whose surface dips line up within a few au of some of the `wiggles' seen in the peak intensity. 

\subsection{Comparison with Dust Continuum}\label{subsec:results_dust_cont_compare}

\begin{figure*}
    \centering
    \includegraphics[width=1.0\textwidth]{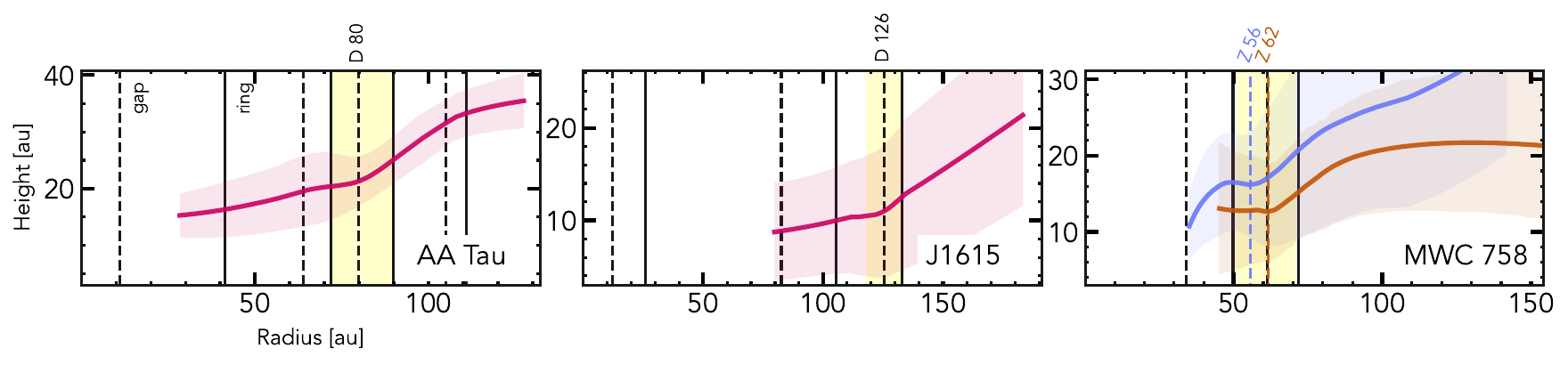}
    \caption{Rolling emission surfaces (CS $J=7-6$, pink; $^{12}$CO $J=3-2$, blue; $^{13}$CO $J=3-2$, orange) with continuum substructures identified by \cite{exoALMA_cont}. The dashed lines show the continuum gaps, and the solid lines show the continuum rings. \maria{The yellow shaded regions highlight the surface structures that overlap with continuum and or gas substructures.} }
    \label{fig:cont_substructure}
\end{figure*}

We compare the emission surface substructure radial locations with the rings and gaps in the continuum reported by \cite{exoALMA_cont}. Unlike the results of the MAPS emission surface study \citep{MAPS_emission_surfaces_2021ApJS..257....4L}, we do not find significant correlation between the molecular emission surfaces and the dust millimeter continuum rings and gaps for our larger sample size.
We find that the disks of AA Tau, J1615, and MWC 758 are the only disks whose molecular emission surface substructure potentially correlates with that of the continuum. We present these structures, along with their continuum counterparts, in Figure \ref{fig:cont_substructure}, with areas of interest highlighted in yellow. For AA Tau and J1615, this correlation is only present in the CS $J=7-6$ emission, while MWC 758 exhibits a substructure match in $^{12}$CO and $^{13}$CO. 

In AA Tau, the dust gap at 80 au, and the subsequent rings surrounding it, align with a slight dip in the CS surface, visualized in the leftmost panel of Figure \ref{fig:cont_substructure}. We do not see this same correlation in $^{12}$CO or $^{13}$CO; there is no gap structure in either line near 80 au. The correlation between the dust substructure and CS emission in J1615 is slight. J1615 exhibits a dust gap 125.5 au, and at this point the CS surface dips slightly and then experiences a slope change, visualized in the middle panel of \ref{fig:cont_substructure}. 

MWC 758 has the clearest correlation between emission surface structure and dust rings/gaps. MWC 758 displays two clear dips in the $^{12}$CO and $^{13}$CO surfaces at 56 au and 62 au, respectively. The emission surface dip in  $^{13}$CO aligns nearly perfectly with a continuum gap at 61.3 au. Additionally, the $^{12}$CO surface appears to increase in height slightly at the innermost dust ring, around 50 au. 
The continuum rings and gaps overlaid with the surfaces that have a substructure match is shown in the rightmost panel of Figure \ref{fig:cont_substructure}. These surface dips may be due to the continuum subtraction, but a physical mechanism may be at play given that we do not see them at the locations of the other continuum gaps. 

\maria{\cite{law_2024ApJ...964..190L} notes a slope change in $^{12}$CO $J=3-2$ emission present in PDS 70 at the location of the dust cavity wall, which is reminiscent of what we see in AA Tau, J1615, and MWC 758. Of these disks, AA Tau and MWC 758 are transition disks, and the observed slope changes may be due to the decreased gas surface density within the cavity, as seen in \cite{law_2024ApJ...964..190L}. A more thorough investigation of the inner radial regions of gas in exoALMA sources and their relation to dust cavity walls is reserved for future work.}

The substructures identified in the peak intensity align with several continuum rings and gaps. Figure \ref{fig:peak_intensity_structure} shows the radial profiles of the peak intensities, with the dust gaps as black dashed lines and the dust rings as black solid lines. In AA Tau, we see a dip in $^{13}$CO that aligns within a beamsize of the dust gap at 79 au. We see similar structure in J1615, where gaps identified in $^{13}$CO and CS align with the dust gap at 125 au. This is also seen in J1842; dips in the intensity in $^{12}$CO and $^{13}$CO align with the dust gap at 63 au. There are several instances where dust rings align with the brightest intensity peak, as well as substructures interior to this peak, but many of these are within one beam, and we exclude them from this analysis.

\subsection{Comparison with Velocity Signatures}\label{subsec:substructure_veloc}

Several of the exoALMA sources display kinematic perturbations in the individual channel maps, which are discussed and characterized in \cite{exoALMA_channel_maps}; azimuthally averaged velocity perturbations are derived in \cite{exoALMA_rotation}. The signatures characterized as ``velocity kinks" may be indicative of embedded protoplanets \citep{pinte_2018A&A...609A..47P, pinte_review_2023ASPC..534..645P}, although other explanations are possible \citep{lodato_2023MNRAS.518.4481L}. We find that surface modulations in AA Tau, J1842, and LkCa 15 are radially co-incident within a beamsize of the proposed planet locations listed in Table 2 of \cite{exoALMA_channel_maps}. The radial locations of the proposed planets are plotted as green lines in Figure \ref{fig:rolling_surfaces}. In AA Tau, the velocity kink aligns with a dip in the CS surface at 95 au, which is also within a beamsize of the dust continuum feature at 80 au shown more clearly in Figure \ref{fig:cont_substructure}. In J1842, the proposed planet location of 150 au aligns with two dips in the emission surfaces, co-incident between $^{12}$CO and $^{13}$CO. In LkCa 15, the most prominent surface modulation in $^{12}$CO aligns with the proposed planet location of 370 au.

In AA Tau, J1615, LkCa 15, and SY Cha, peak intensity structures align with the proposed planet locations. The dip identified in $^{13}$CO at 86 au aligns with the proposed planet location of 85 au, which also co-aligns with a continuum gap. For J1615, the bright peak intensity shoulder at 336 au aligns with the proposed planet at 370 au. In LkCa 15, the second `wave' peak is within a beamsize of the proposed planet location. In SY Cha, a $^{13}$CO peak at 150 au aligns with the proposed planet location.

\section{Discussion} \label{sec:discussion}

\subsection{Comparison with Source Properties} \label{sec:discussion_statistics}

Numerous studies have shown that emission height is correlated with some disk and stellar properties. \cite{MAPS_emission_surfaces_2021ApJS..257....4L} showed that the disk-averaged $\langle z / r \rangle$ is  positively correlated with the gas disk size.
This trend was further verified in \citealt{law_2022ApJ...932..114L, law_2023ApJ...948...60L}, and \citealt{paneque_2023A&A...669A.126P}.
Relationships between $\langle z / r \rangle$, stellar mass, and gas temperature were also seen. \maria{Most recently, \cite{paneque_2025arXiv250108294P} demonstrated the correlation between emission height, CO depletion, and disk mass.}
Here, we build off of the data compiled in \citealt{law_2022ApJ...932..114L, law_2023ApJ...948...60L, paneque_2025arXiv250108294P}, adding the exoALMA sources, to further explore these correlations.

\begin{figure*}[t]
    \centering
    \includegraphics[width=1.0\textwidth]{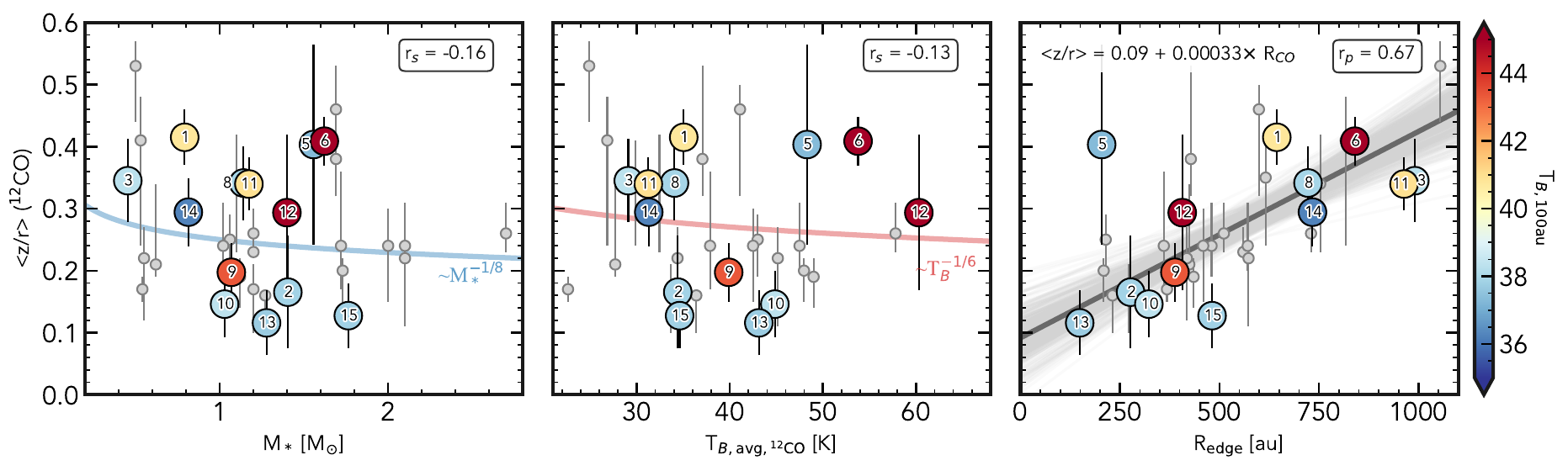}
    \includegraphics[width=1.0\textwidth]{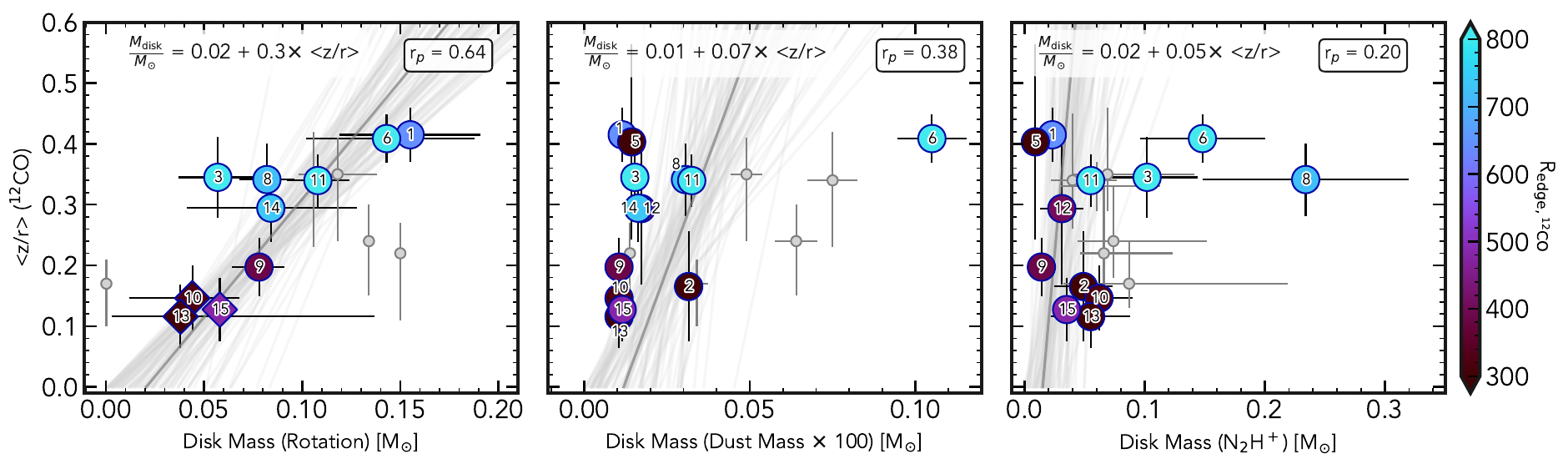}

    \caption{\maria{Top panel: Average z/r in $^{12}$CO for disks with moderate inclination, plotted against M$_*$ (leftmost plot), average $^{12}$CO brightness temperature (middle plot) and the outer radius of the gaseous disk (rightmost plot). The points are colored by brightness temperature at 100 au. Background gray points are literature values from previous studies. 
    Bottom panel: Average z/r in $^{12}$CO for disks with moderate inclination, plotted against disk mass calculated via the rotation curve from \cite{exoALMA_disk_mass} (leftmost plot), disk mass calculated via the dust mass from \cite{exoALMA_cont} and (middle plot), and disk mass calculated using N$_2$H$^+$ from \cite{exoALMA_trapman}. Points that represent exoALMA sources are colored by the $^{12}$CO R$_{\rm edge}$. Smaller gray points are MAPS sources with mass measurements from \cite{martire_2024A&A...686A...9M}, \cite{dsharp_2018ApJ...869L..41A}, and \cite{exoALMA_trapman}. The Pearson correlation coefficient, r$_p$, or the Spearman correlation coefficient, r$_s$, is shown in the top right of each figure. When a linear function is fit, the results are plotted and the equation is shown in the top left of each figure. Disk labels: AA Tau = 1, CQ Tau = 2, DM Tau = 3, HD 143006 = 5, HD 34282 = 6, J1615 = 8, J1842 = 9, J1852 = 10, LkCa 15 = 11, MWC 758 = 12, PDS 66 = 13, SY Cha = 14, V4046 Sgr = 15. Disks without retrievable surfaces are excluded.}}
    \label{fig:zr_variable}
\end{figure*}

Plots showing $^{12}$CO emission surface $\langle z / r \rangle$ versus stellar mass, average gas temperature, and the gas disk size $R_{\rm edge}$ are shown in the top panels of Figure \ref{fig:zr_variable}. Here, we focus our discussion to $^{12}$CO emission surface and briefly comment on $^{13}$CO emission surface. 
The $\langle z / r \rangle$ calculation is described in subsection \ref{subsec:results_surfaces}; we use methods consistent with those used for the compiled background points.  We find consistent results with those observed in \cite{law_2022ApJ...932..114L, law_2023ApJ...948...60L}. The emission surface $\langle z / r \rangle$ weakly decreases with both stellar mass and average $^{12}$CO temperature. To explain these trends, \cite{law_2022ApJ...932..114L} invokes the equation relating the gas pressure scale height with the stellar mass and disk temperature (see Equation 3 in \citealt{law_2022ApJ...932..114L}).
Assuming that the emission surfaces scale with gas pressure scale heights and that the stellar luminosity and mass are related following $L_* \propto M_*^3$, as in \citet{law_2022ApJ...932..114L}, \zr{} $\sim M_*^{-1/8}$ and \zr{} $\sim$ T$^{-1/6}$. We plot these relations on the upper left and upper middle panels in Figure \ref{fig:zr_variable}. We quantify the correlation between these variables using Spearman's rank correlation coefficient (SCC, or r$_{s}$), which is used for non-linear relations. For \zr{} versus stellar mass, we find r$_s$ = -0.16. For \zr{} versus average brightness temperature, we find r$_s$ = -0.13. 
Overall, we find a trend that \zr{} decreases with stellar mass and temperature although we note that these trends are highly scattered, and not reliably correlated, similar to what \cite{law_2023ApJ...948...60L} found (see also \citealt{paneque_2023A&A...669A.126P}).

\maria{Next, we look at the correlation between \zr{} and the gas disk size. We find a linear correlation between them as shown in the top rightmost panel of Figure \ref{fig:zr_variable}. To keep consistency with previous studies \citep[e.g.,][]{law_2022ApJ...932..114L, law_2023ApJ...948...60L}, we use \linmix{} \citep{linmix_2007ApJ...665.1489K} to fit a line to the data. We find a linear relation quantified by \zr{} = 0.090$^{+0.030}_{-0.030}$ + 0.00033$^{+0.000053}_{-0.000052}$ $\times$R$_{\rm edge}$. This is steeper than previous relations found in  \cite{law_2022ApJ...932..114L, law_2023ApJ...948...60L}, but overall consistent. 
We use the Pearson correlation coefficient (PCC, or r$_p$) to quantify the linear correlation between \zr{} and R$_{CO}$.
We find a PCC of r$_p$ = 0.67, indicating a strong positive correlation. We repeated the \zr{} measurement, but only taking emission surface points at 125 au for each disk, $\pm$ the beamsize of 0\farcs15, essentially measuring the \zr{} at the same radius for each disk. We found that the measured \zr{} values did not change significantly, and still resulted in a correlation with R$_{CO}$. \maria{We note that HD 143006 (labeled as point 5) is an outlier, likely due to the face-on nature of this disk and subsequent associated \zr{} error. With this point removed, the PCC becomes r$_p$ = 0.76.}}

We have also explored trends in stellar mass, average temperature, and outer edge for $^{13}$CO emission. In general, we find \zr{}$_{13_{CO}}$ decreases with mass and mean brightness temperature, consistent with \cite{law_2023ApJ...948...60L}. Similar to what we found for $^{12}$CO, we retrieve a linear relationship (albeit much steeper) between \zr{}$_{13_{CO}}$ and R$_{CO}$ compared to previous studies, although this linear relation suffers from a larger amount of scatter.

\maria{\cite{exoALMA_interpret_height} predicts emission surface height should positively correlate with gas surface density and hence disk mass. \cite{paneque_2025arXiv250108294P} also recently presented a relation between \zr{} and disk mass via chemical modeling; they find that the disk mass and volatile carbon abundance are primarily responsible for setting the observed \zr{}.
Here, we check whether the observations are in agreement with the predictions. The bottom left panel of Figure \ref{fig:zr_variable} shows \zr{} versus disk mass derived from the rotation curves in \cite{exoALMA_disk_mass} (left panel), disk mass derived from dust mass measurements from \cite{exoALMA_cont} (middle panel), and disk mass derived using N$_2$H$^+$ from \cite{exoALMA_trapman} (rightmost panel), along with literature values shown in gray.}
\maria{As shown in the figure, we find linear relations between \zr{} and disk mass derived from multiple methods. For \zr{} vs rotation disk mass, we find the strongest linear relation, with a PCC of r$_p$ = 0.64. Emission height versus disk mass via dust mass has a PCC of r$_p$ = 0.38, and \zr{} versus disk mass via N$_2$H$^+$ has the weakest linear relation with a PCC of r$_p$ = 0.20. In the former two plots, HD 34282 (6) and J1615 (8) stand as outliers, respectively. This may hint that the \zr{}/disk mass relation flattens out at higher masses, or that the relation is better quantified by a non-linear equation. \cite{paneque_2025arXiv250108294P} finds that the 
\zr{}/disk mass relation differs between T Tauri and Herbig stars, \mariatwo{which may at least partially contribute to HD 34282 being an outlier, although we note that MWC 758 and a few literature sources are also Herbig stars and do not exhibit outlier behavior.}
The strong correlation between  \zr{} and the disk mass measured with rotation curves may be in part due to the fact that this method uses the emission surface as a prior. However, independent of the mass measurements, we find a strong linear relation between \zr{} and R$_{edge}$. This implies that the relation between \zr{} and dynamical disk mass is not entirely due to the use of emission surface. The results presented in Figure \ref{fig:zr_variable} imply that the disk mass and \zr{} are physically related, as predicted in \cite{exoALMA_interpret_height} and modeled in \cite{paneque_2025arXiv250108294P}.}

\maria{While various methods have been proposed to measure the mass and surface density of protoplanetary disks \citep[e.g.,][see also review by \citealt{miotello_2023ASPC..534..501M}]{exoALMA_disk_mass,exoALMA_interpret_height,exoALMA_trapman, andrews_2024ApJ...970..153A}, these methods require observations of multiple molecular lines \citep{exoALMA_trapman} or sophisticated velocity analyses \citep{exoALMA_disk_mass, andrews_2024ApJ...970..153A}. Compared with other approaches, extraction of $^{12}$CO emission surface is relatively straightforward, offering an alternative way to estimate disk mass and surface density, as in \cite{paneque_2025arXiv250108294P}. We also calculate \zr{} versus gas surface density $\Sigma$ at 100~au, inferred based on the rotational disk masses assuming the surface density profile used in \cite{exoALMA_disk_mass}, where $\gamma$ = 1 (see their equation A1). We find a positive relationship between emission height and surface density, with a PCC of r$_p$ = 0.75, in agreement with the prediction by \citet{exoALMA_interpret_height}. 
To this end, we provide fits between \zr{} and disk mass and surface density, with the caveat that this relation is specific to dynamical masses and other methods will not follow this relation as closely: }

\begin{equation}
\frac{M_{\rm disk}}{M_\odot} = 0.020_{-0.0056}^{+0.047} + 0.26_{-0.089}^{+0.23}~\langle z/r \rangle
\end{equation}
\begin{equation}
\frac{\Sigma(100~{\rm au})}{{\rm g~cm}^{-2}} = 1.010_{-0.29}^{+2.26} + 11.41_{-7.22}^{+14.19}~\langle z/r \rangle
\end{equation}

\maria{The remaining linear relations derived using disk mass via dust and disk mass via N$_2$H$^+$ are shown in the upper right corners of Figure \ref{fig:zr_variable}.}
We use \verb|emcee| \citep{emcee_2013PASP..125..306F} to obtain these linear fits with 100 walkers and 5000 iterations. This allows us to place priors such that the model cannot be below zero. The errors are propagated from \zr{} and M$_{\rm disk}$.

\maria{Additionally, we explore the relationship between \zr{} and disk mass for $^{13}$CO. We find similar correlations to that seen with $^{12}$CO. For \zr{} versus dynamical disk mass, r$_p$ = 0.88; for \zr{} versus disk mass via dust mass,  r$_p$ = 0.27; for \zr{} versus disk mass via N$_2$H$^+$,  r$_p$ = 0.34. We note that we do not include archival estimates here. }

\subsection{Comparison to Previous Results}\label{sec:discussion_comparision}

Several disks from the exoALMA sample have previous gas component observations at different spatial and angular resolution, and with other molecular transitions. In the following sections, we compare our results to previous studies of the same disks when available.

\subsubsection{Emission Surfaces}\label{subsec:compare_emission_surfaces}
Previous work by \cite{law_2023ApJ...948...60L} derived the emission heights, temperature profiles, and outer radii for DM Tau ($^{12}$CO $J=3-2$ and $^{13}$CO $J=2-1$ emission), HD 34282 ($^{12}$CO $J=2-1$ and $^{12}$CO $J=3-2$ emission), LkCa 15 ($^{12}$CO $J=2-1$, $^{12}$CO $J=3-2$, $^{13}$CO $J=2-1$,  and $^{13}$CO $J=3-2$ emission). Additionally, \cite{law_2022ApJ...932..114L} derived the emission surface of V4046 Sgr ($^{12}$CO $J=3-2$ emission). Below, we compare the emission surfaces between this work and the previous work.

The previous DM Tau emission surfaces for $^{12}$CO and $^{13}$CO J=2-1 were extracted at angular resolutions of $\sim$0$\farcs$13 and $\sim$0$\farcs$19, respectively. \cite{law_2023ApJ...948...60L} finds a $\langle z / r \rangle$ of 0.41$^{+0.7}_{-0.17}$ in $^{12}$CO $J=2-1$ and 0.22$^{+0.8}_{-0.04}$ in $^{13}$CO $J=2-1$. The surface was fit out to 2$\farcs$7, or 388.8 au. These results differ to ours; we find shallower $^{12}$CO and $^{13}$CO surfaces, with a $\langle z / r \rangle$ of 0.30 and 0.13, respectively. When compared to our parametric surfaces, that of \citet{law_2023ApJ...948...60L} is consistent, albeit slightly higher and with a much shorter cutoff radius. 
The differences in emission surfaces can be partially explained by the difference in angular and spectral resolution between the observations, as well as the implementation of the \verb|iterative| function in \disksurf{}, which can lead to slight differences in observed surface structure. In particular, the iterative surface process neglects the outer diffuse emission seen in DM Tau, and instead finds a tapered power-law surface. We have included the diffuse outer points in Figures \ref{fig:binned_surfaces} and \ref{fig:surfaces_12co}; if we include these in our surface fit, we find results more consistent with that of \citet{law_2023ApJ...948...60L}. 

\maria{The $^{12}$CO and $^{13}$CO emission of LkCa 15 was first mapped by \citet{leemker_2022A&A...663A..23L}, and
\citet{law_2023ApJ...948...60L} derived the emission height and 2-D temperature profile using the $J=2-1$ and $J=3-2$ transitions of $^{12}$CO and $^{13}$CO.} \citet{law_2023ApJ...948...60L} finds $\langle z / r \rangle$ values of 0.26 and 0.13 for the respective $J=2-1$ transitions, and $\langle z / r \rangle$ values of 0.23 and 0.15 for the $J=3-2$ transitions. In comparison, we find elevated $\langle z / r \rangle$ values of 0.34 and 0.21 for $^{12}$CO and $^{13}$CO. When overplotted, our derived surface is consistent with that from \citet{law_2023ApJ...948...60L}, the main difference being their parametric surface is slightly below the one we find. These differences may be due to the resolution ($\sim$0\farcs27 versus 0\farcs15), or due to the iterative function in \disksurf{}, which was not implemented in  \citet{law_2023ApJ...948...60L}.

The final disk from \cite{law_2023ApJ...948...60L} that overlaps with the exoALMA sample is HD 34282. They find $\langle z / r \rangle$ values of 0.46 and 0.38 for the $^{12}$CO $J=2-1$ and $J=3-2$ transitions. 
In comparison, we find a $\langle z / r \rangle$ value of 0.41, consistent with the 0.38 value Law et al. finds. When comparing emission surface morphology, our surface more closely matches that of their derived $J=3-2$ surface, as expected, although we find our surface has a higher peak and a sharper taper. 

The V4046 Sgr disk emission surface and properties were previously analyzed in \cite{law_2022ApJ...932..114L}. Our emission surface differs significantly from the previously derived one; \cite{law_2022ApJ...932..114L} finds a maximum outer radius of $\sim$160 au, while our surface extends out to $\sim$400 au.
Their reported outer radius of $\sim$160 au matches closely with the large gap observed at 173 au in this work; it is likely that the difference in sensitivities allowed us to pick up the emission in the outer portion of this disk past the gap. 

In general, our results agree with those previously reported. Some of the biggest differences in emission surface morphology and outer radius are likely due to differences in angular resolution, spectral resolution, and sensitivity.

\subsubsection{Temperature Structure}\label{subsec:compare_temp_profiles}

Both DM Tau and LkCa 15 have previously derived 2-D temperature profiles. For DM Tau, we find mostly consistent temperature profiles to that of \cite{law_2023ApJ...948...60L}. We find an atmospheric temperature of 37 K and a midplane temperature of 20 K; \cite{law_2023ApJ...948...60L} finds T$_{\rm atm}$ = 38 and T$_{\rm mid}$ = 26 K. The 2-D shape of the profile is also mostly consistent, with the edge of the $^{13}$CO surface having a temperature of $\sim$20 K, and the rising portion of the $^{12}$CO surface at $\sim$30 K. 

For the temperature structure of LkCa 15, we find a hotter atmospheric temperature (T$_{\rm atm}$) of 48 K, compared with 35 K from \cite{law_2023ApJ...948...60L}. The midplane temperatures are consistent; we report 20 K, and they report 21 K. The discrepancy between atmospheric temperatures may be because \citealt{law_2023ApJ...948...60L} included all temperature data points, even those below 20 K, whereas we exclude these points; additionally, we use different equations to connect the temperature structure. 

In \cite{law_2023ApJ...948...60L}, the temperature profile of HD 34282 in $^{12}$CO $J=2-1$ and $J=3-2$ is derived. We find generally consistent radial profile shapes, identifying the same bump as \cite{law_2023ApJ...948...60L} at around 295 au. Our data extends out another $\sim$300 au due to our achieved sensitivity. However, the overall temperature we measure for this peak intensity profile is considerably higher (by about 15 K), which is likely due to the difference in beamsizes between the datasets. 

Any differences in the temperature profiles can be attributed to several factors: the CO transition used is different, the spectral and angular resolution of the cubes is not the same, and the functional form used to fit the 2-D temperatures is not the same. Despite these differences, it seems that we have independently come to the same answers, implying that our temperature prescriptions are comparable.

\subsection{Diffuse Emission} \label{sec:diffuse_surfaces}
Some disks exhibit diffuse emission at large radii, visible in the channel maps (see \citealt{exoALMA_channel_maps}), moment maps, and azimuthally averaged emission surfaces. We see a sharp drop-off in AA Tau and SY Cha in $^{12}$CO emission, and a similar morphology in LkCa 15 $^{13}$CO emission. In DM Tau, the outer surface does not taper, and instead extends outwardly as diffuse emission.

One explanation is that these disks may be experiencing UV photo-desorption. When a strong enough external radiation field is present, gas may be irradiated in the outer disk, which would allow for CO to emit from regions at and near the midplane, extending out radially. The emission morphology of IM Lup is is similar to that of the disks described here. \cite{cleeves_2016ApJ...832..110C} modeled IM Lup and the surrounding radiation field and found that UV photo-desorption can explain the extended emission in IM Lup. This was further confirmed by \cite{pinte_2018A&A...609A..47P}, who modeled IM Lup and found that a disk experiencing ISM radiation and UV chemistry most closely matched the observations. CO gas in the midplane indicative of photo-desorption has also been observed in the edge-on Flying Saucer disk \citep{dutrey_2017A&A...607A.130D}, which resides in the $\rho$ Ophiuchus star-forming region. Photo-desorption can be due to stellar irradiation, or from the surrounding interstellar medium. AA Tau, LkCa 15, and DM Tau are all located in the Taurus star forming region, which is lacking in massive stars compared to other star forming regions \citep{luhman_2004ApJ...617.1216L}. This means that Taurus has a lower far-ultraviolet (FUV) radiation field as compared to other SFRs, such as Lupus; Taurus is generally thought to have no photoionizing sources.

However, the diffuse emission we observe suggest it may be possible that even in low FUV areas like Taurus, disks still experience some photo-desorption. The disks that we observe the diffuse emission are some of the most massive in the exoALMA sample, with R$_{\rm edge}$ ranging from 644 au (AA Tau) to 990 au (DM Tau). It may be the case that the outer regions of the disks are low-density enough to be photo-desorbed, even in a FUV field like Taurus's. \maria{Recent work in \cite{anania_2025arXiv250118752A} indicate that regions with late-type B and early-type A stars such as Taurus can reach non-negligible levels of FUV radiation, in agreement with the preliminary findings here.} However, more work is needed to model these disks to determine if they are truly experiencing photo-desorption, or if there are other potential causes (see Section \ref{sec:projection_effects}).

\subsection{Origins and Implications of Substructures in Emission Surfaces and Intensity Profiles} \label{sec:discussion_substructure}

In Section \ref{sec:substucture}, we detailed the substructures identified in the emission surfaces and peak intensities, and compared their radial locations to continuum rings/gaps and velocity kink signatures. Here, we discuss the potential origin of these substructures, as well as their implications.

Many of the emission surface dips are co-located between molecular tracers. This suggests that these features are likely physical and coherent over a range of heights. 
Several of the surface gaps also coincide within a beamsize with peak intensity modulations in AA Tau, J1842, LkCa 15, and SY Cha. Planets forming within the disk are expected to open gaps, whose signatures could be imprinted on the azimuthally averaged emission surfaces. As shown in Figure \ref{fig:rolling_surfaces}, surface dips in the disks of AA Tau, J1842, and LkCa 15 coincide with the radial locations of velocity kinks. 

\maria{A few peak intensity bright spots coincide with proposed planet locations in AA Tau, J1615, LkCa 15, and SY Cha, which could be attributed to processes induced by planet formation. As a planet forms, there are several methods by which it is expected to alter the disk. Forming protoplanets will heat the surrounding gas \citep{cleeves_2015ApJ...807....2C, bae_2022ApJ...934L..20B}. This can happen via accretion heating, as the planet accretes during its formation \citep{szulagyi_2016MNRAS.460.2853S}, or shock heating as the disk material infalls onto the planet and/or circumplanetary disk \citep{szulagyi_2017MNRAS.465L..64S}. Despite this, the planet and subsequent CPD are expected to exist at the midplane, and their ability to cause localized heating in the molecular layers above this is unclear. The radial profiles are an azimuthal average, implying that the bright spots persist throughout the disk and do not just stem from a singular azimuthal location.} 
\maria{Additionally, it is expected that a planet will carve a gap in the gas as it forms, which would likely manifest as a temperature \textit{deficit} in the radial intensity profiles \citep{facchini_2018A&A...612A.104F}. Whether or not the signatures identified here are the result of forming planets is yet to be seen. Further efforts to model the impact of planet formation on the gaseous components of thermally stratified disks is needed to characterize these peak intensity signatures.}

Surface gaps could also be `artificially' induced when we make the assumption that the disk is rotating at the Keplerian velocity. If a disk is experiencing velocity perturbations, this could cause artifacts to the surface when the isovelocity curve is drawn (see also \citealt{paneque_2023A&A...669A.126P}, who discusses the same possibility). This could be another explanation as to why several of the surface gaps coincide with the proposed planet locations. However, even if the signatures are induced by non-Keplerian velocity deviations, their characterization is still helpful in pinpointing potential sites of planet formation.

\textit{LkCa 15.} Besides the peak intensity bright spot coinciding with the proposed planet location, the LkCa 15 disk exhibits `wiggles' in the $^{12}$CO $J=3-2$ peak intensity profile. The intensity peaks show a striking wave-like pattern that extends radially, with the bright spots being placed every $\sim$175 au. Additionally, one of the peak intensity bright spots at 333 au coincides within a beamsize with an emission surface gap at 342 au, and both of these signatures nearly coincide with the proposed planet location at 370 au (see Figures \ref{fig:rolling_surfaces}, \ref{fig:peak_intensity_structure}). \citet{exoALMA_line_profiles} explores the 2-D residual maps of several exoALMA sources; we find consistent radial peak intensity profiles, and the residual map of LkCa 15 clearly displays the peaks and bumps that persist azimuthally throughout the disk. In this work, we also see evenly spaced peak intensity `wiggles' to a lesser extent in AA Tau and SY Cha, but we focus on the clearest case of LkCa 15.

LkCa 15 is thought to have several planets embedded within its disk. First, there are several rings and a larger cavity in continuum emission \citep{andrews_2011ApJ...742L...5A, facchini_2020A&A...639A.121F, exoALMA_cont}, with dust trapping along two Lagrangian points \citep{long_2022ApJ...937L...1L}, all indicative of potential ongoing planet formation processes impacting the disk structure. Several studies within the exoALMA collaboration find further evidence of ongoing planet formation in LkCa 15. The disk exhibits a high degree of \dvphi{} structure, derived in \cite{exoALMA_rotation}, implicative of pressure variations that could be linked to planet formation. \cite{exoALMA_lkca15} models the continuum and gaseous disk components, concluding that the observed signatures are consistent with massive planets orbiting between 10 and 30 au. Although there are no direct detections of any planet in the disk, it remains likely that LkCa 15 has at least one embedded protoplanet. This raises the possibility that the peak intensity wiggles are linked to the planet formation process.

One theory is that the signatures seen in the LkCa 15 peak intensity profiles are the result of tightly-wound spirals. Several exoALMA sources display large-scale spirals (e.g., HD 135344B, MWC 758; \citealt{exoALMA_line_profiles, exoALMA_channel_maps}). While those sources have more obvious spirals, it is possible that the peak/bump structure in LkCa 15 actually stems from spirals, which could explain their even spacing. Simulations have established that planets can launch spiral arms throughout the disk \citep{kley_1999astro.ph..9394K, dong_2015ApJ...809...93D, bae_2016ApJ...829...13B}, through Linblad resonances \citep{ogilvie_2002MNRAS.330..950O,bae_2018ApJ...859..118B,bae_2018ApJ...859..119B} or buoyancy resonances \citep{zhu_2012ApJ...758L..42Z, bae_2021ApJ...912...56B}. 
Observationally, temperature spirals have been characterized in TW Hya and MWC 480 \citep{teague_2019ApJ...884L..56T, wolfer_2021A&A...648A..19W, teague_2021ApJS..257...18T, discminer2_2023A&A...674A.113I}, where the authors postulate that a giant planet may have induced the signatures. However, if the temperature structure in LkCa 15 is driven by buoyancy spirals, we would expect the velocity residuals to be dominated by vertical gas motions, $v_z$. Based on the residuals in \citealt{exoALMA_line_profiles} (see Figure 10), this may not be the case in LkCa 15. Because the peak intensity structure aligns with some of the structure seen in $v_{\phi}$ \citep{exoALMA_rotation}, it is more likely that these peaks are driven by pressure variations. More work is needed to determine if the disk structure is truly a spiral, or rather concentric peaks and troughs. If a planet is driving the temperature variations, it likely exists at large radii; modeling the impacts of giant planet formation on the outer gaseous disk remains a topic of interest.

Thermal wave instability, or irradiation instability, could also produce wave-like features. The central star is responsible for much of the disk heating; this is ly related to the disk thickness and orientation. If parts of the disk receive more thermal radiation, this can lead to a growing perturbation that will produce peaks and troughs in the surface density \citep{dalessio_1999ApJ...511..896D, siebenmorgen_2012A&A...539A..20S, ueda_2021ApJ...914L..38U, wu_2021ApJ...923..123W}.  
However, these studies have generally focused on the inner disk; this instability is known to operate at radii closer to the star because of the thermal requirement. The wave structure in LkCa 15 extends out to 900 au, which may be too far for irradiation instability to effectively operate.

Of course, other physical mechanisms are expected to cause disk perturbations as well. Magnetorotational instability (MRI), which generates turbulence, can excite density perturbations which could manifest as spirals \citep{flock_2015A&A...574A..68F}. However, MRI is not expected to produce symmetric features, which makes it an unlikely candidate. Vertical shear instability (VSI, \citealt{nelson_2013MNRAS.435.2610N}) can also induce surface density perturbations. Due to angular velocity, the surface layers of protoplanetary disks rotate more slowly than the midplane, which naturally induces VSI. Numerical simulations have shown that VSI can produce antisymmetric structures. \cite{exoALMA_turbulence} and \cite{barraza_2021A&A...653A.113B} find that VSI can manifest as concentric rings in the velocity residuals (i.e. the `corrugation mode'), but these signatures are not seen in the modeled peak intensity maps. Concentric patterns in the surface density and velocity structure are also retrieved in simulations by \cite{zhang_2024ApJ...968...29Z}, who investigates VSI in a thermally stratified disk. They find that perturbations in the gas surface density can develop, sometimes exhibiting wave-like structure. However, due to the high inclination of LkCa 15, it is difficult to definitively say that the signatures are driven by VSI. It is likely that multiple physical process are impacting the disk, and further investigation and modeling work is needed.

\subsubsection{Projection Effects} \label{sec:projection_effects}

Throughout Section \ref{sec:substucture}, we have alluded to the fact that some of the observed substructures can be explained by projection effects, and may not be physical. Here, we discuss the ways in which our viewing angle, optical depth effects, \maria{and temperature disparities} could cause artificially induced dips. 

\textit{Diffuse Emission.} First, the `diffuse' emission detected in AA Tau, LkCa 15, and SY Cha all has a characteristic shape, seen in Figures \ref{fig:surfaces_12co} and \ref{fig:surfaces_13co}. All of these disks are highly inclined ($i$ \textgreater~ 50$^{\circ}$), and all of the surfaces suffer an abrupt drop, at which point the more diffuse emission from closer to the midplane continues. In the channel maps, this emission does appear as a combination between of emission from the front, back, and midplane, which implies that CO is not frozen at the midplane and indicitave of photo-desorption. However, at high inclinations, from our viewing the back surface will appear to extend beyond the front surface, and in some cases can appear just as bright. With real observations, this projection effect could mimic diffuse emission, particularly in the presence of noise or non-Keplerian rotation. One hint that this effect may be at play is that we only see a sharp surface drop in LkCa 15 $^{13}$CO emission, not $^{12}$CO. If the morphology is due to photo-desorption, it is unlikely that it would only impact $^{13}$CO.

\begin{figure}[h]
\centering
\begin{tabular}{ll}
\includegraphics[scale=0.5]{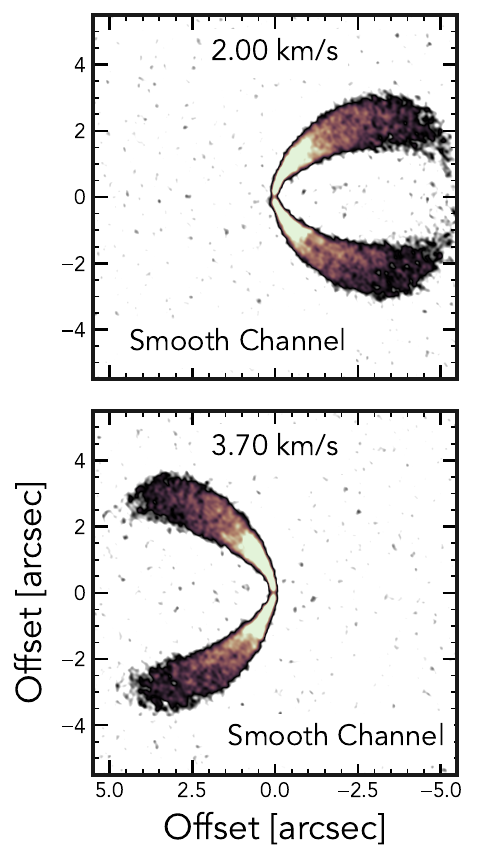}
&
\includegraphics[scale=0.5]{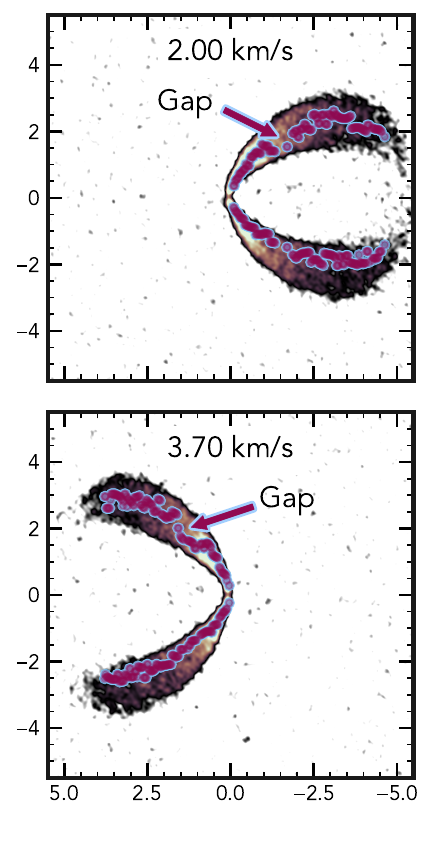}
\end{tabular}
\caption{Two channel maps of V4046 Sgr at 2.00 km/s and 3.70 km/s. The left column shows two channels with no points overlaid, and the right column shows the channels with the raw emission surface points identified by \disksurf{}. There is a notable gap in the emission surface points that is not visible in the channels.}
\label{Fig:v4046}
\end{figure}

\textit{V4046 Sgr.} The V4046 Sgr disk exhibits one of the most prominent \maria{emission surface substructures} in both width and depth out of the exoALMA sample in both $^{12}$CO and $^{13}$CO at 174 au, \maria{where a large decrease in surface height can be seen in Figure \ref{fig:binned_surfaces}}. Curiously, the channel maps and peak intensity map of this disk appear relatively smooth, and no gap can be identified by eye. The intensity does appear to decrease more sharply past $\sim$150 au (see Figure \ref{fig:radial_temp_profile}), but other disks that have similar intensity profile slope changes do not show large gaps. \cite{exoALMA_channel_maps} presents the channel maps of V4046 Sgr, and notes that there are no structures visible by eye. Figure \ref{Fig:v4046} presents two channels in $^{12}$CO; the left column shows the channels without anything overlaid, and the right column shows the channels with the \disksurf{} emission surface points for the respective channels. There is no gap visible by eye in the left columns, but the emission surface points do have a gap at the same radial location in multiple channels. It is uncommon that such a large surface gap identified via \disksurf{} is not visible in the channel maps.

It is evident from Figure \ref{Fig:v4046} that the intensity of the disk drops off dramatically at the location of the gap. This intensity drop can also be seen in the peak intensity at around 150 au, shown in Figure \ref{fig:radial_temp_profile}. The gap could be related to the intensity drop; decreases in column density or optical depth could result in a decreased intensity. However, other disks that show the same level of intensity change do not exhibit large gaps as V4046 Sgr does. Additionally, although the intensity decreases, in $^{12}$CO it remains above 20 K, which means the gas likely remains moderately optically thick.

The gap in the V4046 Sgr surface may be caused by projection effects. Without a visible gap in the channel or moment maps, it is unlikely that one actually exists, \maria{although there are other possibilities}. \disksurf{} works by identifying peak points along an isovelocity curve \citep{pinte_2018A&A...609A..47P}; in general, this process can accurately identify real gaseous emission gaps \citep{MAPS_emission_surfaces_2021ApJS..257....4L, law_2022ApJ...932..114L, law_2023ApJ...948...60L,  galloway_2023ApJ...950..147G, paneque_2023A&A...669A.126P}. In the case of V4046 Sgr, we hypothesize that this gap is actually caused by low optical depth (e.g. $\tau$ $\simeq$ 1) or \maria{low temperature} of the front surface. If the optical depth \maria{and or temperature} of the CO emission begins to fall off, this would occur at different radial locations from our viewing angle, due to the inclined angle of the disk. At a certain radial location where the front and back surface are best separated, the back surface would appear slightly brighter than the front. The fact that we only see the gap in the upper channels, or the near-side of disk, lends more credence to this theory; the opposite effect would be at play for the far side of the disk and we wouldn't see this effect. Inspecting the isovelocity contours from \discminer{} models reveals that the gap falls over the predicted back surface. Therefore, we conclude that this gap is \maria{likely} not physically real, but rather an effect caused by the projection of the disk, \maria{although other explanations remain possible.}

We believe that the gap in V4046 Sgr is the only obvious case of projection effects. The other surface modulations, discussed in Section \ref{sec:substucture}, are found in disks that are not smoothly varying. Many of the substructures can be identified visually on the channel maps, and are of true physical origin. For future high-resolution imaging campaigns, it will be critical to identify and understand the role that projection plays on our observations.

\section{Conclusions} \label{sec:conclusions}
This work presents the emission surfaces, 2-D temperature profiles, and radial intensity profiles of the exoALMA sources. Our analysis leads to the following conclusions:

\begin{enumerate}
  \item Nearly every disk in the sample exhibits stratified molecular layers, with $^{12}$CO $J=3-2$ tracing elevated regions of the disk, with an average $\langle z / r \rangle$ of $\approx$ 0.28. $^{13}$CO $J=3-2$ traces layers below that of $^{12}$CO, with an average $\langle z / r \rangle$ of $\approx$ 0.16. CS $J=7-6$ lies closest to the midplane, but still exhibits elevated layers in several disks, with an average $\langle z / r \rangle$ of $\approx$ 0.18.
  \item We search for `substructure', i.e. gaps and dips, in the emission surfaces and radial profiles. Most disks exhibit substructure in both the emission surfaces and temperature profiles, often co-incident between molecules and in some cases coinciding with proposed planet locations.
    \begin{itemize}
        \item In AA Tau, DM Tau, LkCa 15, and SY Cha we see evidence of diffuse emission at large radii, which generally manifest in the emission surfaces as a sharp drop with subsequent midplane emission. This structure may be due to photo-desorption, which implys that this effect may be common even in low FUV star forming regions. 
        \item In CQ Tau, DM Tau, J1842, J1852, LkCa 15, MWC 758, and V4046 Sgr, we find correlation of emission surface substructures between separate molecular layers, potentially indicative of physical processes which impact the vertical height of the disk at these locations.
        \item In AA Tau, DM Tau, J1615, J1842, LkCa 15, MWC 758, and SY Cha, we find correlation between peak intensity substructures and emission surface substructures. 
        \item We do not find statistically significant correlation between the continuum rings and gaps and any surface substructure. The notable exceptions are AA Tau, J1615, and MWC 758. In the CS surfaces of AA Tau and J1615, we see hints of surface height changes at the location of gaps in the continuum. In MWC 758, the continuum rings and gaps overlap with those identified in the $^{12}$CO and $^{13}$CO surfaces. 
        \item For some disks, structures in the the emission surfaces and peak intensities overlap with velocity perturbations seen 
        in the channel maps, discussed in \citealt{exoALMA_channel_maps}. Whether the surface substructures are truly physical decreases of column density or intensity within the disk, or are instead an `artificial' detection driven by the velocity perturbations themselves, is still unclear. Future work to characterize the origin of these substructures is needed.
        \item Projection effects due to inclination and viewing angle may cause artificial structures in our observations. More work is needed to characterize the interplay between disk position and \maria{identified substructures}.
    
    \end{itemize}

  \item The 2-D temperature profiles derived suggest the average T$_{\rm atm}$ temperature is $\sim$43 K, whereas the average T$_{\rm mid}$ temperature is found to be $\sim$24 K. This midplane temperature is hotter than that predicted by the scale height equation and simulations, and suggests a need for an improved temperature profile prescription and methods to better constrain these temperatures.
  \item We compare our work to previous studies and find consistent results. \maria{Stellar} mass and average $^{12}$CO temperature is loosely correlated with $^{12}$CO surface height, but no definite relations can be identified. Surface height and the outer disk radius R$_{\rm CO}$ show a strong linear relation as seen in \cite{MAPS_emission_surfaces_2021ApJS..257....4L, law_2022ApJ...932..114L, law_2023ApJ...948...60L, paneque_2023A&A...669A.126P}. Additional linear relations are found between \zr{}, disk mass, and surface density, suggesting that the emission height is inherently related to these quantities. We propose that \zr{} could be used to infer the disk mass.
  \item The data discussed in this paper is provided as a publicly available Value-Added-Data-Product (VADP). The non-parametric emission surfaces (raw, binned, and rolling points, along with their temperatures), parametric surfaces, peak intensity radial profiles, integrated intensity radial profiles, and the radial locations of the surface and intensity modulations are all included in the VADP release.
\end{enumerate}

\section*{Acknowledgments}
The authors thank the referee for the insightful report that significantly improved the quality of the paper.

This paper makes use of the following ALMA data: ADS/JAO.ALMA\#2021.1.01123.L. ALMA is a partnership of ESO (representing its member states), NSF (USA) and NINS (Japan), together with NRC (Canada), MOST and ASIAA (Taiwan), and KASI (Republic of Korea), in cooperation with the Republic of Chile. The Joint ALMA Observatory is operated by ESO, AUI/NRAO and NAOJ. The National Radio Astronomy Observatory is a facility of the National Science Foundation operated under cooperative agreement by Associated Universities, Inc. We thank the North American ALMA Science Center (NAASC) for their generous support including providing computing facilities and financial support for student attendance at workshops and publications. 
The National Radio Astronomy Observatory and Green Bank Observatory are facilities of the U.S. National Science Foundation operated under cooperative agreement by Associated Universities, Inc.

JB acknowledges support from NASA XRP grant No. 80NSSC23K1312. Support for AFI was provided by NASA through the NASA Hubble Fellowship grant No. HST-HF2-51532.001-A awarded by the Space Telescope Science Institute, which is operated by the Association of Universities for Research in Astronomy, Inc., for NASA, under contract NAS5-26555. 
JS has received funding from the European Research Council (ERC) under the European Union’s Horizon 2020 research and innovation programme (PROTOPLANETS, grant agreement No. 101002188). Computations have been done on the ’Mesocentre SIGAMM’ machine, hosted by Observatoire de la Cote d’Azur. 
CL has received funding from the European Union's Horizon 2020 research and innovation program under the Marie Sklodowska-Curie grant agreement No. 823823 (DUSTBUSTERS) and by the UK Science and Technology research Council (STFC) via the consolidated grant ST/W000997/1. 
AJW has received funding from the European Union’s Horizon 2020
research and innovation programme under the Marie Skłodowska-Curie grant
agreement No 101104656. 
MB has received funding from the European Research Council (ERC) under the European Union’s Horizon 2020 research and innovation programme (PROTOPLANETS, grant agreement No. 101002188). 
S.F. is funded by the European Union (ERC, UNVEIL, 101076613). Views and opinions expressed are however those of the author(s) only and do not necessarily reflect those of the European Union or the European Research Council. Neither the European Union nor the granting authority can be held responsible for them. S.F. acknowledges financial contribution from PRIN-MUR 2022YP5ACE. 
GR acknowledges funding from the Fondazione Cariplo, grant no. 2022-1217, and the European Research Council (ERC) under the European Union’s Horizon Europe Research \& Innovation Programme under grant agreement no. 101039651 (DiscEvol). Views and opinions expressed are however those of the author(s) only, and do not necessarily reflect those of the European Union or the European Research Council Executive Agency. Neither the European Union nor the granting authority can be held responsible for them. 
Support for BZ was provided by The Brinson Foundation. 
C.P. acknowledges Australian Research Council funding  via FT170100040, DP18010423, DP220103767, and DP240103290. 
DF has received funding from the European Research Council (ERC) under the European Union’s Horizon 2020 research and innovation programme (PROTOPLANETS, grant agreement No. 101002188). 

NC has received funding from the European Research Council (ERC) under the European Union Horizon Europe research and innovation program (grant agreement No. 101042275, project Stellar-MADE).
PC acknowledges support by the Italian Ministero dell'Istruzione, Universit\`a e Ricerca through the grant Progetti Premiali 2012 – iALMA (CUP C52I13000140001) and by the ANID BASAL project FB210003. 
MF is supported by a Grant-in-Aid from the Japan Society for the Promotion of Science (KAKENHI: No. JP22H01274).
MF is supported by a Grant-in-Aid from the Japan Society for the Promotion of Science (KAKENHI: No. JP22H01274). 
CHG acknowledges support from the National Aeronautics and Space Administration under grant No. 80NSSC18K0828. CHG gratefully acknowledges the support by an LANL/CSES Student Fellow project.
CH acknowledges support from NSF AAG grant No. 2407679.
JDI acknowledges support from an STFC Ernest Rutherford Fellowship (ST/W004119/1) and a University Academic Fellowship from the University of Leeds. 
CL has received funding from the European Union's Horizon 2020 research and innovation program under the Marie Sklodowska-Curie grant agreement No. 823823 (DUSTBUSTERS).
FMe acknowledges funding from the European Research Council (ERC) under the European Union's Horizon Europe research and innovation program (grant agreement No. 101053020, project Dust2Planets).
D.P. acknowledges Australian Research Council funding via DP18010423, DP220103767, and DP240103290.
GWF acknowledges support from the European Research Council (ERC) under the European Union Horizon 2020 research and innovation program (Grant agreement no. 815559 (MHDiscs)). GWF was granted access to the HPC resources of IDRIS under the allocation A0120402231 made by GENCI.
H.-W.Y.\ acknowledges support from National Science and Technology Council (NSTC) in Taiwan through grant NSTC 113-2112-M-001-035- and from the Academia Sinica Career Development Award (AS-CDA-111-M03).
TCY acknowledges support by Grant-in-Aid for JSPS Fellows JP23KJ1008.

\clearpage

\newpage
\appendix
\label{appendix}
\section{Emission Surfaces}\label{sec:emission_surf_extra}

\begin{figure*}[!ht]
    \centering
    \includegraphics[width=1.0\textwidth]{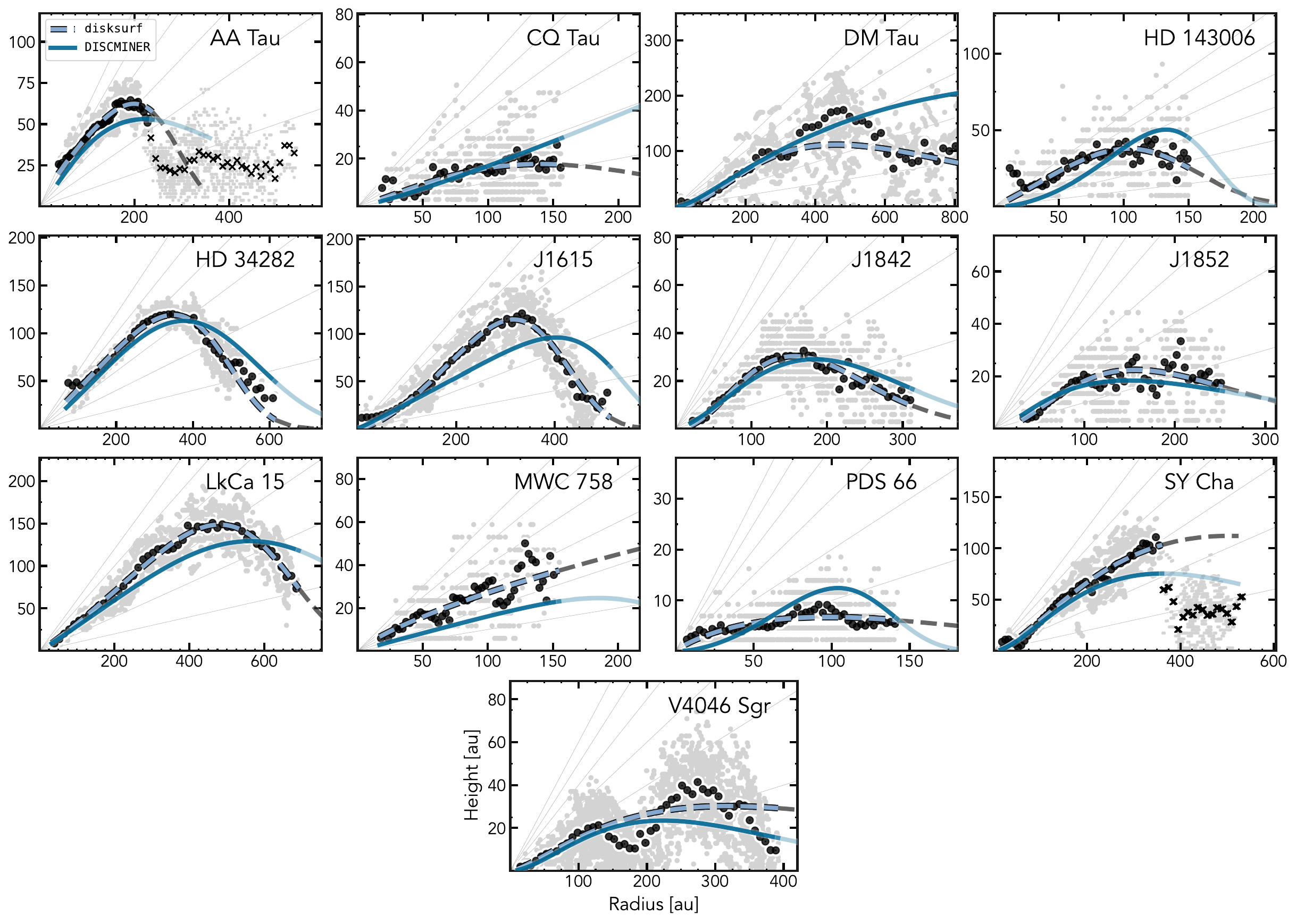}
    \caption{Non-parametric (data points) and parametric surfaces (solid and dashed curves) for $^{12}$CO $J=3-2$. The larger black points represent radially binned data. We exclude two disks \maria{(HD 135344B and J1604)} whose surfaces could not be extracted due to a face-on geometry. Background gray lines show scale heights $(z/r)$ from 0.1 to 0.6 with 0.1 increments. The darker blue solid line shows the tapered power law surface fit from \discminer{}, and the light blue and gray dashed line shows the tapered power law fit from \disksurf{}, which uses the background data points. The gray cross markers for AA Tau and SY Cha show the diffuse outer surface points that we do not include in the parametric fits.}
    \label{fig:surfaces_12co}
\end{figure*}

\begin{figure*}[!ht]
    \centering
    \includegraphics[width=1.0\textwidth]{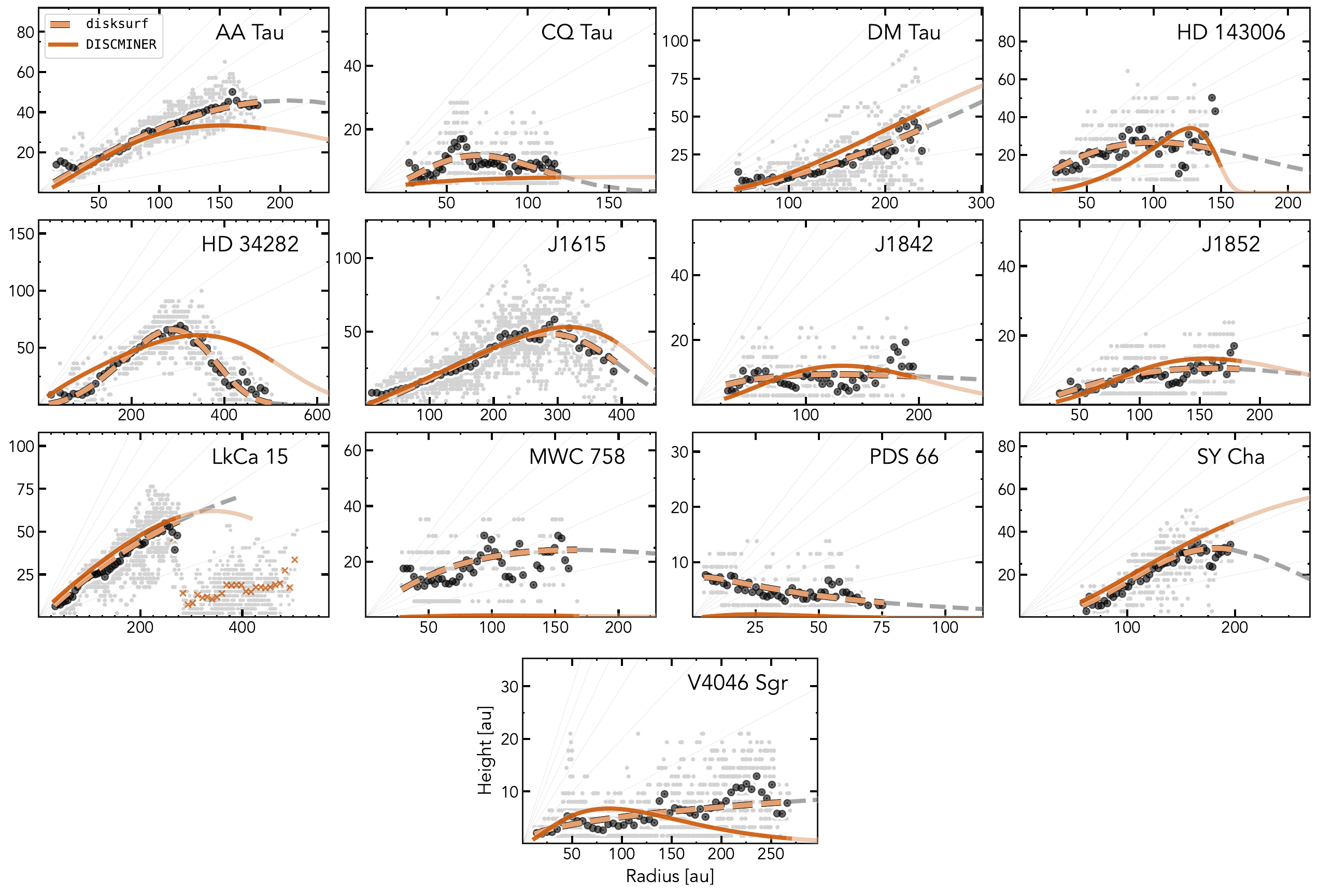}
    \caption{The same as Figure \ref{fig:surfaces_12co}, but for $^{13}$CO $J=3-2$. The orange solid line shows the tapered power law surface fit from \discminer{}, and the gray and peach dashed line shows the tapered power law fit from \disksurf{}.}
    \label{fig:surfaces_13co}
\end{figure*}

\begin{figure*}[t]
    \centering
    \includegraphics[width=1.0\textwidth]{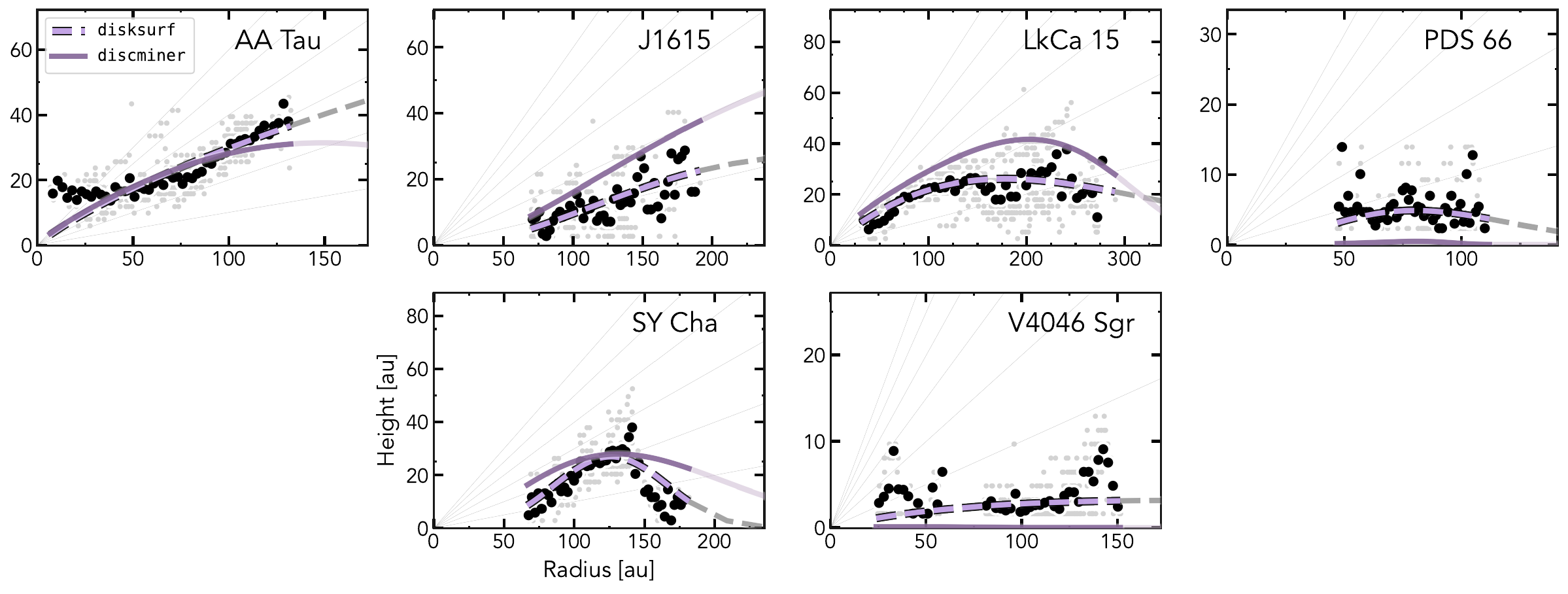}
    \caption{The same as Figure \ref{fig:surfaces_12co}, but for CS $J=7-6$. The purple solid line shows the tapered power law surface fit from \discminer{}, and the gray and lavender dashed line shows the tapered power law fit from \disksurf{}.}
    \label{fig:surfaces_cs}
\end{figure*}

Figures \ref{fig:surfaces_12co}, \ref{fig:surfaces_13co}, and \ref{fig:surfaces_cs} depict the emission surfaces for $^{12}$CO $J=3-2$, $^{13}$CO $J=3-2$, and CS $J=7-6$, respectively. In all three figures, the `raw' emission surface $r-z$ points are shown in gray. The larger black points show these raw surfaces binned by a quarter of the beamsize. The dotted lines show the tapered power law fit from \disksurf{}, fit to the background gray points. The solid lines show the tapered power law fit from \discminer{}, derived via line profile modeling. 
Table \ref{tab:parametric_surfaces} presents the tapered power-law fits from both codes for all molecules.

\begin{figure*}[!ht]
    \centering
    \includegraphics[width=1.0\textwidth]{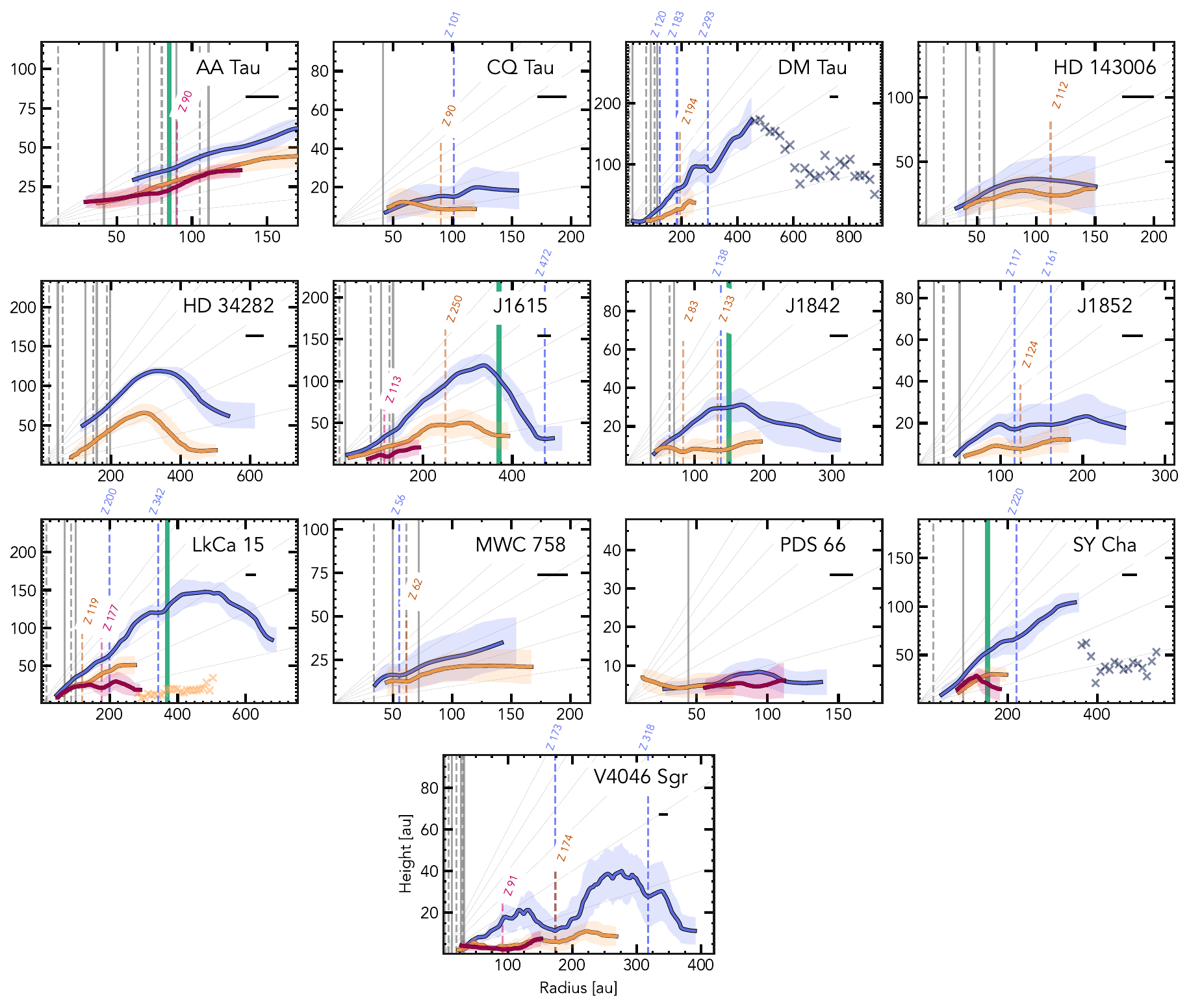}
    \caption{Rolling surfaces made from the raw $r-z$ \disksurf{} emission surface points for $^{12}$CO $J=3-2$ (blue), $^{13}$CO $J=3-2$ (orange), and CS $J=7-6$ (pink). Vertical dashed lines denote emission surface substructure, whose identification is discussed in Section \ref{sec:substucture}. We label dips in the emission surfaces with the nomenclature first used in \cite{MAPS_emission_surfaces_2021ApJS..257....4L}, where `Z' denotes a dip in the emission surface, followed by the radial location in au. We also plot the continuum rings (solid gray lines) and gaps (dashed gray lines) identified in \cite{exoALMA_cont}.
    Thin background gray lines show scale heights $(z/r)$ from 0.1 to 0.6 with 0.1 increments. The beamsize of $0\farcs15$ is shown in the upper right corner for each disk. AA Tau is zoomed in to better show structure.}
    \label{fig:rolling_surfaces}
\end{figure*}

Figure \ref{fig:rolling_surfaces} shows the rolling surfaces derived from the non-parametric surfaces. These rolling surfaces were used to identify substructures in the emission surfaces as discussed in Section \ref{sec:substucture}. The surface substructures are plotted as colored lines and denoted with a Z followed by the radial location in au, following the nomenclature from \cite{MAPS_emission_surfaces_2021ApJS..257....4L}. The continuum rings and gaps from \cite{exoALMA_cont} are also plotted as gray solid and dashed lines.

\begin{table*}[!ht]
    \centering
    \caption{Exponentially tapered power law fits for all disks and all molecules from \disksurf{} and \discminer{}; characteristic $\langle z / r \rangle$ values determined from \disksurf{} data points.}
    \label{tab:parametric_surfaces}
    \begin{tabular}{c c c  c c c c| c c c c c}
       \hline \hline

    \multicolumn{1}{c}{Source} & \multicolumn{1}{c}{Line} & \multicolumn{1}{c}{$z/r$} & \multicolumn{4}{c}{\disksurf{}} & \multicolumn{4}{c}{\discminer{}} \\
    \cline{4-7} 
     \cline{7-11}
    \multicolumn{2}{c}{} &  & z$_0$ (au) & $\psi$  &  r$_{taper}$ (au) & q$_{taper}$ &z$_0$ (au) & $\psi$  & r$_{taper}$ (au) & q$_{taper}$  \\ \hline

    DM Tau & $^{12}$CO $J=3-2$ & 0.34 & 25.2 & 1.8 & 366 & 3.2 & 87 & 1.9 & 79.6 & 0.48 \\
    ~ & $^{13}$CO $J=3-2$ & 0.14 & 21.5 & 1.9 & 262 & 0.35 & 20 & 2.3 & 242 & 0.93 \\
    ~ & CS $J=7-6$ & - & - & - & - & - & 6.5 & 3.2 & 94.1 & 0.92 \\ \hline
    AA Tau & $^{12}$CO $J=3-2$ & 0.41 & 42.5 & 0.79 & 289 & 4.8 & 50 & 1.2 & 240 & 1.3 \\
    ~ & $^{13}$CO $J=3-2$ & 0.32 & 33.4 & 0.92 & 309 & 2.6 & 52 & 1.4 & 151 & 1.4 \\
    ~ & CS $J=7-6$ & 0.32 & 70.9 & 1 & 164 & 0.28 & 33 & 0.83 & 232 & 2.2 \\ \hline
    LkCa 15 & $^{12}$CO $J=3-2$ & 0.34 & 32.2 & 1.1 & 649 & 4.8 & 29 & 1.1 & 795 & 3.2 \\
    ~ & $^{13}$CO $J=3-2$ & 0.21 & 36.4 & 1.2 & 399 & 0.65 & 27 & 0.87 & 511 & 3.5 \\
    ~ & CS $J=7-6$ & 0.21 & 28.3 & 0.94 & 244 & 1.5 & 29 & 0.72 & 303 & 4.6 \\ \hline
    HD 34282 & $^{12}$CO $J=3-2$ & 0.41 & 41.3 & 1 & 480 & 4.9 & 34 & 1.2 & 512 & 3.2 \\
    ~ & $^{13}$CO $J=3-2$ & 0.15 & 12.8 & 2 & 347 & 4.6 & 27 & 0.79 & 510 & 4.4 \\
    ~ & CS $J=7-6$ & - & - & - & - & - & 16 & 1.7 & 268 & 1.8 \\ \hline
    MWC 758 & $^{12}$CO $J=3-2$ & 0.29 & 64.9 & 0.96 & 176 & 0.24 & 16 & 0.97 & 254 & 5.3 \\
    ~ & $^{13}$CO $J=3-2$ & - & 34.1 & 0.89 & 204 & 1.1 & 67 & 3.2 & 12.5 & 0.71 \\
    ~ & CS $J=7-6$ & - & - & - & - & - & 6.1 & 5 & 120 & 3.3 \\ \hline
    CQ Tau & $^{12}$CO $J=3-2$ & 0.17 & 21.7 & 1.1 & 187 & 1.9 & 42 & 1.2 & 346 & 0.087 \\
    ~ & $^{13}$CO $J=3-2$ & 0.14 & 53.8 & 1.8 & 73.8 & 2 & 39 & 1.1 & 16.6 & 0.41 \\
    ~ & CS $J=7-6$ & - & - & - & - & - & 28 & 4.6 & 80.2 & 1.5 \\ \hline
    SY Cha & $^{12}$CO $J=3-2$ & 0.29 & 40.1 & 1.6 & 335 & 1 & 43 & 1.8 & 210 & 1 \\
    ~ & $^{13}$CO $J=3-2$ & 0.16 & 21.3 & 2.3 & 185 & 2.4 & 73 & 2.4 & 66.1 & 0.7 \\
    ~ & CS $J=7-6$ & 0.16 & 34.1 & 3.1 & 128 & 3.2 & 50 & 1.9 & 124 & 1.7 \\ \hline
    PDS 66 & $^{12}$CO $J=3-2$ & 0.12 & 12.4 & 0.81 & 143 & 1.4 & 17 & 1.8 & 127 & 4.5 \\
    ~ & $^{13}$CO $J=3-2$ & - & 13.5 & 0.14 & 43.2 & 0.79 & 7.5 & 1.2 & 29 & 1.5 \\
    ~ & CS $J=7-6$ & - & 38.1 & 2.5 & 62.6 & 1.7 & 1.2 & 2.9 & 91.7 & 8.4 \\ \hline
    HD 135344B & $^{12}$CO $J=3-2$ & - & - & - & - & - & 14 & 1.4 & 226 & 10 \\
    ~ & $^{13}$CO $J=3-2$ & - & - & - & - & - & - & - & - & - \\
    ~ & CS $J=7-6$ & - & - & - & - & - & - & - & - & - \\ \hline
    HD 143006 & $^{12}$CO $J=3-2$ & 0.4 & 49.6 & 1.1 & 146 & 3.4 & 40 & 1.9 & 161 & 6 \\
    ~ & $^{13}$CO $J=3-2$ & 0.29 & 43.4 & 0.91 & 147 & 1.8 & 24 & 2.2 & 146 & 13 \\
    ~ & CS $J=7-6$ & - & - & - & - & - & 17 & 0.72 & 103 & 4.5 \\ \hline
    RXJ1604.3-2130 A & $^{12}$CO $J=3-2$ & - & - & - & - & - & - & - & - & - \\
    ~ & $^{13}$CO $J=3-2$ & - & - & - & - & - & - & - & - & - \\
    ~ & CS $J=7-6$ & - & - & - & - & - & - & - & - & - \\ \hline
    RXJ1615.3-3255 & $^{12}$CO $J=3-2$ & 0.34 & 27.9 & 1.5 & 403 & 5 & 26 & 1 & 530 & 6.9 \\
    ~ & $^{13}$CO $J=3-2$ & 0.17 & 17.7 & 1.2 & 381 & 4.9 & 19 & 1 & 425 & 5.9 \\
    ~ & CS $J=7-6$ & 0.17 & 14.9 & 2.3 & 185 & 1.3 & 38 & 2.2 & 121 & 0.8 \\ \hline
    V4046 Sgr & $^{12}$CO $J=3-2$ & 0.13 & 26.4 & 1.5 & 196 & 0.94 & 26 & 1.8 & 151 & 1.2 \\
    ~ & $^{13}$CO $J=3-2$ & 0.052 & 12.9 & 0.54 & 257 & 0.093 & 33 & 1.6 & 65.6 & 1.1 \\
    ~ & CS $J=7-6$ & 0.052 & 5.52 & 1 & 153 & 0.86 & 0.086 & 0.0039 & 85.9 & 7.1 \\ \hline
    RXJ1842.9-3532 & $^{12}$CO $J=3-2$ & 0.2 & 47.4 & 2.2 & 124 & 1.5 & 26 & 1.5 & 211 & 1.9 \\
    ~ & $^{13}$CO $J=3-2$ & 0.091 & 15 & 0.61 & 207 & 0.98 & 17 & 1.7 & 143 & 2 \\
    ~ & CS $J=7-6$ & - & - & - & - & - & 29 & 1.4 & 145 & 4.5 \\ \hline
    RXJ1852.3-3700 & $^{12}$CO $J=3-2$ & 0.15 & 63.1 & 2.2 & 82.1 & 1.1 & 75 & 1.8 & 60.9 & 0.84 \\
    ~ & $^{13}$CO $J=3-2$ & 0.079 & 15.7 & 1.4 & 155 & 1.3 & 31 & 2.7 & 90.3 & 1.3 \\
    ~ & CS $J=7-6$ & - & - & - & - & - & 1 & 3.6 & 108 & 1.4 \\ \hline

    \end{tabular}
\end{table*}

\clearpage
\section{Temperature Profiles}\label{sec:app_temp}

\begin{figure*}[!ht]
    \centering
    \includegraphics[width=1.0\textwidth]{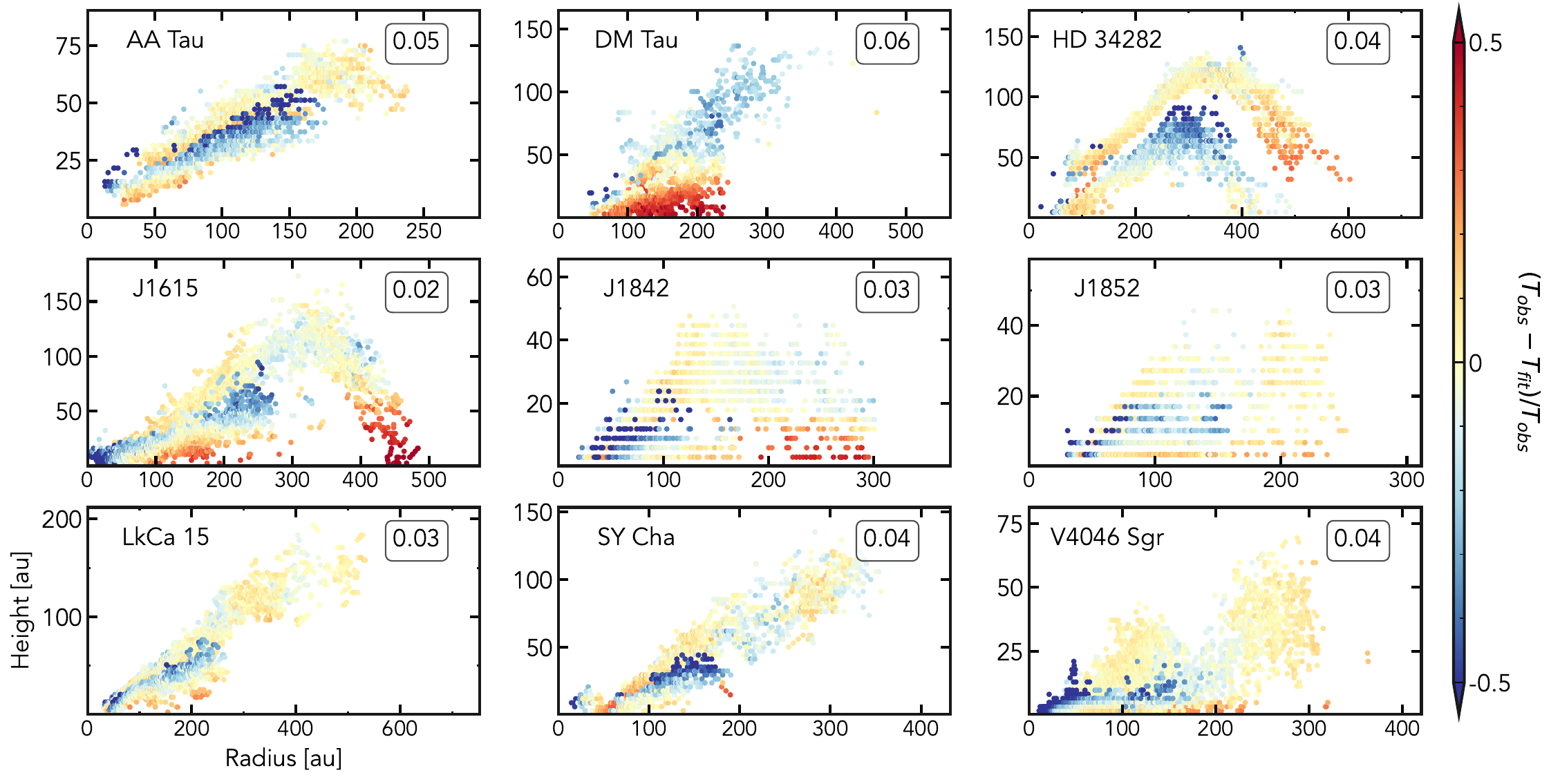}
    \caption{Non-parametric surface points used to calculate the 2-D temperature in $r-z$ space. The color of the points corresponds to the difference between the observed brightness temperature (T$_{\rm obs}$) and the temperature found by the 2-D fit (T$_{\rm fit}$). The boxed numbers in the upper right corners represent the absolute value of the average $\frac{T_{\rm obs} - T_{\rm fit}}{T_{\rm obs}}$; values closer to zero indicate the 2-D fit more closely matches the observed temperature structures.}
    \label{fig:temp_profile_errors}
\end{figure*}

Figure \ref{fig:temp_profile_errors} shows the non-parametric surface points used to calculate the 2-D temperature profiles. The corresponding color of each point is calculated using $\frac{T_{\rm obs} - T_{\rm fit}}{T_{\rm obs}}$, where T$_{\rm obs}$ is the observed brightness temperature of the point, and T$_{\rm fit}$ is the corresponding value that the 2-D temperature fit finds. Values closer to zero indicate the 2-D fit more closely matches the observed temperature structures. 
Colors closer to blue imply that the temperature fit is underestimating the observed temperature, and colors closer to red imply that the temperature fit is overestimating the observed temperature.
It is evident that the 2-D temperature profiles cannot account for areas where molecular layers overlap; these areas are visible particularly where the top of the $^{13}$CO emission layer overlaps with the bottom of the $^{12}$CO layer. Additionally, many of the points close to the midplane are overestimated in the 2-D temperature fits.

These results highlight a need for a better analytical temperature profile. Neither of most commonly-used temperature profile equations, i.e. the sinusoidal form from \cite{dartois_2003A&A...399..773D}, and the tangential form from \cite{dullemond_2020A&A...633A.137D}, cannot adequately describe the temperature change between overlapping molecular layers. Additionally, both functional forms tend to be highly sensitive to the imposed priors, which means much of the results are imparted from our initial assumptions of the atmospheric and midplane temperature functions.

\section{Radial Profiles}\label{sec:app_radprofiles}
\begin{figure*}[!ht]
    \centering
    \includegraphics[width=1.0\textwidth]{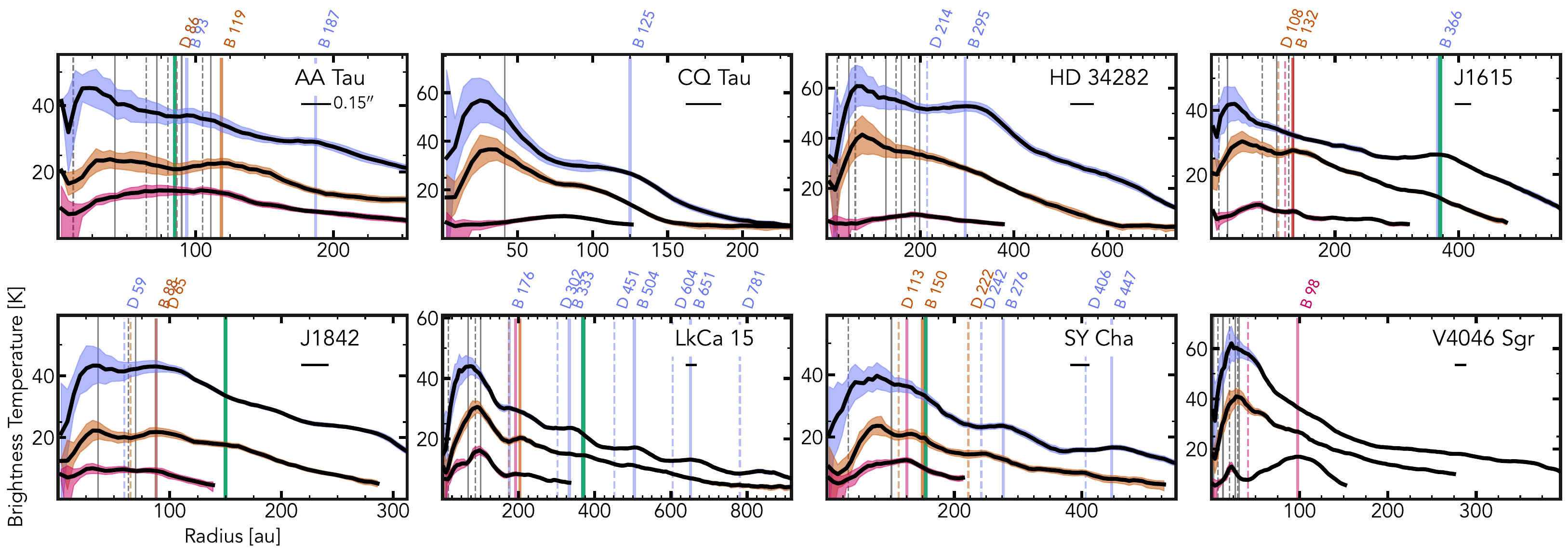}
    \caption{Peak intensities derived from \gofish{} for disks with identified intensity structure. Intensity bright spots are marked with solid lines and labeled with a `B', and intensity dips are marked with dashed lines and labeled with a 'D'. \maria{Several bright spots and gaps are unlabeled due to overlap; we refer the reader to Table \ref{tab:substructures} for the full list of radial locations.} Blue corresponds to $^{12}$CO, orange corresponds to $^{13}$CO, and pink corresponds to CS. Proposed planet locations, discussed in \cite{exoALMA_channel_maps}, are shown at their suggested radial locations in green. Continuum rings and gaps, identified and discussed in \cite{exoALMA_cont}, are shown as gray solid lines (rings) and dashed lines (gaps).}
    \label{fig:peak_intensity_structure}
\end{figure*}

\begin{figure*}[!ht]
    \centering
    \includegraphics[width=1.0\textwidth]{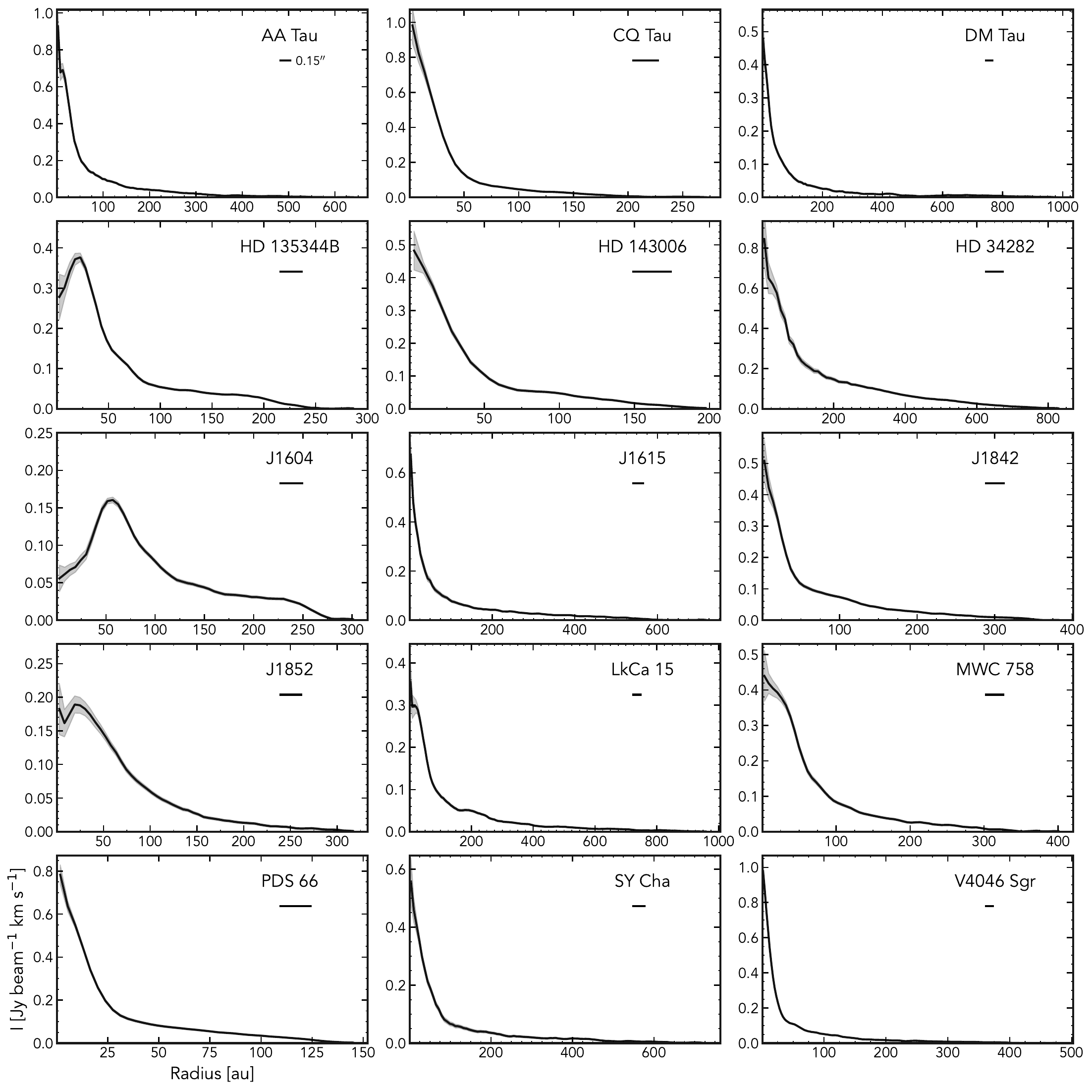}
    \caption{Integrated intensity profiles for $^{12}$CO $J=3-2$. \maria{The shaded regions represent the upper and lower uncertainties. The beamsize of 0$\farcs$15 is shown in the top right corner.}}
    \label{fig:integrated_intensity_12co}
\end{figure*}

\begin{figure*}[!ht]
    \centering
    \includegraphics[width=1.0\textwidth]{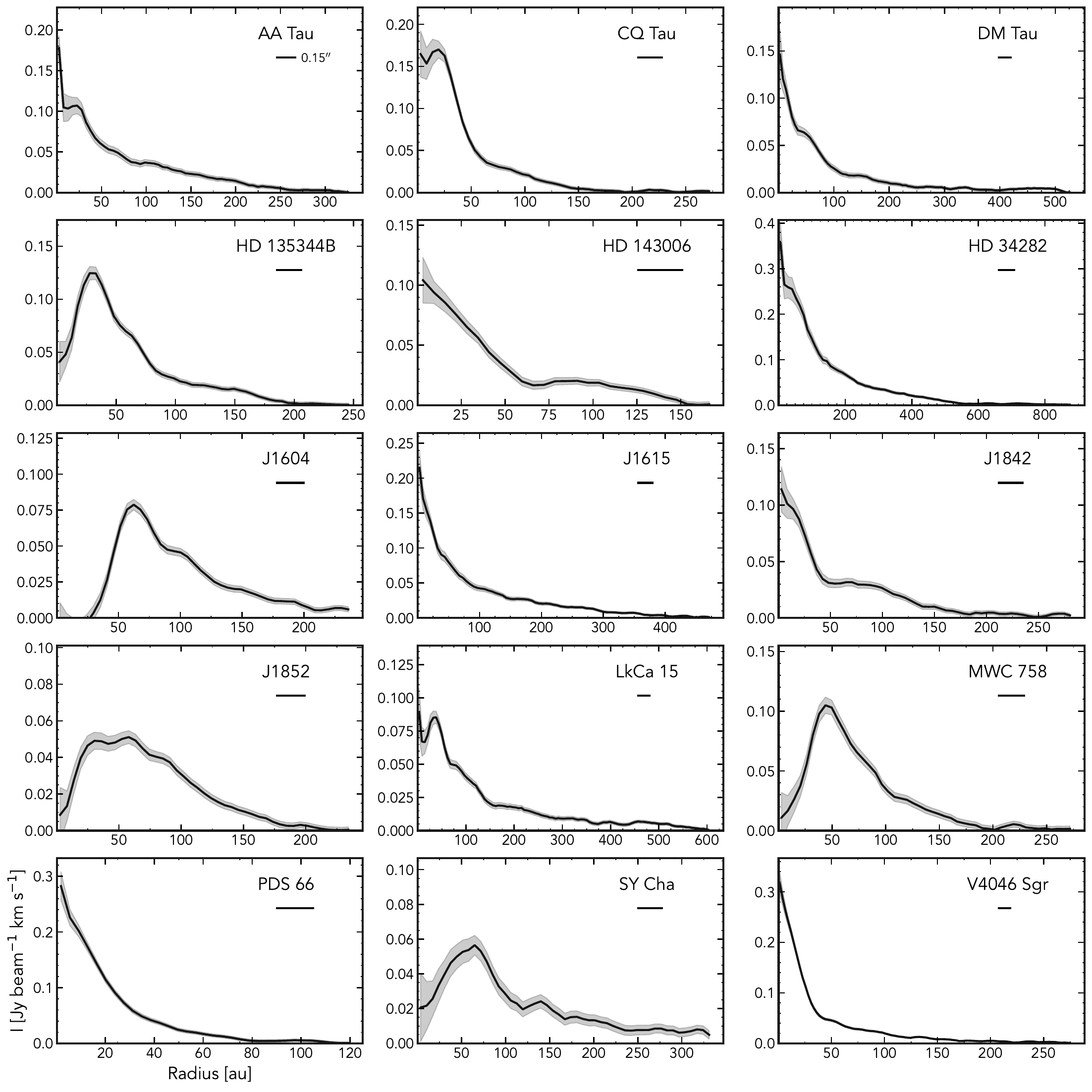}
    \caption{The same as Figure \ref{fig:integrated_intensity_12co}, but for $^{13}$CO $J=3-2$.}
    \label{fig:integrated_intensity_13co}
\end{figure*}

\begin{figure*}[!ht]
    \centering
    \includegraphics[width=1.0\textwidth]{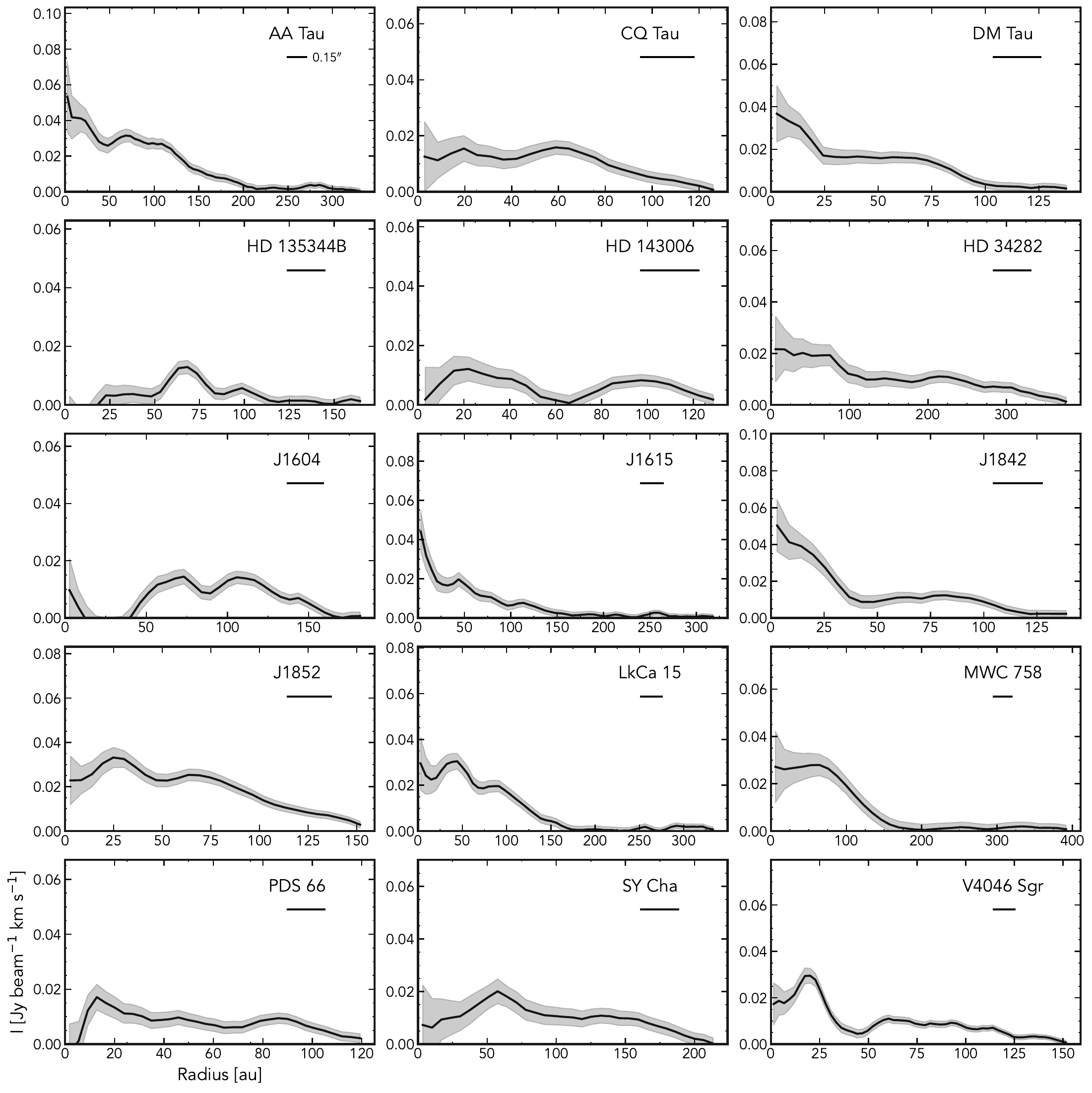}
    \caption{ The same as Figure \ref{fig:integrated_intensity_12co}, but for CS $J=7-6$.}
    \label{fig:integrated_intensity_cs}
\end{figure*}

\begin{figure*}
    \centering
    \includegraphics[width=1.0\textwidth]{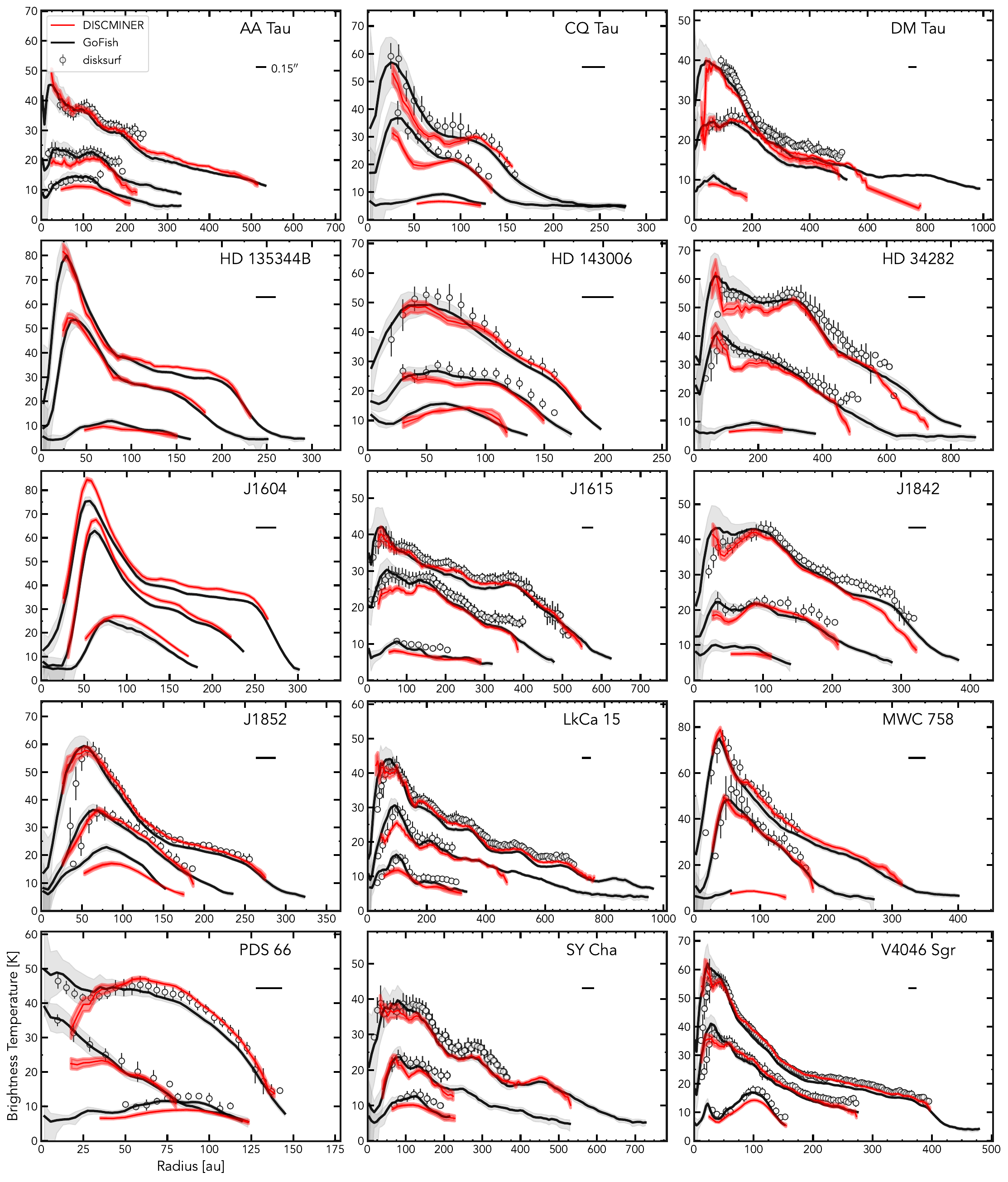}
    \caption{Peak intensities derived from \gofish{} (black lines), \disksurf{} (white circles), and \discminer{} (red lines). The beamsize of $0\farcs15$  is shown in the upper right of each plot.}
    \label{fig:peak_intensity_compare}
\end{figure*}

Figure \ref{fig:peak_intensity_structure} shows the sources with identified peak intensity substructures. 
Figures \ref{fig:integrated_intensity_12co}, \ref{fig:integrated_intensity_13co}, and \ref{fig:integrated_intensity_cs} show the integrated intensity in Jy beam$^{-1}$ km s$^{-1}$, calculated using the fiducial image cubes with \gofish{}. We leave an analysis of the integrated intensity profiles for future work. Figure \ref{fig:peak_intensity_compare} compares the peak intensities retrieved from \gofish{} (black lines), \disksurf{} (black points), and \discminer{} (red lines). All peak intensities in this figure were derived using the Fiducial images (see \citealt{exoALMA_main} for image details), but the \discminer{} intensities were calculated using the continuum-subtracted cubes. The three codes used to derive the peak intensities are in general agreement, and nearly all are within error bars. Most the the intensities diverge in the inner disk, where beam-dilution causes the observed intensity to drop.  \cite{MAPS_emission_surfaces_2021ApJS..257....4L} hypothesized that the decrease in intensity could be due to the lines becoming optically thin, unresolved substructures, or emission being absorbed by dust. A combination of these effects is a possibility for each disk, but we save an in-depth of diagnosis of these individual disks for future work. \maria{A few of the disks presented in Figure \ref{fig:peak_intensity_structure} exhibit non-Kelperian motion, as discussed in \citealt{exoALMA_channel_maps}, \citealt{exoALMA_line_profiles}, and \citealt{exoALMA_rotation}. Nearly all non-Keplerian motion can impact the disk temperature structure; likewise any temperature variation can induce some \maria{non-Keplerian} motion. Because of this, we do not definitively associate any of the peak intensity substructures with the observed non-Keplerian motion. Additionally, because these radial profiles are an azimuthal average, only large-scale peak intensity perturbations will be detected. We refer the reader to future exoALMA papers for discussion of the signatures of velocity perturbations on peak intensity maps.}

\begin{table*}[!ht]
    \centering
    \label{tab:substructures}
    \caption{Emission surface substructures (gaps only) and peak intensity substructures (peaks only). The bolded substructures represent the radial location of the larger emission surface drops seen in AA Tau, SyCha, and LkCa 15, discussed in Section \ref{sec:diffuse_surfaces}. Errors are generically defined as the beamsize (0$\farcs$15). We exclude HD 135344 B and J1604, which we do not derive emission surfaces for.}
    \begin{tabular}{l c c | c c}
        \hline \hline

    Source  & Line & Emission Surface Dips (au) & Peak Intensity Bumps (au)  & Peak Intensity Dips (au) \\ \hline

    DM Tau & $^{12}$CO $J=3-2$ & 120$\pm{22}$, 183$\pm{22}$, 293$\pm{22}$ & -&-\\&$^{13}$CO $J=3-2$ & 194$\pm{22}$ & 127$\pm{22}$&89$\pm{22}$\\&CS $J=7-6$ & - & -&-\\ \hline
    AA Tau & $^{12}$CO $J=3-2$ & \textbf{235}$\pm{\textbf{20}}$ & 93$\pm{20}$, 187$\pm{20}$&86$\pm{20}$\\&$^{13}$CO $J=3-2$ & - & 119$\pm{20}$&86$\pm{20}$\\&CS $J=7-6$ & 90$\pm{20}$ & -&-\\ \hline

    LkCa 15 & $^{12}$CO $J=3-2$ & 
    200$\pm{24}$, 342$\pm{24}$  & 
    \makecell{ 176$\pm{24}$, 333$\pm{24}$, \\ 504$\pm{24}$, 651$\pm{24}$} & 
    \makecell{ 302$\pm{24}$, 451$\pm{24}$ \\ 604$\pm{24}$, 781$\pm{24}$}
    \\    
    &$^{13}$CO $J=3-2$ & 119$\pm{24}$, \textbf{282}$\pm{\textbf{24}}$ & 203$\pm{24}$&174$\pm{24}$\\&CS $J=7-6$ & 177$\pm{24}$ & 192$\pm{24}$&-\\ \hline
    
    
    HD 34282 & $^{12}$CO $J=3-2$ & - & 295$\pm{46}$&214$\pm{46}$\\&$^{13}$CO $J=3-2$ & - & -&-\\&CS $J=7-6$ & - & -&-\\ \hline
    MWC 758 & $^{12}$CO $J=3-2$ & 56$\pm{23}$ & -&-\\&$^{13}$CO $J=3-2$ & 62$\pm{23}$ & -&-\\&CS $J=7-6$ & 62$\pm{23}$ & -&-\\ \hline
    CQ Tau & $^{12}$CO $J=3-2$ & 101$\pm{22}$ & 125$\pm{22}$&-\\&$^{13}$CO $J=3-2$ & 90$\pm{22}$ & -&-\\&CS $J=7-6$ & - & -&-\\ \hline
    SY Cha & $^{12}$CO $J=3-2$ & 220$\pm{27}$, \textbf{354}$\pm{\textbf{27}}$ & 276$\pm{27}$, 447$\pm{27}$&242$\pm{27}$, 406$\pm{27}$\\&$^{13}$CO $J=3-2$ & - & 150$\pm{27}$&113$\pm{27}$, 222$\pm{27}$\\&CS $J=7-6$ & - & 125$\pm{27}$&-\\ \hline
    PDS 66 & $^{12}$CO $J=3-2$ & - & -&-\\&$^{13}$CO $J=3-2$ & - & -&-\\&CS $J=7-6$ & - & -&-\\ \hline
    HD 135344B & $^{12}$CO $J=3-2$ & - & -&-\\&$^{13}$CO $J=3-2$ & - & -&-\\&CS $J=7-6$ & - & -&-\\ \hline
    HD 143006 & $^{12}$CO $J=3-2$ & - & -&-\\&$^{13}$CO $J=3-2$ & 112$\pm{25}$ & -&-\\&CS $J=7-6$ & - & -&-\\ \hline
    RXJ1604.3-2130 A & $^{12}$CO $J=3-2$ & - & -&-\\&$^{13}$CO $J=3-2$ & - & -&-\\&CS $J=7-6$ & - & -&-\\ \hline
    RXJ1615.3-3255 & $^{12}$CO $J=3-2$ & 472$\pm{23}$ & 366$\pm{23}$&-\\&$^{13}$CO $J=3-2$ & 250$\pm{23}$ & 132$\pm{23}$&108$\pm{23}$\\&CS $J=7-6$ & 113$\pm{23}$ & 132$\pm{23}$&120$\pm{23}$\\ \hline
    V4046 Sgr & $^{12}$CO $J=3-2$ & 173$\pm{11}$, 318$\pm{11}$ & -&-\\&$^{13}$CO $J=3-2$ & 174$\pm{11}$ & -&-\\&CS $J=7-6$ & 91$\pm{11}$ & 98$\pm{11}$&42$\pm{11}$\\ \hline
    RXJ1842.9-3532 & $^{12}$CO $J=3-2$ & 138$\pm{23}$ & 88$\pm{23}$&59$\pm{23}$\\&$^{13}$CO $J=3-2$ & 83$\pm{23}$, 133$\pm{23}$ & 88$\pm{23}$&65$\pm{23}$\\&CS $J=7-6$ & - & -&-\\ \hline
    RXJ1852.3-3700 & $^{12}$CO $J=3-2$ & 117$\pm{22}$, 161$\pm{22}$ & -&-\\&$^{13}$CO $J=3-2$ & 124$\pm{22}$ & -&-\\&CS $J=7-6$ & - & -&-\\ \hline

    \end{tabular}
\end{table*}

\clearpage

\begin{table}[!ht]
    \centering
    \caption{Outer disk radii, R$_{\rm edge}$, and the radii which encompass 95\%, 90\%, and 68\% of the total flux.}
    \label{tab:gas_disk_radii}
    \begin{tabular}{cccccc}
    \hline \hline
     Source  & Line & R$_{\rm edge}$ (au) & R$_{95}$ (au) & R$_{90}$ (au) &  R$_{68}$ (au) \\ \hline

    DM Tau & $^{12}$CO $J=3-2$ & 991$\pm$92 & 856$^{+54}_{-49}$ & 786$^{+32}_{-27}$ & 580$^{+27}_{-32}$ \\ 
     ~ & $^{13}$CO $J=3-2$ & 532$\pm$22 & 483$^{+27}_{-22}$ & 462$^{+16}_{-22}$ & 310$^{+22}_{-22}$ \\ 
     ~ & CS $J=7-6$ & 143$\pm$22 & 121$^{+16}_{-32}$ & 111$^{+27}_{-27}$ & 78$^{+5}_{-11}$ \\ \hline
    AA Tau & $^{12}$CO $J=3-2$ & 644$\pm$10 & 498$^{+141}_{-56}$ & 442$^{+56}_{-51}$ & 265$^{+20}_{-20}$ \\ 
     ~ & $^{13}$CO $J=3-2$ & 331$\pm$20 & 265$^{+61}_{-30}$ & 235$^{+30}_{-20}$ & 164$^{+10}_{-5}$ \\ 
     ~ & CS $J=7-6$ & 336$\pm$141 & 285$^{+45}_{-20}$ & 265$^{+20}_{-45}$ & 154$^{+5}_{-10}$ \\ \hline
    LkCa 15 & $^{12}$CO $J=3-2$ & 964$\pm$12 & 799$^{+53}_{-59}$ & 699$^{+41}_{-29}$ & 456$^{+18}_{-18}$ \\ 
     ~ & $^{13}$CO $J=3-2$ & 946$\pm$18 & 634$^{+65}_{-83}$ & 533$^{+35}_{-24}$ & 386$^{+12}_{-24}$ \\ 
     ~ & CS $J=7-6$ & 339$\pm$206 & 309$^{+24}_{-35}$ & 292$^{+29}_{-88}$ & 127$^{+12}_{-12}$ \\ \hline
    HD 34282 & $^{12}$CO $J=3-2$ & 840$\pm$0 & 678$^{+151}_{-81}$ & 597$^{+81}_{-46}$ & 422$^{+23}_{-23}$ \\ 
     ~ & $^{13}$CO $J=3-2$ & 886$\pm$23 & 701$^{+174}_{-185}$ & 562$^{+174}_{-104}$ & 342$^{+35}_{-23}$ \\ 
     ~ & CS $J=7-6$ & 388$\pm$162 & 342$^{+35}_{-35}$ & 319$^{+35}_{-23}$ & 261$^{+12}_{-23}$ \\ \hline
    MWC 758 & $^{12}$CO $J=3-2$ & 406$\pm$0 & 301$^{+99}_{-47}$ & 266$^{+53}_{-29}$ & 172$^{+18}_{-18}$ \\ 
     ~ & $^{13}$CO $J=3-2$ & 278$\pm$12 & 219$^{+53}_{-58}$ & 172$^{+58}_{-23}$ & 114$^{+12}_{-12}$ \\ 
     ~ & CS $J=7-6$ & 403$\pm$58 & 357$^{+35}_{-70}$ & 333$^{+58}_{-94}$ & 146$^{+105}_{-35}$ \\ \hline
    CQ Tau & $^{12}$CO $J=3-2$ & 277$\pm$0 & 204$^{+67}_{-78}$ & 170$^{+101}_{-56}$ & 109$^{+34}_{-34}$ \\ 
     ~ & $^{13}$CO $J=3-2$ & 277$\pm$11 & 221$^{+50}_{-95}$ & 165$^{+106}_{-50}$ & 92$^{+22}_{-11}$ \\ 
     ~ & CS $J=7-6$ & 131$\pm$129 & 109$^{+17}_{-17}$ & 103$^{+22}_{-17}$ & 81$^{+6}_{-11}$ \\ \hline
    SY Cha & $^{12}$CO $J=3-2$ & 734$\pm$7 & 604$^{+123}_{-61}$ & 543$^{+61}_{-55}$ & 378$^{+20}_{-27}$ \\ 
     ~ & $^{13}$CO $J=3-2$ & 529$\pm$96 & 461$^{+61}_{-130}$ & 358$^{+109}_{-48}$ & 242$^{+20}_{-20}$ \\ 
     ~ & CS $J=7-6$ & 222$\pm$34 & 188$^{+27}_{-14}$ & 174$^{+27}_{-7}$ & 147$^{+7}_{-7}$ \\ \hline
    PDS 66 & $^{12}$CO $J=3-2$ & 149$\pm$0 & 119$^{+26}_{-22}$ & 112$^{+33}_{-22}$ & 78$^{+15}_{-11}$ \\ 
     ~ & $^{13}$CO $J=3-2$ & 123$\pm$4 & 97$^{+22}_{-44}$ & 83$^{+37}_{-33}$ & 50$^{+18}_{-15}$ \\ 
     ~ & CS $J=7-6$ & 123$\pm$22 & 112$^{+7}_{-11}$ & 105$^{+11}_{-7}$ & 86$^{+4}_{-4}$ \\ \hline
    HD 135344B & $^{12}$CO $J=3-2$ & 291$\pm$0 & 215$^{+71}_{-25}$ & 195$^{+30}_{-15}$ & 144$^{+15}_{-10}$ \\ 
     ~ & $^{13}$CO $J=3-2$ & 251$\pm$5 & 175$^{+71}_{-20}$ & 159$^{+30}_{-15}$ & 114$^{+10}_{-15}$ \\ 
     ~ & CS $J=7-6$ & 170$\pm$20 & 154$^{+10}_{-35}$ & 139$^{+25}_{-30}$ & 99$^{+5}_{-10}$ \\ \hline
    HD 143006 & $^{12}$CO $J=3-2$ & 204$\pm$0 & 160$^{+38}_{-44}$ & 147$^{+50}_{-38}$ & 103$^{+19}_{-25}$ \\ 
     ~ & $^{13}$CO $J=3-2$ & 173$\pm$13 & 135$^{+31}_{-19}$ & 129$^{+38}_{-19}$ & 97$^{+13}_{-13}$ \\ 
     ~ & CS $J=7-6$ & 135$\pm$100 & 122$^{+6}_{-13}$ & 116$^{+13}_{-13}$ & 97$^{+6}_{-0}$ \\ \hline
    RXJ1604.3-2130 A & $^{12}$CO $J=3-2$ & 306$\pm$11 & 247$^{+22}_{-11}$ & 236$^{+11}_{-11}$ & 176$^{+5}_{-5}$ \\ 
     ~ & $^{13}$CO $J=3-2$ & 241$\pm$11 & 214$^{+22}_{-16}$ & 192$^{+22}_{-11}$ & 138$^{+11}_{-5}$ \\ 
     ~ & CS $J=7-6$ & 187$\pm$195 & 149$^{+33}_{-11}$ & 144$^{+11}_{-11}$ & 122$^{+0}_{-5}$ \\ \hline
    RXJ1615.3-3255 & $^{12}$CO $J=3-2$ & 722$\pm$0 & 553$^{+99}_{-35}$ & 506$^{+29}_{-23}$ & 365$^{+18}_{-12}$ \\ 
     ~ & $^{13}$CO $J=3-2$ & 477$\pm$12 & 395$^{+64}_{-41}$ & 348$^{+35}_{-23}$ & 249$^{+12}_{-12}$ \\ 
     ~ & CS $J=7-6$ & 325$\pm$64 & 290$^{+29}_{-41}$ & 260$^{+58}_{-41}$ & 155$^{+41}_{-23}$ \\ \hline
    V4046 Sgr & $^{12}$CO $J=3-2$ & 481$\pm$0 & 369$^{+48}_{-27}$ & 334$^{+21}_{-21}$ & 215$^{+13}_{-11}$ \\ 
     ~ & $^{13}$CO $J=3-2$ & 278$\pm$62 & 240$^{+35}_{-29}$ & 210$^{+29}_{-21}$ & 130$^{+8}_{-8}$ \\ 
     ~ & CS $J=7-6$ & 154$\pm$13 & 138$^{+8}_{-8}$ & 130$^{+5}_{-8}$ & 101$^{+5}_{-3}$ \\ \hline
    RXJ1842.9-3532 & $^{12}$CO $J=3-2$ & 388$\pm$0 & 314$^{+68}_{-40}$ & 280$^{+45}_{-28}$ & 195$^{+17}_{-17}$ \\ 
     ~ & $^{13}$CO $J=3-2$ & 286$\pm$11 & 258$^{+23}_{-51}$ & 224$^{+45}_{-40}$ & 133$^{+17}_{-11}$ \\ 
     ~ & CS $J=7-6$ & 144$\pm$23 & 127$^{+11}_{-28}$ & 110$^{+28}_{-17}$ & 88$^{+11}_{-11}$ \\ \hline
    RXJ1852.3-3700 & $^{12}$CO $J=3-2$ & 322$\pm$0 & 256$^{+61}_{-44}$ & 223$^{+44}_{-28}$ & 140$^{+11}_{-11}$ \\ 
     ~ & $^{13}$CO $J=3-2$ & 240$\pm$33 & 179$^{+55}_{-22}$ & 157$^{+28}_{-17}$ & 113$^{+6}_{-6}$ \\ 
     ~ & CS $J=7-6$ & 157$\pm$22 & 141$^{+11}_{-17}$ & 130$^{+17}_{-11}$ & 96$^{+6}_{-6}$ \\ \hline

            \end{tabular}
\end{table}

\clearpage
\section{Comparing Disksurf to DISCMINER}\label{sec:app_disksurfvsdiscminer}

The exoALMA collaboration utilized two open-source python codes for the analysis of the emission surfaces: \disksurf{} \citep{disksurf_2021JOSS....6.3827T} and \discminer{} \citep{discminer1_2021A&A...650A.179I, discminer2_2023A&A...674A.113I}. As discussed in Section \ref{sec:surfaces}, the emission surfaces retrieved between the two are generally consistent, but there are notable differences (see Figures \ref{fig:surfaces_12co}, \ref{fig:surfaces_13co}, and \ref{fig:surfaces_cs}). Below, discuss and demonstrate the differences between \disksurf{} and \discminer{} on simulated and real data.

\begin{figure}[!ht]
    \centering
    \includegraphics[width=1.0\textwidth]{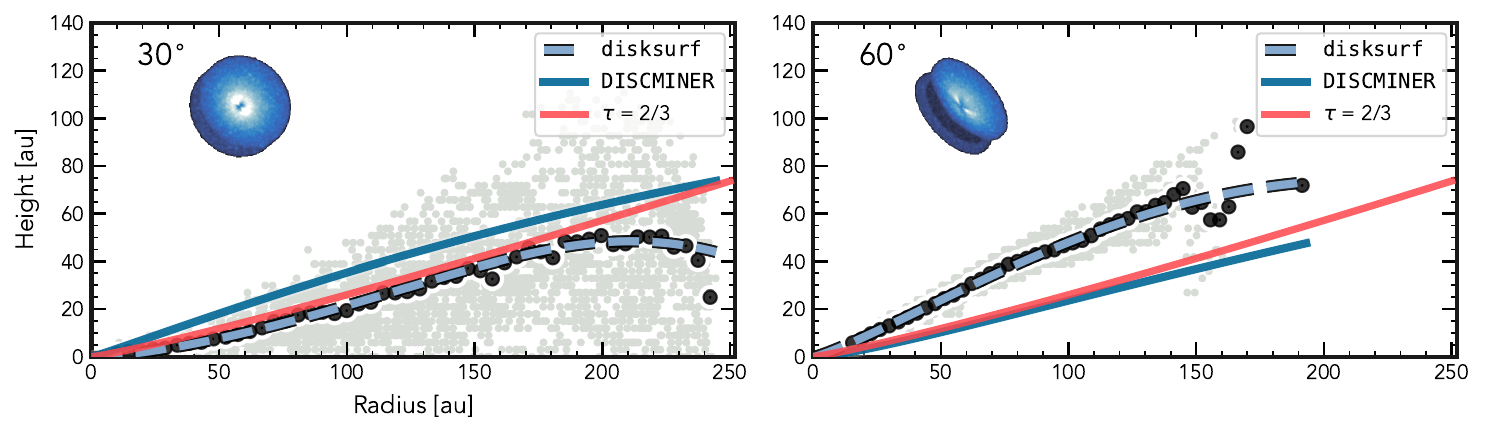}
    \includegraphics[width=1.0\textwidth]{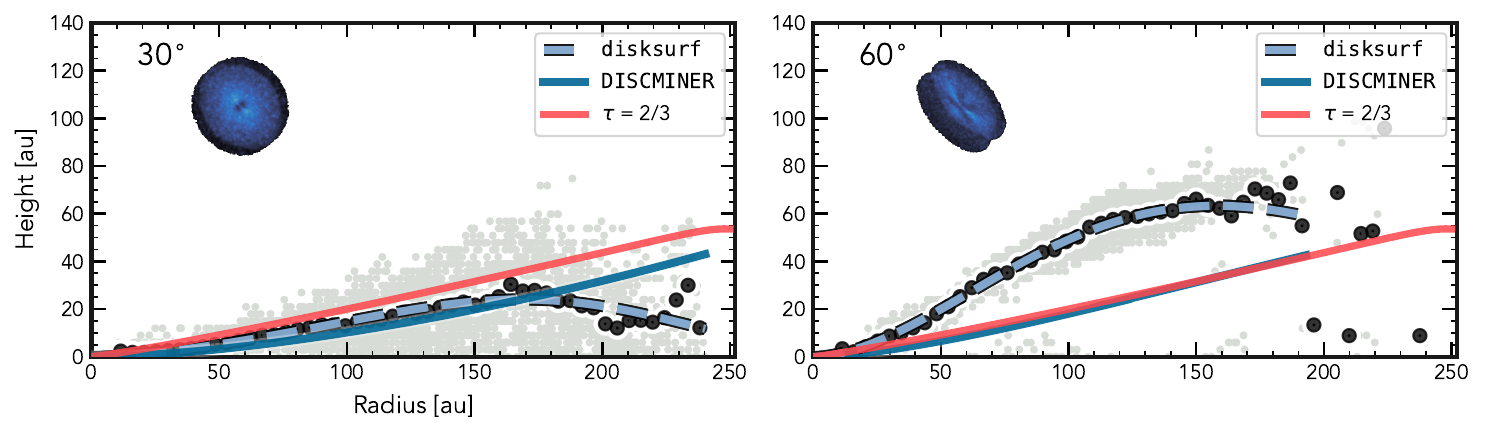}
    \caption{Emission surface comparison between \disksurf{} (dashed line and background points) and \discminer{} (solid blue line) for two model cubes at 30 degrees and 60 degrees, respectively. The model cubes are identical besides their inclination. The red line denotes the known $\tau$ = 2/3 surface. The top two panels show the results for $^{12}$CO $J=3-2$, and the bottom two panels show the results for $^{13}$CO $J=3-2$.}
    \label{fig:emission_surface_MCFOST}
\end{figure}

Figure \ref{fig:emission_surface_MCFOST} presents the results of this test on two test cubes with varying inclination and molecular line emission. The left panels shows the results from the 30 degree inclined simulated disk, and the right panel shows the same results from the same procedure, but on a 60 degree inclined simulated disk. The top two plots are for $^{12}$CO $J=3-2$ emission, and the bottom two plots show the results for $^{13}$CO $J=3-2$ emission.
The red line in both of the plots depicts the $\tau$ = 2/3 surface at an inclination of zero; therefore, the $\tau$ = 2/3 surface in these plots represents more of an upper limit. The simulated cubes were made with MCFOST \citep{mcfost1_2006A&A...459..797P, mcfost2_2009A&A...498..967P} and have identical parameters apart from the inclination and molecular composition. It is immediately evident that the inclination has an impact on the derived surface for both codes. For the 30 degree inclined disk, the \disksurf{} surface aligns more closely with the $\tau$ = 2/3 surface; for the 60 degree inclined disk, the opposite is true, with the \discminer{} surface falling closely in-line with the $\tau$ = 2/3 surface. The \disksurf{} surface scatter in the z-direction also changes between the two cubes, as discussed in Section \ref{subsec:results_surfaces}. More highly inclined disks will have a larger surface scatter when using the method implemented in \disksurf{}, due to the fact that the front and back emission layers are less separated, causing some points that belong below the front surface to be picked up. 

It is critical to interpret these results from simulated cubes carefully. For the 30 degree inclined disk, it appears that \disksurf{} does a better job at finding the $\tau$ = 2/3 surface, and vice versa for the 60 degree inclined disk. However, the parametric surfaces for the real data, shown in Figures \ref{fig:surfaces_12co}, \ref{fig:surfaces_13co}, and \ref{fig:surfaces_cs}, do not show large differences between \disksurf{} and \discminer{} derived tapered power laws. Some of the largest differences in parametric surfaces are seen in J1615 $^{12}$CO emission, where radial location of the tapering between \disksurf{} and \discminer{} differ by $\sim$130 au, and the vertical height of the emission surface found by 
\discminer{} is systematically lower. The other disks that show the largest differences between the parametric surfaces are those that suffer from midplane, backside, or diffuse contamination, namely AA Tau, DM Tau, and SY Cha. Finally, disks with mostly flat surfaces, like those of MWC 758 ($^{13}$CO), PDS 66 ($^{13}$CO, CS), and V4046 Sgr (CS), also suffer from some differences between the two codes. However, unlike the results from the MCFOST cubes, nearly all of the \discminer{} parametric surfaces align with the \disksurf{} surfaces and are within the area shown by the raw $r-z$ points for the real data. This may indicate that the two codes function differently with simulated cubes, but have similar performance on real data.

The root cause of the underlying differences between the surfaces, both for the real data and simulated cubes, is still unknown. It is likely that there are several factors at play, and disentangling their individual impact on the retrieved surfaces is beyond the scope of this paper. As discussed in Section \ref{subsec:parametric_surfaces}, there are several factors that may contribute to the differences. First is the innate differences between how \disksurf{} and \discminer{} work; \disksurf{} utilizes the method outlined in \cite{pinte_2018A&A...609A..47P} to derive $r-z$ points, which relies on finding peaks along isovelocity contours. If the front and back surfaces are not well separated, this procedure can easily confuse points belonging to regions below the emission surface as part of the upper surface, which can lead to artificially lower surfaces. Additionally, since \disksurf{} is reliant on intensity peaks, the shape of the line profile can induce biases. On the other hand, \discminer{} uses line-profile analysis to distinguish whether emission belongs to the front or back surface (see Figure 3 from \citealt{discminer1_2021A&A...650A.179I}), so the surfaces derived by \discminer{} should not be as contaminated with midplane and back-side emission. Another potential cause for differences is the velocity; \disksurf{} assumes purely Keplerian rotation, whereas \discminer{} takes into account any velocity deviations during modeling. This could potentially explain some of the differences seen in the surfaces, but even so, J1615, which has one of the biggest differences in parametric surfaces, does not have velocity deviations that are significantly larger than those of other disks (see \citealt{exoALMA_rotation}).

\begin{figure}[!ht]
    \centering
    \includegraphics[width=1.0\textwidth]{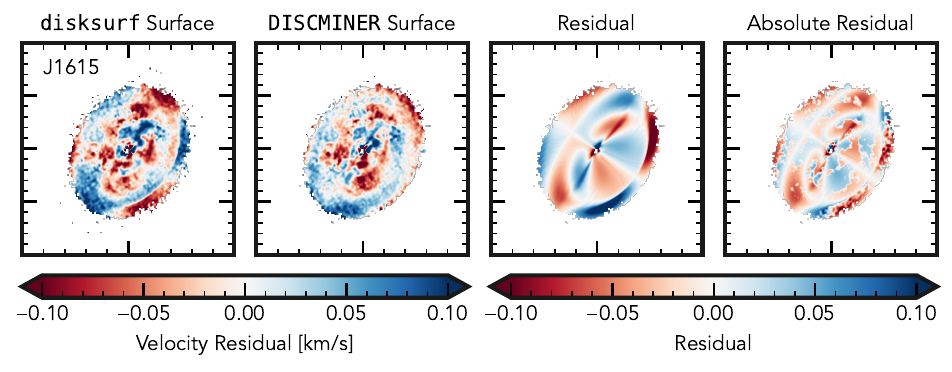}
    \includegraphics[width=1.0\textwidth]{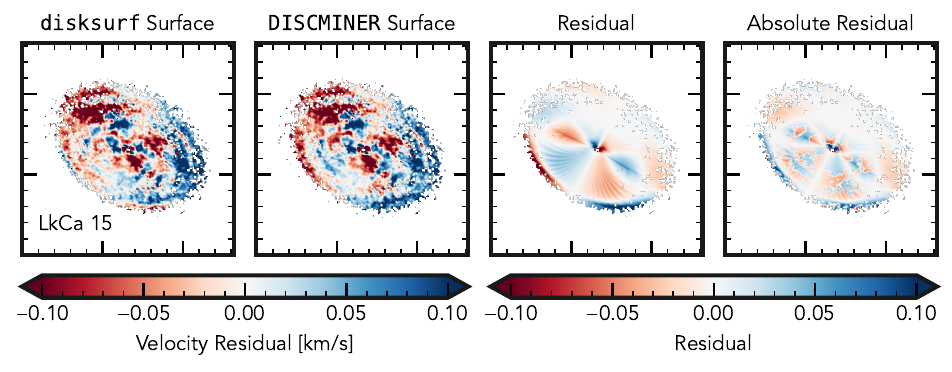}
    \caption{Velocity residuals calculated with \discminer{} for J1615 (top panels) and LkCa 15 (bottom panels), derived with the \disksurf{} parametric emission surface and the \discminer{} parametric emission surface. We subtract the velocity residuals (labeled `Residual'), and subtract the absolute value of the velocity residuals (labeled `Absolute Residual').}
    \label{fig:velocity_residuals}
\end{figure}

Figure \ref{fig:velocity_residuals} shows velocity residuals for J1615 (top panels) and LkCa 15 (bottom panels) calculated with \discminer{}, one using the \disksurf{} parametric emission surface and the other using the \discminer{} surface. We take two residuals to compare the resulting differences; the first residual is simply a subtraction between residuals, and the second residual is a subtraction of the absolute values of the residuals. The differences between the results can be seen in several areas of the disks. The J1615 emission surfaces differ significantly (as seen in Figure \ref{fig:surfaces_12co}), and this effect can be seen in the velocity residuals. The \disksurf{} surface tapers more sharply, resulting in stronger velocity residuals in the outer portions of the disk. Additionally, the \disksurf{} surface is shifted compared to that of \discminer{}, resulting in overall stronger residuals throughout the disk. The strongest residuals can be seen in the far side of the disk. The J1615 surfaces are some of the most discrepant between \disksurf{} and \discminer{}; clearly, the velocity residuals are sensitive to the chosen emission surfaces. For LkCa 15, there are slight differences between the \disksurf{} and \discminer{} surfaces, but not as large as those seen in J1615. Subsequently, the differences in the residuals of LkCa 15 are not as severe. The sharper tapering of the \disksurf{} surface again leads to larger residuals towards the farthest minor axis of the disk.

The question of which surface to utilize will depend on the individual applications. The \disksurf{} non-parametric surfaces allow for the characterization of substructures within the surfaces such as bumps and dips; these can be used in chemical models to account for areas of lowered column density, and are helpful visual indicators for gaseous emission structures that may not have been seen on a visual inspection of the datacube. Additionally, the \disksurf{} surfaces can be more readily used for the calculation of surface errors because they provide \textit{scatter} in the z-direction, which is more representative of physical reality.
The \discminer{} surfaces are useful when parametric surfaces are needed, particularly for velocity residual analysis. Because \discminer{} models multiple components of the disk at once, including the emission surfaces, using these parametric surfaces derived during the modeling is the most consistent method when utilizing other \discminer{} outputs. If the \disksurf{} surface is artificially imposed in \discminer{}, this can lead to nonphysical residuals, because \discminer{} relies on the surface modeling to minimize these residuals.

\clearpage
\bibliography{sample631}{}
\bibliographystyle{aasjournal}

\end{document}